\documentclass[aps,preprint,floats,epsf,epsfig,nofootinbib,letter]{revtex4}

\usepackage{graphicx}% Include figure files
\usepackage{dcolumn}% Align table columns on decimal point
\usepackage{bm}% bold math

\begin{document}
\def\be{\begin{eqnarray}}
\def\en{\end{eqnarray}}
\def\non{\nonumber}
\def\la{\langle}
\def\ra{\rangle}
\def\nc{N_c^{\rm eff}}
\def\vp{\varepsilon}
\def\ep{\epsilon}
\def\drho{\bar\rho}
\def\deta{\bar\eta}
\def\A{{\cal A}}
\def\B{{\cal B}}
\def\c{{\cal C}}
\def\d{{\cal D}}
\def\e{{\cal E}}
\def\p{{\cal P}}
\def\t{{\cal T}}
\def\CP{{\it CP}~}
\def\up{\uparrow}
\def\dw{\downarrow}
\def\vma{{_{V-A}}}
\def\vpa{{_{V+A}}}
\def\smp{{_{S-P}}}
\def\spp{{_{S+P}}}
\def\J{{J/\psi}}
\def\ov{\overline}
\def\Lqcd{{\Lambda_{\rm QCD}}}
\def\pr{{Phys. Rev.}~}
\def\prl{{Phys. Rev. Lett.}~}
\def\pl{{Phys. Lett.}~}
\def\np{{Nucl. Phys.}~}
\def\zp{{Z. Phys.}~}
\def\lsim{ {\ \lower-1.2pt\vbox{\hbox{\rlap{$<$}\lower5pt\vbox{\hbox{$\sim$}
}}}\ } }
\def\gsim{ {\ \lower-1.2pt\vbox{\hbox{\rlap{$>$}\lower5pt\vbox{\hbox{$\sim$}
}}}\ } }

%\font\el=cmbx10 scaled \magstep2{\obeylines\hfill October, 2010}

\begin{flushright}
{\small
CYCU-HEP-10-16 \\[0.1cm]
October, 2010}
\end{flushright}

%\vskip 0.5 cm

\centerline{\large\bf Charmless Hadronic $B$ Decays into a Tensor Meson}
\bigskip
\centerline{\bf Hai-Yang Cheng$^{1,2}$ and Kwei-Chou Yang$^{3}$}
\bigskip
\centerline{$^1$ Institute of Physics, Academia Sinica}
\centerline{Taipei, Taiwan 115, Republic of China}
\medskip
\centerline{$^2$ C.N. Yang Institute for Theoretical Physics,
State University of New York} \centerline{Stony Brook, New York
11794}
\medskip
\centerline{$^3$ Department of Physics, Chung Yuan Christian
University} \centerline{Chung-Li, Taiwan 320, Republic of China}
\bigskip
%\bigskip

\centerline{\bf Abstract}
\bigskip
\small

Two-body charmless hadronic $B$ decays involving a tensor meson
in the final state are studied within the framework of QCD factorization (QCDF). Due to the $G$-parity of the tensor meson, both the chiral-even and chiral-odd two-parton light-cone distribution amplitudes of the tensor meson are antisymmetric under the interchange of momentum fractions of the quark and anti-quark in the SU(3) limit. Our main results are:
(i) In the naive factorization approach, the decays such as $B^-\to \bar K_2^{*0}\pi^-$ and $\ov B^0\to K_2^{*-}\pi^+$ with a tensor meson emitted are prohibited owing to the fact that a tensor meson cannot be created from the local $V-A$ or tensor current. Nevertheless, they receive nonfactorizable contributions in QCDF from vertex, penguin and hard spectator corrections. The experimental observation of $B^-\to \bar K_2^{*0}\pi^-$ indicates the importance of nonfactorizable effects.
%%%
(ii) For penguin-dominated $B\to TP$ and $TV$ decays, the predicted rates in naive factorization are usually too small by one to two orders of magnitude. In QCDF, they are enhanced by power corrections from penguin annihilation and nonfactorizable contributions.
(iii) The dominant penguin contributions to $B\to K_2^*\eta^{(')}$ arise from the processes:  (a) $b\to ss\bar s\to s\eta_s$ and (b) $b\to s q\bar q\to q \bar K_2^*$ with $\eta_q=(u\bar u+d\bar d)/\sqrt{2}$ and $\eta_s=s\bar s$. The interference, constructive for $K_2^*\eta'$ and destructive for $K_2^*\eta$, explains why $\Gamma(B\to K_2^*\eta')\gg\Gamma(B\to K_2^*\eta)$.
(iv) We use the measured rates of $B\to K_2^*(\omega,\phi)$ to extract the penguin-annihilation parameters $\rho_A^{TV}$ and $\rho_A^{VT}$ and the observed longitudinal polarization fractions $f_L(K_2^*\omega)$ and $f_L(K_2^*\phi)$ to fix the phases $\phi_A^{VT}$ and $\phi_A^{TV}$.
(v) The experimental observation that $f_T/f_L\ll 1$ for $B\to  K_2^*(1430)\phi$, whereas $f_T/f_L\sim 1$ for $B\to  K_2^*(1430)\omega$ with $f_T$ being the transverse polarization fraction, can be
{\it accommodated}  in QCDF, but cannot be {\it dynamically explained} at first place. For penguin-dominated $B\to TV$ decays, we find $f_L(K_2^*\rho)\sim f_L(K_2^*\omega)\sim 0.65$\,, whereas $f_L(K^*f_2)\sim 0.93$. It will be of great interest to measure $f_L$ for these modes to test QCDF. Theoretically,
transverse polarization is expected to be small in tree-dominated $\ov B\to TV$ decays except for the $a_2^-\rho^0,~a_2^-\rho^+,~K_2^{*0}K^{*-}$ and $K_2^{*0}\bar
K^{*0}$ modes.
 (vi) For tree-dominated decays, their rates are usually very small except for the $a_2^0(\pi^-,\rho^-),~a_2^+(\pi^-,\rho^-)$ and $f_2(\pi^-,\rho^-)$ modes with branching fractions of order $10^{-6}$ or even bigger.

\pagebreak

\section{Introduction}

In the past few years, BaBar and Belle
\cite{BaBar:etapK2p,BaBar:etaK2p,BaBar:omegaK2p,BaBar:f2Kp,BaBar:f2pKp,BaBar:phiK2p,BaBar:f2pip,BaBar:K20pi0,BaBar:f2K0,BaBar:phiK20,Belle:f2Kp,Belle:f2pKp,Belle:K20Kp,Belle:f2K0}
have begun to measure several charmless $B$ decay modes involving a light tensor meson $T$ in the final states with the results summarized in Table \ref{tab:expt}. From the theoretical point of view, the hadronic decays $B\to TM$ with $M=P,V,A$ are of great interest for two reasons: rate deficit and polarization puzzles. First, these decays have been studied in the naive factorization approach \cite{Katoch,Munoz97,Munoz:99,Kim:2002a,Kim:2002b,Kim:2003,Munoz,Verma,Sharma}.
The predicted rates are in general too small by one to two orders of magnitude.  This implies the importance of $1/m_b$ power corrections.
Since the nonfactorizable amplitudes such as vertex and penguin corrections, spectator interactions cannot be tackled in naive factorization, it is necessary to go beyond the naive factorization framework.  The
theoretical frameworks suitable for this purpose include QCD factorization \cite{BBNS}, perturbative QCD (pQCD) \cite{pQCD} and soft-collinear effective theory (SCET) \cite{SCET}

Second, it is known that an unexpectedly large fraction of transverse polarization has been observed in the penguin-dominated $B\to VV$ channels, such as $B\to \phi K^*, \rho K^*$, contrary to the naive expectation of the longitudinal polarization dominance (for a review, see \cite{ChengSmith}).  However, while the polarization measurement in $B\to \omega K_2^*(1430)$ indicates a large fraction of transverse polarization $f_T$ (see Table  \ref{tab:expt}), the measurement in $B\to \phi K_2^*(1430)$ is consistent with the longitudinal polarization dominance. Therefore, it is important to understand why $f_T/f_L\ll 1$ for $B\to \phi K_2^*(1430)$, whereas $f_T/f_L\sim 1$ for $B\to \omega K_2^*(1430)$, even though both are penguin-dominated.
The polarization studies for $B\to TV, TA, TT$  will further shed light on the underlying helicity structure of the decay mechanism.

In the present work we shall study charmless $B\to TM$  decays within the framework of QCD factorization. One unique feature of the tensor meson is that it cannot be created from the $V-A$ or tensor current.  Hence, the decay with a tensor meson emitted, for example, $B^-\to \bar K_2^{*0}\pi^-$, is prohibited in naive factorization. The experimental observation of this penguin-dominated mode with a sizable rate implies the importance of nonfactorizable effects which will be addressed in QCDF.

The layout of this work is as follows. We study the physical properties of tensor mesons such as decay constants, form factors, light-cone distribution amplitudes and helicity projection operators in Sec. 2 and specify various input parameters. Then we work out in details the next-to-leading order (NLO) corrections to $B\to TP,TV$ decays  in Sec. 3 and present numerical results and discussions in Sec. 4.  Conclusions are given in Sec. 5. Appendix A is devoted to a recapitulation of the ISGW model. Decay amplitudes and  explicit expressions of helicity-dependent annihilation amplitudes are shown in Appendices B and C, respectively. A mini review of of the $\eta-\eta'$ mixing is given in Appendix D.

\begin{table}[t]
\caption{Experimental branching fractions (in units of $10^{-6}$) and the longitudinal polarization fractions $f_L$ for
$B$ decays to final states containing a tensor meson. Data are taken from
\cite{BaBar:etapK2p,BaBar:etaK2p,BaBar:omegaK2p,BaBar:f2Kp,BaBar:f2pKp,BaBar:phiK2p,BaBar:f2pip,BaBar:K20pi0,BaBar:f2K0,BaBar:phiK20,Belle:f2Kp,Belle:f2pKp,Belle:K20Kp,Belle:f2K0}.}
\label{tab:expt}
\begin{ruledtabular}
\begin{tabular}{l c c|l  c c}
Mode & $\B$ & $f_L$ & Mode & $\B$ & $f_L$ \\
\hline
 $\B(B^+\to K_2^*(1430)^+\omega)$ & $21.5\pm4.3$ & $0.56\pm0.11$ & $\B(B^0\to K_2^*(1430)^0\omega)$ & $10.1\pm2.3$ & $0.45\pm0.12$ \\
 $\B(B^+\to K_2^*(1430)^+\phi)$ & $8.4\pm2.1$ &  $0.80\pm0.10$ & $\B(B^0\to K_2^*(1430)^0\phi)$ & $7.5\pm1.0$ & $0.901^{+0.059}_{-0.069}$ \\
 $\B(B^+\to K_2^*(1430)^+\eta)$ & $9.1\pm3.0$ &  & $\B(B^0\to K_2^*(1430)^0\eta)$ & $9.6\pm2.1$ \\
 $\B(B^+\to K_2^*(1430)^+\eta')$ & $28.0^{+5.3}_{-5.0}$ & & $\B(B^0\to K_2^*(1430)^0\eta')$ & $13.7^{+3.2}_{-3.1}$ \\
 $\B(B^+\to K_2^*(1430)^0\pi^+)$ & $5.6^{+2.2}_{-1.4}$ & & $\B(B^0\to K_2^*(1430)^+\pi^-)$ & $<6.3$  \\
 $\B(B^+\to K_2^*(1430)^0 K^+)$ & $<1.1$ &  & $\B(B^0\to K_2^*(1430)^0\pi^0)$ & $<4.0$ \\
 $\B(B^+\to f_2(1270)K^+)$ &
 $1.06^{+0.28}_{-0.29}$  & &  $\B(B^0\to f_2(1270)K^0)$ & $2.7^{+1.3}_{-1.2}$  \\
 $\B(B^+\to f_2(1270)\pi^+)$ & $1.57^{+0.69}_{-0.49}$ &  \\
 $\B(B^+\to f'_2(1525)K^+)$ & $<7.7$~\footnotemark[1] &  \\
 $\B(B^+\to a_2(1320)^0K^+)$ & $<45$ & \\
\end{tabular}
\end{ruledtabular}
 \footnotetext[1]{From the
measurement of $\B(B^+\to f'_2(1525)^0K^+\to K^+K^+K^-)<3.4\times 10^{-6}$ \cite{BaBar:f2pKp}.}
\end{table}

\section{Physical properties of tensor mesons}

\subsection{Tensor mesons}

The observed $J^P=2^+$ tensor mesons $f_2(1270)$, $f_2'(1525)$, $a_2(1320)$ and $K_2^*(1430)$ form an SU(3) $1\,^3P_2$ nonet.  The $q\bar q$ content for isodoublet and isovector tensor resonances is obvious.  Just as the $\eta$-$\eta'$ mixing in the pseudoscalar case, the isoscalar tensor states $f_2(1270)$ and $f'_2(1525)$ also have a mixing, and their wave functions are defined by
 \be
 f_2(1270) &=&
{1\over\sqrt{2}}(f_2^u+f_2^d)\cos\theta_{f_2} + f_2^s\sin\theta_{f_2} ~, \non \\
 f'_2(1525) &=&
{1\over\sqrt{2}}(f_2^u+f_2^d)\sin\theta_{f_2} - f_2^s\cos\theta_{f_2} ~,
 \en
with $f_2^u\equiv u\bar u$ and likewise for $f_2^{d,s}$.   Since $\pi\pi$ is the dominant decay mode of $f_2(1270)$ whereas $f_2'(1525)$ decays predominantly into $K\ov K$ (see Ref.~\cite{PDG}), it is obvious that this mixing angle should be small.  More precisely, it is found that $\theta_{f_2}=7.8^\circ$ \cite{Li} and $(9\pm1)^\circ$ \cite{PDG}.  Therefore, $f_2(1270)$ is primarily an $(u\bar u+d\bar d)/\sqrt{2}$ state, while $f'_2(1525)$ is dominantly $s\bar s$.

For a tensor meson, the polarization tensors $\epsilon_{(\lambda)}^{\mu\nu}$ with helicity $\lambda$ can be constructed in terms of the polarization vectors of a massive vector state moving along the $z$-axis \cite{Berger:2000wt}
\begin{eqnarray}
\epsilon(0)^{*\mu} = (P_3,0,0,E)/m_T,
\quad
\epsilon(\pm1)^{*\mu} = (0,\mp1,+i,0)/\sqrt{2},
\end{eqnarray}and are given by
\begin{eqnarray}
\epsilon^{\mu\nu}_{(\pm2)} &\equiv& \epsilon(\pm1)^\mu \epsilon(\pm1)^\nu,
\\
\epsilon^{\mu\nu}_{(\pm1)} &\equiv& \sqrt{\frac{1}{2}}
[\epsilon(\pm1)^\mu \epsilon(0)^\nu + \epsilon(0)^\mu \epsilon(\pm1)^\nu],
\\
\epsilon^{\mu\nu}_{(0)} &\equiv& \sqrt{\frac{1}{6}}
 [\epsilon(+1)^\mu \epsilon(-1)^\nu + \epsilon(-1)^\mu \epsilon(+1)^\nu]
 + \sqrt{\frac{2}{3}}  \epsilon(0)^\mu \epsilon(0)^\nu.
\end{eqnarray}
The polarization $\epsilon_{\mu\nu}^{(\lambda)}$ can be
decomposed in the frame formed by the two light-like vectors, $z_\mu$ and $p_\nu\equiv P_\nu - z_\nu m_T^2/(2pz)$ with $P_\nu$ and $m_T$ being the momentum and mass of the tensor meson, respectively, and their orthogonal plane~\cite{Ball:1998sk,Ball:1998ff}. The transverse component that we use thus reads
\begin{eqnarray}\label{eq:polprojectiors}
 && \epsilon^{(\lambda)}_{\perp\, \mu\nu} z^\nu
 =\epsilon^{(\lambda)}_{\mu\nu} z^\nu-\epsilon^{(\lambda)}_{\parallel\mu\nu} z^\nu
        = \epsilon^{(\lambda)}_{\mu\nu} z^\nu -
     \frac{\epsilon^{(\lambda)}_{\alpha\nu} z^\alpha z^\nu }{p z} \left( p_\mu-\frac{m_T^2}{2 p z} \,z_\mu\right)\,.
\end{eqnarray}
The polarization tensor $\epsilon^{(\lambda)}_{\alpha\beta}$ satisfies the relations
 \be \label{eq:poltensor}
 \ep_{\mu\nu}^{(\lambda)}=\ep_{\nu\mu}^{(\lambda)}, \qquad \ep^{\mu}_{(\lambda)\mu}=0, \qquad P_\mu
 \ep_{(\lambda)}^{\mu\nu}=P_\nu\ep_{(\lambda)}^{\mu\nu}=0, \quad \epsilon^{(\lambda)}_{\mu\nu}\big(\epsilon^{(\lambda') \mu\nu}\big)^*=\delta_{\lambda\lambda'}.
 \en
The completeness relation reads
\begin{eqnarray} \label{eq:polarization}
 \sum_\lambda \epsilon^{(\lambda)}_{\mu\nu}\left(\epsilon^{(\lambda)}_{\rho\sigma}\right)^\ast
  = \frac12 M_{\mu\rho} M_{\nu\sigma}+\frac12 M_{\mu\sigma} M_{\nu\rho}
  -\frac13 M_{\mu\nu} M_{\rho\sigma}\,,
\end{eqnarray}
where $M_{\mu\nu} = g_{\mu\nu} - P_\mu P_\nu/m_T^2$.

\subsection{Decay constants}

 Decay constants of the vector  meson are defined as
 \be \label{eq:decayc}
 \la V(P,\epsilon)|\bar q_2\gamma_\mu q_1|0\ra &=& -if_V m_V\epsilon^*_\mu, \non \\
   \langle V(P,\epsilon) |\bar q_2 \sigma_{\mu\nu}q_1 |0\rangle
 &=& -f_V^\perp (\epsilon_{\mu}^* P_\nu -
\epsilon_{\nu}^* P_\mu)\,.
   \label{eq:tensor-1p1-2}
\end{eqnarray}
Contrary to the vector meson case, a $^3P_2$ tensor meson with $J^{PC}=2^{++}$ cannot be produced through the local $V-A$ and tensor currents. To see this, we notice that
\begin{eqnarray}
 \langle T(P, \lambda)|V_\mu|0\rangle &=&
 a\epsilon^{*(\lambda)}_{\mu\nu}P^\nu+b\epsilon^{*(\lambda)\nu}_{~\nu} P_\mu=0, \\
 \langle T(P,\lambda)|A_\mu |0\rangle &=& \varepsilon_{\mu\nu\rho\sigma} P^\nu \epsilon_{(\lambda)}^{\rho\sigma *}=0,
\end{eqnarray}
where use of Eq. (\ref{eq:poltensor}) has been made.
Nevertheless, a tensor meson can be created from these local currents involving covariant derivatives:
\be \label{eq:fT}
\la T(P,\lambda)|J_{\mu\nu}(0)|0\ra &=& f_Tm_T^2\epsilon^{*(\lambda)}_{\mu\nu}, \non \\
    \langle T(P,\lambda) |J^\perp_{\mu\nu\alpha}(0) |0 \rangle
 &=& -i  f_{T}^\perp m_T
 (\epsilon_{\mu\alpha}^{(\lambda)*} P_\nu-\epsilon_{\nu\alpha}^{(\lambda)*} P_\mu ) ,
\en
where
\be
J_{\mu\nu}(0) &=& \frac{1}{2}
\left( \bar q_1(0)\gamma_\mu i\stackrel{\leftrightarrow}{D}_\nu q_2(0)
     + \bar q_1(0)\gamma_\nu i\stackrel{\leftrightarrow}{D}_\mu q_2(0) \right),   \nonumber\\
J^\perp_{\mu\nu\alpha}(0) &=& \bar q_1(0) \sigma_{\mu\nu} i\stackrel{\leftrightarrow}{D}_\alpha q_2(0),
\en
and  $\stackrel {\leftrightarrow}{D}_\mu=\stackrel {\rightarrow}{D}_\mu-\stackrel {\leftarrow}{D}_\mu$ with $\stackrel{\rightarrow}{D}_\mu=\stackrel{\rightarrow}\partial_\mu +ig_s A^a_\mu \lambda^a/2$ and $\stackrel{\leftarrow}{D}_\mu=\stackrel{\leftarrow}\partial_\mu -ig_s A^a_\mu \lambda^a/2$.

The decay constant $f_T$ of the tensor meson has been estimated using QCD sum rules for the tensor mesons $f_2(1270)$ \cite{Aliev:1981ju} and $K^*_2(1430)$ \cite{Aliev:2009nn} and  the tensor-meson-dominance hypothesis for $f_2(1270)$ \cite{Aliev:1981ju,Terazawa:1990es,Suzuki:1993zs}. The previous sum rule predictions are \cite{Aliev:1981ju,Aliev:2009nn} \footnote{The dimensionless decay constant $f_T$ defined in \cite{Aliev:1981ju,Aliev:2009nn} differs from ours by a factor of $2m_T$. The factor of 2 comes from a different definition of $\stackrel{\leftrightarrow}{D}_\mu$ there.}
\begin{eqnarray}\label{eq:f-value-1}
    f_{f_2(1270)}(\mu = 1~{\rm GeV}) & \simeq & 0.08\,m_{f_2(1270)}=102~{\rm MeV}\,, \nonumber\\
    f_{K^*_2(1430)}(\mu = 1~{\rm GeV}) & \simeq & (0.10\pm0.01)\,m_{K^*_2(1430)}=(143\pm14)\,{\rm MeV}\,.
\end{eqnarray}
Recently, we have derived a sum rule for $f_T(\mu)f_T^\bot(\mu)$ and revisited the sum-rule analysis for $f_T(\mu)$. Our results of $f_T$ and $f_T^\bot$ for various tensor mesons are shown in Table \ref{tab:input} below \cite{CKY}. Our sum rule results are in good agreement with \cite{Aliev:1981ju} for $f_{f_2(1270)}$, but smaller than that of \cite{Aliev:2009nn} for $f_{K_2^*(1430)}$. The decay constants for $f_2(1270)$ and $f_2^\prime(1525)$ also can be extracted based on the hypothesis of tensor meson dominance together with the data of $\Gamma(f_2\to \pi\pi)$ and $\Gamma(f'_2\to K\bar K)$. We found that  \cite{CKY}
\begin{eqnarray}
   && f_{f_2(1270)} \simeq (0.085\pm 0.001) m_{f_2(1270)}=(108\pm1)\,{\rm MeV}\,, \non \\
   &&  f_{f'_2(1525)} \simeq (0.089\pm 0.003) m_{f'_2(1525)}=(136\pm5)\,{\rm MeV}\,.
\end{eqnarray}
They are in accordance with the sum rule predictions shown in Table \ref{tab:input}.

\subsection{Form factors}

Form factors for $B\to P,V,T$ transitions are defined by
\cite{BSW,Hatanaka:2009gb,Wei}
 \be \label{m.e.}
 \la P(P)|V_\mu|B(p_B)\ra &=& \left(P_\mu+(p_B)_\mu-{m_B^2-m_P^2\over q^2}\,q_ \mu\right)
F_1^{BP}(q^2)+{m_B^2-m_P^2\over q^2}q_\mu\,F_0^{BP}(q^2), \non \\
\langle{V}(P, \lambda)|V_\mu|{ B} (p_B)\rangle
 &=& -i \frac{2}{m_B + m_{V}} \varepsilon_{\mu\nu\alpha\beta}
 \epsilon_{(\lambda)}^{*\nu}
 p_B^\alpha P^{\beta} V^{BV}(q^2), \nonumber \\
 \langle V (P,\lambda)|A_\mu|{B}(p_B)\rangle
 &=& 2 m_{V} \frac{\epsilon^{(\lambda)*}\cdot p_B}{q^2} q_\mu
A_0^{BV}(q^2)+
 (m_B + m_{V})\left[ \epsilon^{(\lambda)*}_{\mu}-\frac{e^{(\lambda)*}\cdot p_B}{q^2} q_\mu\right] A_1^{BV}(q^2) \non \\
&& - {\epsilon^{(\lambda)*} \cdot p_B\over m_B+m_V}\left[P_\mu+(p_B)_\mu-{m_B^2-m_V^2\over q^2}q_\mu\right]
A_2^{BV}(q^2)
\nonumber \\
 \langle{T}(P, \lambda)|V_\mu|{ B} (p_B)\rangle
 &=& -i \frac{2}{m_B + m_{T}} \varepsilon_{\mu\nu\alpha\beta}
 e_{(\lambda)}^{*\nu}
 p_B^\alpha P^{\beta} V^{BT}(q^2), \nonumber \\
 \langle T (P,\lambda)|A_\mu|{ B}(p_B)\rangle
 &=& 2 m_{T} \frac{e^{(\lambda)*}\cdot p_B}{q^2} q_\mu
A_0^{BT}(q^2)+
 (m_B + m_{T})\left[ e^{(\lambda)*}_{\mu}-\frac{e^{(\lambda)*}\cdot p_B}{q^2} q_\mu\right] A_1^{BT}(q^2) \non \\
&& - {e^{(\lambda)*} \cdot p_B\over m_B+m_T}\left[P_\mu+(p_B)_\mu-{m_B^2-m_T^2\over q^2}q_\mu\right]
A_2^{BT}(q^2),
 \en
where $q_\mu=(p_B-P)_\mu$ and $e^{*\mu}_{(\lambda)}\equiv \epsilon^{*\mu\nu}_{(\lambda)}p_{_{B\nu}}/m_B$.  Throughout the paper we have adopted the convention
 $\varepsilon^{0123}=-1$.

In the Isgur-Scora-Grinstein-Wise (ISGW) model \cite{ISGW}, the general expression for the $B\to T$ transition is parametrized as
 \be \label{eq:BTff}
 \la T(P,\lambda)|(V-A)_\mu|B(p_B)\ra &=&
 ih(q^2)\varepsilon_{\mu\nu\rho\sigma}\epsilon^{*\nu\alpha}p_{B\alpha}(p_B+P)^\rho
 q^\sigma-k(q^2)\epsilon^*_{\mu\nu}p_B^\nu  \non \\
 &-& b_+(q^2)\epsilon^*_{\alpha\beta}p_B^\alpha p_B^\beta (p_B+P)_\mu
 -b_-(q^2)\epsilon^*_{\alpha\beta}p_B^\alpha p_B^\beta q_\mu,
 \en
where the form factor $k$ is dimensionless, and the canonical dimension of $h, b_+$ and $b_-$ is $-2$.
The relations between these two different sets of form factors are
\be
&& V^{BT}(q^2)=m_B(m_B+m_T)h(q^2), \qquad A_1^{BT}(q^2)={m_B\over m_B+m_T}k(q^2), \\
&& A_2^{BT}(q^2)=-m_B(m_B+m_T)b_+(q^2), \quad A_0^{BT}(q^2)={m_B\over 2m_T}[k^2(q^2)+(m_B^2-m_T^2)b_+(q^2)+q^2 b_-(q^2)]. \non
\en
The $B\to T$ transition form factors have been evaluated in the ISGW model \cite{ISGW} and its improved version, ISGW2 \cite{ISGW2}, the covariant light-front quark model (CLFQ) \cite{CCH}, the light-cone sum rule (LCSR) approach \cite{kcy},  the large energy effective theory (LEET) \cite{Charles,Ebert,Datta} and the pQCD approach \cite{Wei}. In LEET,  form factors are evaluated at large recoil and all the form factors in the LEET limit to be specified below can be parametrized in terms of two independent universal form factors $\zeta_\bot$ and $\zeta_\parallel$ \cite{Hatanaka:2009gb}:
\be
V^{BT}(q^2)&=& {m_{_T}\over |\vec{p}_{_T}|}\left(1+{m_T\over m_B}\right)\zeta_\bot(q^2),  \non \\
A_0^{BT}(q^2)&=& {m_{_T}\over |\vec{p}_{_T}|}\left[ \left(1-{m_T^2\over m_BE_T}\right)\zeta_\parallel(q^2)+{m_T\over m_B}\zeta_\bot(q^2)\right],  \non \\
A_1^{BT}(q^2)&=& {m_{_T}\over |\vec{p}_{_T}|}\left({2E_T\over m_B+m_T}\right)\zeta_\bot(q^2),  \non \\
A_2^{BT}(q^2)&=& {m_{_T}\over |\vec{p}_{_T}|}\left(1+{m_T\over m_B}\right)\left[\zeta_\bot(q^2)-{m_T\over E_T}\zeta_\parallel(q^2)\right],
\en
where  $E_T$ is the energy of the tensor meson
\be
E_T={m_B\over 2}\left(1+{m_T^2-q^2\over m_B^2}\right).
\en
In the LEET limit,
\be
E_T, m_B\gg m_T,\Lambda_{\rm QCD}.
\en

Using the recent analysis of tensor meson distribution amplitudes \cite{CKY}, one of us (KCY) has calculated the form factors of $B$ decays into tensor mesons using the LCSR approach \cite{kcy}. The LCSR results are close to LEET and pQCD calculations.

The $B\to a_2(1320), f_{2q}=(u\bar u+d\bar d)/\sqrt{2}, K_2^*(1430)$ transition form factors calculated in various models at the maximal recoil $q^2=0$ are summarized in Table \ref{tab:FF}.
The ISGW model \cite{ISGW} is based on the non-relativistic constituent
quark picture. In general, the form factors evaluated
in the ISGW model are reliable only at $q^2=q^2_m\equiv
(m_B-m_T)^2$, the maximum momentum transfer. The reason is that
the form-factor $q^2$ dependence in the ISGW model is proportional
to exp[$-(q^2_m-q^2)$] and hence the form factor decreases
exponentially as a function of $(q^2_m-q^2)$ (see Appendix A for details). This has been
improved in the ISGW2 model \cite{ISGW2} in which the form factor has a more
realistic behavior at large $(q^2_m-q^2)$ which is expressed in
terms of a certain polynomial term. As noticed in \cite{Kim:2003}, form factors are increased in the ISGW2 model so that the branching fractions of $B\to TM$ decays are enhanced by about an order of magnitude compared to the estimates based on the ISGW model.

%%%%%%%%%%%%%%%%%%%%%%%%%%%%%%%%%%%%%%%%%%%%%%%%%%%%%
{\squeezetable
\begin{table}[t]
\caption{$B\to T$ transition form factors at $q^2=0$ evaluated in the ISGW2, CLFQ, LCSR, LEET and pQCD models. The CLFQ results are obtained by first calculating the form factors $h(q^2),b_+(q^2)$ and $b_-(q^2)$ using the covariant light-front approach and $k(q^2)$ from the heavy quark symmetry relation Eq. (\ref{eq:k}) and then converted them into the form-factor set $V(q^2)$ and $A_{0,1,2}(q^2)$. To compute the form factors in LEET we have applied $\zeta_\bot(0)=0.28\pm0.04$ and $\zeta_\parallel(0)=0.22\pm0.03$\,. LCSR and pQCD results are taken from \cite{kcy} and \cite{Wei}, respectively.} \label{tab:FF}
\begin{ruledtabular}
\begin{tabular}{| l r r r c r|| c r r r c r |}
~$F$~
    & ISGW2
    & CLFQ
    & LCSR
    & LEET
    & pQCD
&  $F$
    & ISGW2
    & CLFQ
    & LCSR
    & LEET
    & pQCD
 \\
    \hline
$V^{Ba_2}$
    & 0.32
    & 0.28
    & $0.18\pm0.02$
    & $0.18\pm0.03$
    & $0.18^{+0.05}_{-0.04}$
& $A_0^{Ba_2}$
    & $0.20$
    & $0.24$
    & $0.21\pm0.04$
    & $0.14\pm0.02$
    & $0.18^{+0.06}_{-0.04}$ \\
$A_1^{Ba_2}$
    & $0.16$
    & $0.21$
    & $0.14\pm0.02$
    & $0.13\pm0.02$
    & $0.11^{+0.03}_{-0.03}$
& $A_2^{Ba_2}$
    & 0.14
    & 0.19
    & $0.09\pm0.02$
    & $0.13\pm0.02$
    & $0.06^{+0.02}_{-0.01}$ \\
$V^{Bf_{2q}}$
    & 0.32
    & 0.28
    & $0.18\pm0.02$
    & $0.18\pm0.02$
    & $0.12^{+0.03}_{-0.03}$
& $A_0^{Bf_{2q}}$
    & $0.20$
    & $0.25$
    & $0.20\pm0.04$
    & $0.13\pm0.02$
    & $0.13^{+0.04}_{-0.03}$ \\
$A_1^{Bf_{2q}}$
    & $0.16$
    & $0.21$
    & $0.14\pm0.02$
    & $0.12\pm0.02$
    & $0.08^{+0.02}_{-0.02}$
& $A_2^{Bf_{2q}}$
    & 0.14
    & 0.19
    & $0.10\pm0.02$
    & $0.13\pm0.02$
    & $0.04^{+0.01}_{-0.01}$ \\
$V^{BK_2^*}$
    & 0.38
    & 0.29
    & $0.16\pm0.02$
    & $0.21\pm0.03$
    & $0.21^{+0.06}_{-0.05}$
& $A_0^{BK_2^*}$
    & 0.27
    & $0.23$
    & $0.25\pm0.04$
    & $0.15\pm0.02$
    & $0.18^{+0.05}_{-0.04}$ \\
$A_1^{BK_2^*}$
    & $0.24$
    & $0.22$
    & $0.14\pm0.02$
    & $0.14\pm0.02$
    & $0.13^{+0.04}_{-0.03}$
& $A_2^{BK_2^*}$
    & 0.22
    & 0.21
    & $0.05\pm0.02$
    & $0.14\pm0.02$
    & $0.08^{+0.03}_{-0.02}$ \\
\end{tabular}
\end{ruledtabular}
\end{table}
}

%%%%%%%%%%%%%%%%%%%%%%%%%%%%%%%%%
The CLFQ model  is a
relativistic quark model in which a consistent and fully
relativistic treatment of quark spins and the center-of-mass
motion is carried out. This model is very suitable to study hadronic form factors. Especially, as the
recoil momentum increases (corresponding to a decreasing $q^2$),
we need to start considering relativistic effects seriously. In
particular, at the maximum recoil point $q^2=0$ where the
final-state meson could be highly relativistic, it is expected
that the corrections to non-relativistic quark model will be sizable in this case.

The CLFQ and ISGW2 model predictions for $B\to T$ transition form factors differ mainly in two aspects: (i) when $q^2$ increases, $h(q^2)$,  $b_+(q^2)$ and $b_-(q^2)$ increases more rapidly in the former and (ii) the form factor $k$ obtained in both models is quite different, for example, $k^{BK_2^*}(0)=0.015$ in the former and 0.293 in the latter. Indeed, it has been noticed \cite{CCH}
that among the four $B\to
T$ transition form factors, the one $k(q^2)$ is particularly
sensitive to $\beta_T$, a parameter describing the tensor-meson wave function, and that $k(q^2)$ at zero recoil shows a large deviation from the heavy quark symmetry relation.
It is not clear to us if the very complicated
analytic expression for $k(q^2)$ in Eq. (3.29) of \cite{CCH} is complete. To overcome this difficulty, it was pointed out in \cite{CCH} that one may apply the heavy quark
symmetry relation to obtain $k(q^2)$
for $B\to T$ transition
 \be \label{eq:k}
 k(q^2)=\,m_Bm_{T}\left(1+{m_B^2+m_{T}^2-q^2\over
 2m_Bm_{T}}\right)\left[ h(q^2)-{1\over 2}b_+(q^2)+{1\over 2}b_-(q^2)\right].
 \en
In Table \ref{tab:FF} the CLFQ results are obtained by first calculating the form factors $h(q^2),b_+(q^2)$ and $b_-(q^2)$ using the covariant light-front approach \cite{CCH} and $k(q^2)$ from the heavy quark symmetry relation Eq. (\ref{eq:k}) and then converted them into the form-factor set $V(q^2)$ and $A_{0,1,2}(q^2)$.

%%%%%%%%%%%%%%%%%%%%%%%%%%%%%%%%%%%%%%%%%%%%%%%%%%%%%
\begin{table}[t]
\caption{$B\to T$ transition form factors obtained
in the covariant light-front model and fitted to the
3-parameter form Eq. (\ref{eq:FFpara}). } \label{tab:FFinCLFQ}
\begin{ruledtabular}
\begin{tabular}{| l c c c || c c c c |}
~~~$F$~~~~~
    & $F(0)$
    & $a$
    & $b$~~~
&  $F$
    & $F(0)$
    & $a$
    & $b$~~~~~
 \\
    \hline
$V^{Ba_2}$
    & 0.28
    & 2.19
    & $2.22$~~~~~
& $A_0^{Ba_2}$
    & $0.24$
    & $1.28$
    & $0.84$~~~~~ \\
$A_1^{Ba_2}$
    & $0.21$
    & $1.38$
    & $0.47$~~~~~
& $A_2^{Ba_2}$
    & 0.19
    & $1.93$
    & $1.69$~~~~~ \\
$V^{Bf_{2q}}$
    & 0.28
    & 2.19
    & $2.22$~~~~~
& $A_0^{Bf_{2q}}$
    & $0.25$
    & $1.37$
    & $0.95$~~~~~ \\
$A_1^{Bf_{2q}}$
    & $0.21$
    & $1.39$
    & $0.46$~~~~~
& $A_2^{Bf_{2q}}$
    & 0.19
    & $1.93$
    & $1.69$~~~~~ \\
$V^{BK_2^*}$
    & 0.29
    & 2.17
    & $2.22$~~~~~
& $A_0^{BK_2^*}$
    & $0.23$
    & $1.23$
    & $0.74$~~~~~ \\
$A_1^{BK_2^*}$
    & $0.22$
    & 1.42
    & $0.50$~~~~~
& $A_2^{BK_2^*}$
    & 0.21
    & 1.96
    & $1.79$~~~~~ \\
\end{tabular}
\end{ruledtabular}
\end{table}

%%%%%%%%%%%%%%%%%%%%%%%%%%%%%%%%%

Form factors in the CLFQ model  are first calculated in the
spacelike region and their momentum dependence is fitted to a
3-parameter form
  \be \label{eq:FFpara}
 F(q^2)=\,{F(0)\over 1-a(q^2/m_{B}^2)+b(q^2/m_{B}^2)^2}\,.
 \en
The parameters $a$, $b$ and $F(0)$ are first determined in the
spacelike region. This parametrization is then analytically
continued to the timelike region to determine the physical form
factors at $q^2\geq 0$. The results are exhibited in Table \ref{tab:FFinCLFQ}. The momentum dependence of the form factors in the LCSR approach can be found in \cite{kcy}, while
a slightly different parametrization
\be \label{eq:FFpara1}
 F(q^2)=\,{F(0)\over (1-q^2/m_{B}^2)[1-a\,q^2/m_{B}^2+b\,q^4/m_{B}^4]}
\en
is used in the pQCD approach for the calculations of the form-factor $q^2$ dependence \cite{Wei}.

For the calculation in LEET, we have followed \cite{Hatanaka:2009sj} to use $\zeta_\bot(0)=0.28\pm0.04$ and $\zeta_\parallel(0)=0.22\pm0.03$. For the $q^2$ dependence, we shall use
\be
\zeta_{\bot,\parallel}^T(q^2)={ \zeta_{\bot,\parallel}^T(0)\over (1-q^2/m_B^2)^2}.
\en
For the ISGW2 model, the $q^2$ dependence of the form factors is governed by Eq. (\ref{eq:Fn}).

\subsection{Light-cone distribution amplitudes}
The light-cone distribution amplitudes (LCDAs) of the tensor meson are defined as~\cite{CKY} \footnote{The LCDAs of the tensor meson were first studied in \cite{Braun:2000cs}.}
\begin{eqnarray} \label{eq:chial-even-LCDAs}
 \langle T(P,\lambda)|\bar q_1(y)\gamma_\mu {q_2}(x)|0\rangle
 &=&
  -i f_{T} m^2_T
  \int\limits_{0}^1 \! du\,e^{i(uPy+\bar uPx)} \Bigg\{ P_\mu \frac{\epsilon^{(\lambda)*}_{\alpha\beta}z^\alpha z^\beta}
{(Pz)^2} \,\Phi^T_\parallel(u)
 +  \Bigg(\frac{\epsilon^{(\lambda)*}_{\mu\alpha} z^\alpha }{Pz} \non \\
 &-&
 P_\mu \frac{\epsilon^{(\lambda)*}_{\beta\alpha} z^\beta z^\alpha }{(Pz)^2} \Bigg)\, g_v(u)
 - \frac{1}{2} z_\mu \frac{\epsilon^{(\lambda)*}_{\alpha\beta}z^\alpha z^\beta}{(Pz)^3} m_T^2 \,\bar g_3(u)
+ {\cal O} (z^2) \Bigg\} \,, \\
 \langle T(P,\lambda)|\bar q_1(y)\gamma_\mu\gamma_5 q_2 (x)|0\rangle
 &=&
   -i f_{T} m_T^2
  \int\limits_{0}^1 \! du\,e^{i(uPy+\bar u Px)} \varepsilon_{\mu\nu\alpha\beta} {z^\nu P^\alpha} \epsilon_{(\lambda)}^{*\beta\delta}z_\delta\,{1\over 2Pz}\, g_a(u)\,,
\end{eqnarray}
\begin{eqnarray}
 \langle T(P,\lambda)|\bar q_1(y)\sigma_{\mu\nu}  q_2(x)|0\rangle
 &=&
 - f_{T}^\perp m_T
  \int\limits_{0}^1 \! du\,e^{i(uPy+ \bar u Px)} \Bigg\{\left[\epsilon^{(\lambda)*}_{\mu\alpha} z^\alpha P_\nu
 - \epsilon^{(\lambda)*}_{\nu\alpha} z^\alpha P_\mu\right] \frac{1}{Pz} \Phi^T_\perp(u) \nonumber\\
 &+&  (P_\mu z_\nu - P_\nu z_\mu)
 \frac{ m_T^2\epsilon^{(\lambda)*}_{\alpha\beta}z^\alpha z^\beta}{(Pz)^3} \bar h_t(u) \nonumber  \\
 &+&
   \frac{1}{2} \left[\epsilon^{(\lambda)*}_{\mu\alpha} z^\alpha z_\nu
 - \epsilon^{(\lambda)*}_{\nu\alpha} z^\alpha z_\mu\right] \frac{m_T^2}{(Pz)^2} \bar h_3(u) + {\cal O} (z^2)
  \Bigg\} \,,\\
 \langle T(P,\lambda)|\bar q_1(y)  q_2(x)|0\rangle
 &=&
- f_{T}^\perp m_T^3
  \int\limits_{0}^1 \! du\,e^{i(uPy+\bar u Px)} \frac{\epsilon^{(\lambda)*}_{\alpha\beta}z^\alpha z^\beta}{2Pz}h_s(u)\,, \label{eq:chial-odd-LCDAs}
\end{eqnarray}
where $\bar g_3=g_3+\Phi^T_\parallel-2g_v$, $\bar h_t=h_t-{1\over 2}(\Phi^T_\bot+h_3)$, $\bar h_3=h_3-\Phi_\bot^T$, and $z\equiv y-x$. Here
$\Phi^T_\parallel, \Phi^T_\perp$ are twist-2 LCDAs, \footnote{
Since in  the transversity basis, one denotes the corresponding parallel and perpendicular states by $A_\parallel$ and $A_\bot$,
a better notation for the longitudinal and transverse LCDAs will be $\Phi_L$ and $\Phi_T$, respectively, rather than $\Phi_\parallel$ and $\Phi_\perp$.  Indeed, the transverse polarization includes both parallel and perpendicular polarizations.
In the present work we follow the conventional notation for LCDAs.
}
$g_v,g_a, h_t, h_s$  twist-3 ones,
and $g_3, h_3$  twist-4. In the SU(3) limit, due to the $G$-parity of the tensor meson, $\Phi^T_\parallel, \Phi^T_\perp, g_v, g_a, h_t, h_s, g_3$ and $h_3$ are antisymmetric under the replacement $u\to 1-u$ \cite{CKY}.

Using the QCD equations of motion~\cite{Ball:1998sk,Ball:1998ff}, the two-parton distribution amplitudes $g_v, g_a, h_t$ and $h_s$ can be represented in terms of
$\Phi^T_{\parallel,\perp}$ and three-parton distribution amplitudes. Neglecting the three-parton distribution amplitudes containing gluons and terms proportional to light quark masses, twist-3 LCDAs $g_a,g_v,h_t$ and $h_s$ are related to twist-2 ones through the Wandzura-Wilczek (WW) relations:
\begin{eqnarray}\label{eq:WW}
    g_v^{WW}(u) &=& \int\limits_{0}^u dv\, \frac{\Phi^T_\parallel(v)}{\bar v}+
                  \int\limits_{u}^1 dv\, \frac{\Phi^T_\parallel(v)}{v}\,,
\nonumber\\
    g_a^{WW}(u) &=& 2\bar{u}\int\limits_{0}^u dv\, \frac{\Phi^T_\parallel(v)}{\bar v}+
                  2u\int\limits_{u}^1 dv\, \frac{\Phi^T_\parallel(v)}{v}\,,
\nonumber\\
 h_t^{WW}(u) &=& \frac{3}{2} (2u-1)\left(\int\limits_{0}^u dv\, \frac{\Phi^T_\perp(v)}{\bar v} -
                  \int\limits_{u}^1 dv\, \frac{\Phi^T_\perp(v)}{v}\right)\,,
\nonumber\\
    h_s^{WW}(u) &=& 3 \left( \bar{u}\int\limits_{0}^u dv\, \frac{\Phi^T_\perp(v)}{\bar v}+
                  u\int\limits_{u}^1 dv\, \frac{\Phi^T_\perp(v)}{v} \right)\,.
\end{eqnarray}
These WW relations further give us
\begin{eqnarray}\label{eq:T-ww}
  && {1\over 4}{g'_a(u)}+{1\over 2}g_v(u)= \int_u^1
\frac{\Phi^T_\parallel (v)}{v}dv \equiv \Phi^T_+(u)\,,\nonumber\\
&& {1\over 4}{g'_a(u)}-{1\over 2}g_v(u)=- \int_0^u
\frac{\Phi^T_\parallel (v)}{\bar v}dv \equiv -\Phi^T_-(u)\,,\nonumber\\
 &&
h'_s(u)= -3\Bigg[ \int_0^u
\frac{\Phi^T_\perp(v)}{\bar v}dv -\int_u^1
\frac{\Phi^T_\perp(v)}{v}dv \Bigg] \equiv -3 \Phi_t(u), \\
 &&\int_0^u dv \big( \Phi^T_\perp (v) -{2\over 3}h_t
(v))= u\bar u\Bigg[ \int_0^u \frac{\Phi^T_\perp(v)}{\bar v}dv
-\int_u^1
\frac{\Phi^T_\perp(v)}{v}dv\Bigg] = u \bar u\Phi_t(u), \nonumber\\
&&\int_0^u dv \big( \Phi^T_\parallel (v) -{1\over 2}g_v(v))=
 \frac{1}{2}\Bigg[ \bar u\int_0^u \frac{\Phi^T_\parallel(v)}{\bar v}dv
 -u \int_u^1 \frac{\Phi^T_\parallel(v)}{v}dv\Bigg]
 =\frac{1}{2}\bigg(\bar u \Phi^T_-(u) -u \Phi^T_+(u)\bigg). \nonumber
\end{eqnarray}

The LCDAs $\Phi^T_{\parallel,\perp}(u,\mu)$ and $\Phi_t(u,\mu)$ can be expanded as
\begin{eqnarray}\label{eq:conformal-partial-wave-t2}
\Phi_{\parallel,\perp}^T(u,\mu) &=& 6u(1-u)
\sum_{\ell=0}^\infty a_\ell^{(\parallel,\perp),T}(\mu) C^{3/2}_\ell(2u-1), \non \\
\Phi_t(u,\mu) &=& 3
\sum_{\ell=0}^\infty a_\ell^{\perp,T}(\mu) P_{\ell+1}(2u-1),
\end{eqnarray}
where the Gegenbauer moments $a_\ell^{(\parallel,\perp),T}$ with $\ell$ being even vanish in the SU(3) limit, $\mu$ is the normalization scale and $P_n(x)$ are the Legendre polynomials.
In the present study the distribution amplitudes are normalized to be
\begin{eqnarray}\label{eq:norm1}
  \int\limits_{0}^{1} du\, (2u-1)\,\Phi^T_{\parallel}(u)
  = \int\limits_{0}^{1} du\, (2u-1)\,\Phi^T_{\perp}(u) = 1 \,,  \qquad \int^1_0 du\,\Phi_t(u) = 0 \,.
\end{eqnarray}
Consequently, the first Gegenbauer moments are fixed to be $a_1^{\parallel,T}=a_1^{\perp,T}={5\over 3}$. Moreover, we have
\begin{eqnarray}\label{eq:norm}
  3 \int\limits_{0}^{1} du\, (2u-1)\,g_a(u) &=&
  \int\limits_{0}^{1} du\, (2u-1)\,g_v(u) = 1\,, \non \\
  2 \int\limits_{0}^{1} du\, (2u-1)\,h_s(u) &=&
  \int\limits_{0}^{1} du\, (2u-1)\,h_t(u) = 1\,,
\end{eqnarray}
which hold even if the complete leading twist DAs and corrections from the three-parton distribution amplitudes containing gluons are included.  The asymptotic wave function is therefore
\begin{eqnarray}\label{eq:phi-as}
    \Phi_{\parallel,\perp}^{T,\rm as}(u) = 30 u(1-u)(2u-1),
\end{eqnarray}
and the corresponding expressions for the twist-3 distributions are
\begin{eqnarray}\label{eq:g-h-as}
 g_v^{\rm as}(u) &=& 5 (2u-1)^3\,,\qquad g_a^{\rm as}(u) =
    10 u(1-u) (2u-1) \,,\nonumber\\
 h_t^{\rm as}(u) &=& \frac{15}{2} (2u-1) (1-6u+6u^2)\,,
 \qquad h_s^{\rm as}(u) = 15 u(1-u) (2u-1)\,,
\end{eqnarray}
and
\be
\Phi_t^{\rm as}(u)=5(1-6u+6u^2).
\en
Note that, contrary to the twist-2 LCDA $\Phi^T_{\parallel,\bot}(u)$, the twist-3 one $\Phi_t(u)$ is even under the replacement $u\to 1-u$ in the SU(3) limit.

For vector mesons, the
general expressions of LCDAs are
 \be
 \Phi_V(x,\mu)=6x(1-x)\left[1+\sum_{n=1}^\infty
 a_n^{\parallel,V}(\mu)C_n^{3/2}(2x-1)\right],
 \en
and
 \be
 \Phi_v(x,\mu)=3\left[2x-1+\sum_{n=1}^\infty
 a_{n}^{\bot,V}(\mu)P_{n+1}(2x-1)\right].
 \en
Likewise, for pseudoscalar mesons,
 \be
 \Phi_P(x,\mu)=6x(1-x)\left[1+\sum_{n=1}^\infty
 a_n^P(\mu)C_n^{3/2}(2x-1)\right], \qquad \Phi_p(x,\mu)=1.
 \en

\subsection{Helicity projection operators}
In the QCDF calculation, we need to know the helicity projection operators in the momentum space. To do this, using the identity
\be
\bar q^1_\alpha(y)q^2_\delta(x) &=& {1\over 4}\Big\{\mathbf{1} [\bar q^1(y)q^2(x)] +\gamma_5[\bar q^1(y)\gamma_5 q^2(x)]+\gamma^\rho[\bar q_1(y)\gamma_\rho q^2(x)] \non \\
&& -\gamma^\rho\gamma_5[\bar q^1(y)\gamma_\rho\gamma_5 q^2(x)]+{1\over 2}\sigma^{\rho\lambda}[\bar q^1(y)\sigma_{\rho\lambda}q^2(x)]\Big\}_{\delta\alpha}
\en
and Eqs. (\ref{eq:chial-even-LCDAs})-(\ref{eq:chial-odd-LCDAs}), we obtain
\begin{eqnarray}
 &&\langle T(P,\lambda)|\bar q^1_{\alpha}(y) \, q^2_\delta(x)|0\rangle
= -\frac{i}{4} \, \int_0^1 du \,  e^{i (u  P y +
    \bar u P x)}
   \Bigg\{ f_T m_T^2 \Bigg[
    \not\!P \, \frac{\epsilon^{*(\lambda)}_{\mu\nu} z^\mu z^\nu}{(Pz)^2} \,
    \Phi^T_\parallel(u) -{1\over 2}\not\!z{\epsilon^{*(\lambda)}_{\mu\nu}z^\mu z^\nu \over (Pz)^3} m_T^2\bar g_3(u)  \non \\
    && \qquad +  \left({\epsilon^{*(\lambda)}_{\mu\nu}z^\nu \over Pz}-P_\mu{\epsilon^{*(\lambda)}_{\nu\beta}z^\nu z^\beta\over (Pz)^2}\right)\gamma^\mu\, g_v(u)
     +  \frac{1}{2} \epsilon_{\mu\nu\rho\sigma}\gamma^\mu
    \epsilon^{*\nu\beta}_{(\lambda)} z_\beta P^\rho z^\sigma\gamma_5{1\over Pz}\, g_a(u)  \Bigg] \non \\
  &&\qquad - \,{i\over 2}f^{\perp}_T m_T \Bigg[ \sigma^{\mu\nu}\left(\epsilon^{(\lambda)*}_{\mu\beta} z^\beta P_\nu
 - \epsilon^{(\lambda)*}_{\nu\beta} z^\beta P_\mu\right) \frac{1}{Pz} \Phi^T_\perp(u)
   +\sigma^{\mu\nu}(P_\mu z_\nu - P_\nu z_\mu)
 \frac{ m_T^2\epsilon^{(\lambda)*}_{\rho\beta}z^\rho z^\beta}{(Pz)^3} \,\bar h_t(u) \nonumber  \\
 && \qquad +
   \frac{1}{2} \sigma^{\mu\nu}\left(\epsilon^{(\lambda)*}_{\mu\beta} z^\beta z_\nu
 - \epsilon^{(\lambda)*}_{\nu\beta} z^\beta z_\mu\right) \frac{m_T^2}{(Pz)^2}\, \bar h_3(u)+ \epsilon^{*(\lambda)}_{\mu\nu}z^\mu z^\nu {m_T^2\over Pz}\,h_s(u)\Bigg]+{\cal O}[(x-y)^2]\Bigg\}_{\delta\alpha}\,. \label{eq:DAs}
 \end{eqnarray}

Since any four momentum can be split into light-cone and transverse components as $k^\mu=k_-^\mu+k_+^\mu+k_\bot^\mu$,
we shall assign the momenta
 \begin{eqnarray}
 k_1^\mu = u E n_-^\mu
 + \frac{k_\perp^2}{4 uE}n_+^\mu+k_\perp^\mu \,,
 \qquad
 k_2^\mu = \bar u E n_-^\mu
 + \frac{k_\perp^2}{4 \bar uE}n_+^\mu- k_\perp^\mu \,,
\end{eqnarray}
to the quark and antiquark, respectively, in an energetic light final-state meson
with the momentum $P^\mu$ and mass $m$, satisfying the relation
$P^\mu =En_-^\mu + m^2 n_+^\mu/(4E) \simeq E n_-^\mu$, where we have defined two light-like
vectors $n_\pm^\mu$ with $n_-^\mu\equiv (1,0,0,-1)$ and $n_+^\mu\equiv
(1,0,0,1)$ and assumed
that the meson moves along the $n_-^\mu$ direction.
To obtain
the light-cone projection operator of the meson in the momentum
space, we take the Fourier transformation of Eq. (\ref{eq:DAs}) and apply the following substitution in the calculation
\begin{equation}
z^\mu \to -i \frac{\partial}{\partial k_{1\, \mu}}\simeq -i \Bigg(
\frac{n_+^\mu}{2E}\frac{\partial}{\partial u} +
\frac{\partial}{\partial k_{\perp\, \mu}}\Bigg)\,,
\end{equation}
where terms of order $k_\perp^2$ have been omitted.
The longitudinal projector reads
 \begin{eqnarray} \label{eq:VproL}
 M^T_\parallel(\lambda=0) &=& -i\frac{f_T}{4} E \,
 \Bigg\{
 \left[\epsilon^{(\lambda)*}_{\alpha\beta} n_+^\alpha n_+^\beta \left(m_T\over 2E\right)^2\right]
 \not\! n_- \,\Phi^T_\parallel(u)+ \frac{f_T^\perp}{f_T} \frac{m_T}{E}
 \,\left[\epsilon^{(\lambda)*}_{\alpha\beta}
  n_+^\alpha n_+^\beta \left(m_T\over 2E\right)^2\right]
  \non \\
& &\times \Bigg[ -\frac{i}{2}\,\sigma_{\mu\nu} \,  n_-^\mu  n_+^\nu \,
 h_{t}(u)
 - \,i E\int_0^u dv \,(\Phi^T_\perp(v) -
 h_{t}(v)) \
 \sigma_{\mu\nu} n_-^\mu \, \frac{\partial}{\partial k_\perp{}_\nu}
  +  \frac{h'_{s}(u)}{2} \Bigg]\,
  \non \\
 & &+ \frac{f_T^\perp}{f_T} \frac{m_T}{E} iE\sigma_{\mu\nu} n_-^\nu \epsilon^{(\lambda)*\mu\alpha}\delta_{\lambda,0}{\partial\over\partial k_\bot^\alpha}\int^u_0 dv\Phi^T_\bot(v)
 +{\cal O} \left({m_T^2\over E^2}\right)
 \Bigg\}
 \non \\
  &=& -i \frac{f_T}{4} E \, \left[\epsilon^{(\lambda)*}_{\alpha\beta}
  n_+^\alpha n_+^\beta \left(m_T\over 2E\right)^2\right]
 \Bigg\{
 \not\! n_- \,\Phi^T_\parallel(u)
 +\frac{f_T^\perp}{f_T} \frac{m_T}{E}
  \non \\
 & &\times \Bigg[ -\frac{i}{2}\,\sigma_{\mu\nu} \,  n_-^\mu  n_+^\nu \, h_{t}(u)
 - \,\frac{3i}{2} E\int_0^u dv \,
 \left(\Phi^T_\perp(v) -\frac{2}{3} h_{t}(v)\right)
 \sigma_{\mu\nu} n_-^\mu \, \frac{\partial}{\partial k_\perp{}_\nu}
  +  \frac{h'_{s}(u)}{2} \Bigg]\,
  \non \\
  & & +{\cal O} \left({m_T^2\over E^2}\right) \Bigg\}
  \,,
\end{eqnarray}
and the transverse projectors have the form
 \begin{eqnarray} \label{eq:VproT}
 M^T_\perp(\lambda=\pm1) &=& -i\frac{f^{\perp}_T}{4} E\left[\epsilon^{*(\lambda)}_{\bot\mu\alpha}
  n_+^\alpha  \left(m_T\over 2E\right)\right]
 \Bigg\{
 \gamma^\mu\not\! n_- \,
   \Phi^T_\perp(u)\nonumber\\
& & +  \frac{f_T}{f_T^\perp} \frac{m_T}{E}
\Bigg[
 \gamma^\mu g_v(u) -  \, E\int_0^u dv\, \Big(2\Phi^T_\parallel(v) -g_v(v)\Big)\not\!n_-{\partial\over \partial k_{\bot\mu}} \non \\
  &-&  \,i \varepsilon_{\mu\nu\rho\sigma} \, \gamma^\nu n_-^\rho \gamma_5
  \left(n_+^\sigma \,{g'_a(u)\over 4}- E\,\frac{g_a(u)}{2} \,
  \frac{\partial}{\partial k_\perp{}_\sigma}\right)
 \Bigg]
 +{\cal O} \left({m_T^2\over E^2}\right) \Bigg\}\,  ,
\end{eqnarray}
and
\be
 M^T_\bot(\lambda=\pm2)=-i {f_T^\bot \over 4} E
 \Bigg\{
 \frac{m_T}{E}iE\sigma_{\mu\nu}n_-^\nu \epsilon^{(\lambda)*\mu\alpha}\delta_{\lambda,\pm2}{\partial \over\partial k_\bot^\alpha}\int_0^u dv\Phi^T_\bot(v)
 +{\cal O} \left({m_T^2\over E^2}\right) \Bigg\}\,.
\en

The exactly longitudinal and transverse polarization tensors of the tensor meson, which are independent of the coordinate variable $z=y-x$, have the expressions
\begin{eqnarray}
\epsilon^{*(0)\mu\nu}n^+_\nu &=& \sqrt{2\over 3}\,\frac{2E^2}{m_T^2}
\Bigg[ \Bigg(1-\frac{m_T^2}{4E^2}\Bigg) n_-^\mu
 - \frac{m_T^2}{4 E^2} n_+^\mu\Bigg], \non \\
\epsilon^{*(\lambda)\mu\nu}_\perp n_\nu^+ &=& \Bigg(\epsilon^{*(\lambda)\mu\nu} - \frac{\epsilon^{*(\lambda)\nu\alpha}
n^+_\alpha}{2}\,n_-^\mu- \frac{\epsilon^{*(\lambda)\nu\alpha} n^-_\alpha}{2}\,n_+^\mu \Bigg)n_\nu^+\delta_{\lambda,\pm1}\,,
\end{eqnarray}
which in turn imply that
\be
\epsilon^{*(\lambda)}_{\mu\nu}
  n_+^\mu n_+^\nu \left({m_T\over 2E}\right)^2 &=& \sqrt{2\over 3}\,\delta_{\lambda,0}\,, \non \\
\epsilon^{*(\lambda)}_{\bot\mu\nu}
  n_+^\nu \left({m_T\over 2E}\right) &=& \sqrt{1\over 2}\,\epsilon^*_\mu(\pm1)\delta_{\lambda,\pm1}\,.
\en
The projector on the transverse polarization states in
the helicity basis reads
 \be \label{eq:MT-+}
 M^T_{\mp1}(u) &=& -i {f_T^\perp\over 4\sqrt{2}}\,E
 \Bigg\{
 \not\! \epsilon^{*}(\mp1)\not\! n_-\Phi^T_\perp(u) \non \\
 && + \frac{1}{2} \frac{f_T}{f_T^\perp} \frac{m_T}{E}
   \Bigg[
 \not\! \epsilon^{*}(\mp1)(1-\gamma_5) \left(g_v(u)\pm {g'_a(u)\over 2}\right)+\not\!
\epsilon^{*}(\mp1)(1+\gamma_5)
\left(g_v(u)\mp {g'_a(u)\over 2}\right) \non  \\
&&  -E\not\! n_-(1-\gamma_5)\left(\int^u_0
dv(2\Phi^T_\parallel(v)-g_v(v)) \mp{g_a(u)\over
2}\right)\epsilon^*_{\nu}(\mp1){\partial\over \partial k_{\perp\nu}}   \\
&&  -E\not\! n_-(1+\gamma_5)\left(\int^u_0 dv(2\Phi^T_\parallel(v)-g_v(v))
\pm{g_a(u)\over 2}\right)\epsilon^*_{\nu}(\mp1){\partial\over \partial k_{\perp\nu}}
\Bigg] +{\cal O} \left({m_T^2\over E^2}\right) \Bigg\}. \non
 \en
After applying the Wandzura-Wilczek relations Eq. (\ref{eq:T-ww}), the transverse
helicity projector (\ref{eq:MT-+}) can be
simplified to
  \be \label{eq:MT-+simple}
 M^T_{\mp1}(u) &=& -i {f_T^\perp\over 4\sqrt{2}}\,E
 \Bigg\{
 \not\! \epsilon^{*}(\mp1)\not\! n_-\Phi^T_\perp(u) \nonumber\\
 & & +\frac{f_T}{f_T^\perp} {m_T\over E} \left[
\epsilon^{*}_{\nu}(\mp1)\Phi^T_+(u)\left(\gamma^\nu(1\mp\gamma_5) +
 uE\not\! n_-(1\mp\gamma_5){\partial\over \partial k_{\perp\nu}}\right)\right.
 \nonumber \\
&&
 + \left. \epsilon^{*}_{\nu}(\mp1)\Phi^T_-(u)\left(\gamma^\nu(1\pm\gamma_5) -\bar{u}E\not\! n_-(1\pm\gamma_5){\partial\over \partial k_{\perp\nu}}\right) \right]
 +{\cal O}\left({m_T^2\over E^2} \right) \Bigg\},
 \en
to be compared with
  \be \label{eq:VVMV-+simple}
 M^V_{\mp1}(u) &=& -i {f_V^\perp\over 4}\,E
 \Bigg\{
 \not\! \epsilon^{*}(\mp1)\not\! n_-\Phi_\perp^V(u) \nonumber\\
 & & +\frac{f_V}{f_V^\perp} {m_V\over E} \left[
\epsilon^{*}_{\nu}(\mp1)\Phi_+^V(u)\left(\gamma^\nu(1\mp\gamma_5) +
 uE\not\! n_-(1\mp\gamma_5){\partial\over \partial k_{\perp\nu}}\right)\right.
 \nonumber \\
&&
 + \left. \epsilon^{*}_{\nu}(\mp1)\Phi_-^V(u)\left(\gamma^\nu(1\pm\gamma_5) -\bar{u}E\not\! n_-(1\pm\gamma_5){\partial\over \partial k_{\perp\nu}}\right) \right]
 +{\cal O}\left({m_V^2\over E^2} \right) \Bigg\}
 \en
for the vector meson.
The longitudinal projector for the tensor meson can be recast as
 \begin{eqnarray} \label{eq:VproL--simple}
M^T_\parallel(\lambda=0)
  &=& -i \frac{f_T}{4} \, \sqrt{\frac{2}{3}} E \Bigg\{
 \not\! n_- \,\Phi^T_\parallel(u)
+ \frac{3}{2}\frac{f_T^\perp}{f_T} \frac{m_T}{E} \Phi_t(u) \non \\
 & &\times \Bigg[ -\frac{i}{2}\,\sigma_{\mu\nu} \,  n_-^\mu  n_+^\nu \, (u-\bar{u})
 - \, iE u\bar{u} \
 \sigma_{\mu\nu} n_-^\mu \, \frac{\partial}{\partial k_\perp{}_\nu}
  -\mathbf{1} \Bigg]\,
  +{\cal O} \left({m_T^2\over E^2}\right) \Bigg\}
  \,,
\end{eqnarray}
to be compared with
 \begin{eqnarray} \label{eq:VVproL--simple}
M^V_\parallel(\lambda=0)
  &=& -i\frac{f_V}{4}  E \Bigg\{
 \not\! n_- \,\Phi_\parallel^V(u)
+  \frac{f_V^\perp}{f_V} \frac{m_V}{E} \Phi_v(u) \non \\
 & &\times \Bigg[ -\frac{i}{2}\,\sigma_{\mu\nu} \,  n_-^\mu  n_+^\nu \, (u-\bar{u})
 - \, iE u\bar{u} \
 \sigma_{\mu\nu} n_-^\mu \, \frac{\partial}{\partial k_\perp{}_\nu}
  -\mathbf{1} \Bigg]\,
  +{\cal O} \left({m_V^2\over E^2}\right) \Bigg\}
  \,
\end{eqnarray}
for the vector meson.

\subsection{A summary of input parameters}

It is useful to summarize all the input parameters we have used in this work. Some of the input quantities are collected in Table \ref{tab:input}.

The Wilson coefficients $c_i(\mu)$ at various scales, $\mu=4.4$ GeV, 2.1 GeV,
1.45 GeV and 1 GeV are taken from \cite{Groot}. For the renormalization scale
of the decay amplitude, we choose $\mu=m_b(m_b)$. However, as will be discussed
below, the hard spectator and annihilation contributions will be evaluated at
the hard-collinear scale $\mu_h=\sqrt{\mu\Lambda_h}$ with $\Lambda_h\approx 500
$ MeV.

%%%%%%%%%%%%%%%%%%%%%%%%%%%%%
\begin{table}[tbp!]
\caption{
Input parameters.
The values of the scale dependent quantities $f^\perp_V(\mu)$ and $a^{\bot,V}_{1,2}(\mu)$ are
given for $\mu=1\,\rm{GeV}$. The values of Gegenbauer moments are taken from \cite{Ball2007} and Wolfenstein parameters  from \cite{CKMfitter}.
} \label{tab:input}
\begin{center}
\begin{tabular}{|c||c|c|c|c|c|c|c|c|}
\hline\hline
\multicolumn{7}{|c|}{Light vector mesons \cite{BallfV,Ball2007}} \\
\hline
$V$  & $f_V({\rm MeV})$ &
 $f^\perp_V({\rm MeV})$ & $a^{\parallel,V}_1$ & $a^{\parallel,V}_2$ & $a^{\bot,V}_1$ & $a^{\bot,V}_2$\\
\hline
$\rho$ & $216\pm3$ & $165\pm 9$ & 0 & $0.15\pm0.07$ &  0 & $0.14\pm0.06$ \\
$\omega$ & $187\pm5$ & $151\pm 9$ & 0 & $0.15\pm 0.07$ & 0 & $0.14\pm0.06$ \\
$\phi$ & $215\pm5$ & $186\pm 9$ & 0 & $0.18\pm 0.08$ & 0 & $0.14\pm0.07$\\
$K^*$ & $220\pm5$ & $185\pm 10$ & $0.03\pm 0.02$ & $0.11\pm 0.09$ & $0.04\pm0.03$ & $0.10\pm0.08$ \\
\hline\hline
\multicolumn{7}{|c|}{Light tensor mesons \cite{CKY}}  \\
\hline
 \multicolumn{2}{|c|}{$T$} & \multicolumn{2}{|c|}{$f_T({\rm MeV})$} &
\multicolumn{2}{|c|}{$f_T^\bot({\rm MeV})$} & $a_1^{\parallel,T},a_1^{\bot,T}$ \\
\hline
 \multicolumn{2}{|c|}{$f_2(1270)$} & \multicolumn{2}{|c|}{$102\pm6$} & \multicolumn{2}{|c|}{$117\pm25$} & ${5\over 3}$ \\
 \hline
 \multicolumn{2}{|c|}{$f'_2(1525)$} & \multicolumn{2}{|c|}{$126\pm4$} & \multicolumn{2}{|c|}{$65\pm12$} & ${5\over 3}$ \\
 \hline
 \multicolumn{2}{|c|}{$a_2(1320)$} & \multicolumn{2}{|c|}{$107\pm6$} & \multicolumn{2}{|c|}{$105\pm21$} & ${5\over 3}$ \\
 \hline
 \multicolumn{2}{|c|}{$K^*_2(1430)$} & \multicolumn{2}{|c|}{$118\pm5$} & \multicolumn{2}{|c|}{$77\pm14$} & ${5\over 3}$ \\
\hline\hline
\multicolumn{7}{|c|}{$B$ mesons} \\
\hline
$B$ & \multicolumn{2}{|c|}{$m_B({\rm GeV}$)} & $\tau_B({\rm ps})$ &
\multicolumn{2}{|c|}{$f_B({\rm MeV})$} & $\lambda_B({\rm MeV})$ \\
\hline
$B_u$ & \multicolumn{2}{|c|}{$5.279$} & $1.638$ &
\multicolumn{2}{|c|}{$210\pm 20$} & $300\pm 100$ \\
\hline
$B_d$ & \multicolumn{2}{|c|}{$5.279$} & $1.525$ &
\multicolumn{2}{|c|}{$210\pm 20$} & $300\pm 100$ \\
\hline\hline
%$B_s$ & \multicolumn{2}{|c|}{$5.366$} & $1.472$ &
%\multicolumn{2}{|c|}{$230\pm 20$} & $300\pm 100$ \\
%\hline\hline
\multicolumn{7}{|c|}{Form factors at $q^2=0$ \cite{Ball2007,Ball:BV}} \\
\hline
\multicolumn{2}{|c|}{$F^{BK}_{0,1}(0)$} &
$A^{B K^*}_0(0)$ &
$A^{B K^*}_1(0)$ &
$A^{B K^*}_2(0)$ &
\multicolumn{2}{|c|}{$V^{B K^*}_0(0)$}  \\
\hline
\multicolumn{2}{|c|}{$0.35\pm0.04$} &
$0.374\pm0.033$ &
$0.292\pm0.028$ &
$0.259\pm0.027$ &
\multicolumn{2}{|c|}{$0.411\pm0.033$}  \\
\hline
\multicolumn{2}{|c|}{$F^{B\pi}_{0,1}(0)$} &
$A^{B \rho}_0(0)$ &
$A^{B \rho}_1(0)$ &
$A^{B \rho}_2(0)$ &
\multicolumn{2}{|c|}{$V^{B \rho}_0(0)$}  \\
\hline
\multicolumn{2}{|c|}{$0.25\pm0.03$} &
$0.303\pm0.029$ &
$0.242\pm0.023$ &
$0.221\pm0.023$ &
\multicolumn{2}{|c|}{$0.323\pm0.030$}  \\
\hline
\multicolumn{2}{|c|}{$F^{B\eta_q}_{0,1}(0)$} &
$A^{B \omega}_0(0)$ &
$A^{B \omega}_1(0)$ &
$A^{B \omega}_2(0)$ &
\multicolumn{2}{|c|}{$V^{B \omega}_0(0)$}  \\
\hline
\multicolumn{2}{|c|}{$0.296\pm0.028$} &
$0.281\pm0.030$ &
$0.219\pm0.024$ &
$0.198\pm0.023$ &
\multicolumn{2}{|c|}{$0.293\pm0.029$}  \\
\hline
\hline
\multicolumn{7}{|c|}{Quark masses} \\
\hline
\multicolumn{2}{|c|}{$m_b(m_b)/{\rm GeV}$} &
$m_c(m_b)/{\rm GeV}$ & \multicolumn{2}{|c|}{$m_c^{\rm pole}/m_b^{\rm pole}$} &
\multicolumn{2}{|c|}{$m_s(2.1~{\rm GeV})/{\rm GeV}$}   \\
\hline
\multicolumn{2}{|c|}{$4.2$} &
$0.91$ & \multicolumn{2}{|c|}{$0.3$} &
\multicolumn{2}{|c|}{$0.095\pm0.020$}  \\
\hline\hline
\multicolumn{7}{|c|}{Wolfenstein parameters \cite{CKMfitter}} \\
\hline
\multicolumn{2}{|c|}{$A$} & $\lambda$ &
$\bar\rho$ & $\bar\eta$ &
\multicolumn{2}{|c|}{$\gamma$}  \\
\hline
\multicolumn{2}{|c|}{$0.812$} & $0.22543$ &
$0.144$ & $0.342$ &
\multicolumn{2}{|c|}{$(67.2\pm3.9)^\circ$}  \\
\hline
\hline
\end{tabular}
\end{center}
\end{table}
%%%%%%%%%%%%%%%%%%%%%%%%%%%%%%%

\section{$B\to TP, TV$ decays}
Within the framework of QCD factorization \cite{BBNS}, the effective
Hamiltonian matrix elements are written in the form
\begin{equation}\label{fac}
   \langle M_1M_2 |{\cal H}_{\rm eff}|\overline B\rangle
  \! =\! \frac{G_F}{\sqrt2}\sum_{p=u,c} \! \lambda_p^{(q)}\,
\!   \langle M_1M_2 |{\cal T_A}^{p,h}\!+\!{\cal
T_B}^{p,h}|\overline B\rangle \,,
\end{equation}
where $\lambda_p^{(q)}\equiv V_{pb}V_{pq}^*$ with $q=s,d$,
and the superscript $h $ denotes the helicity of the final-state meson. For decays involving a pseudoscalar in the final state, $h$ is equivalent to zero.
${\cal T_A}^{p,h}$ describes contributions from naive factorization, vertex
corrections, penguin contractions and spectator scattering expressed
in terms of the flavor operators $a_i^{p,h}$, while ${\cal T_B}^{p,h}$
contains annihilation topology amplitudes characterized by  the
annihilation operators $b_i^{p,h}$. In general, ${\cal T_A}^{p,h}$ can be expressed in terms of
$c\,\alpha_i^{p,h}(M_1 M_2)\, X^{(\bar B M_1, M_2)}$ for $M_1=T$ or $c\,\alpha_i^{p,h}(M_1 M_2)\, \overline{X}^{(\bar B M_1, M_2)}$ for $M_2=T$,
where $c$ contains factors arising from flavor structures of final-state mesons,
$\alpha_i$ are functions of the Wilson coefficients (see Eqs. (\ref{eq:alphai--1}) and (\ref{eq:alphai--2})), and we have defined the notations
\be
X^{(\bar B T,P)} &\equiv& \la P|J^{\mu}|0\ra\la T|J'_{\mu}|\ov B\ra= -i2f_{P}A_0^{ B T}(m_{P}^2){m_T\over m_B}\epsilon^{*\mu\nu}(0)p_{B\mu}p_{B\nu},  \label{eq:X-1.1} \\
\overline{X}^{(\bar B P,T)} &\equiv& - 2 i f_T  m_B\,p_cF_1^{BP}(m_T^2)\,, \label{eq:X-1.2}
\en
for the decays $\bar B\to TP$, and
\be
X_h^{( \bar BT,V)} &\equiv & \la V |J^{\mu}|0\ra\la
T|J'_{\mu}|\ov B \ra =- if_{V}m_V\Bigg[
(e^*_T\cdot\epsilon^*_V) (m_{B}+m_{T})A_1^{ BT}(m_{V}^2) \label{eq:X-2.1} \non \\
&-& (e^*_T\cdot p_{_{B}})(\epsilon^*_V \cdot p_{_{B}}){2A_2^{
BT}(m_{V}^2)\over m_{B}+m_{T} } +
i\varepsilon_{\mu\nu\alpha\beta}\epsilon^{*\mu}_V e^{*\nu}_Tp^\alpha_{_{B}}
p_T^\beta\,{2V^{ BT}(m_{V}^2)\over m_{B}+m_{T} }\Bigg] ,
 \\
\overline{X}_h^{( \bar BV,T)} &\equiv &
\cases{ {if_{T}\over 2m_{V}}\left[
 (m_B^2-m_{V}^2-m_{T}^2)(m_B+m_{V})A_1^{BV}(m_T^2)-{4m_B^2p_c^2\over
 m_B+m_{V}}A_2^{BV}(m_T^2)\right]  ~{\rm for}\ h=0, \cr
 -if_{T}m_Bm_{T} \left[
 \left(1+{m_{V}\over m_B}\right)A_1^{BV}(m_T^2)\mp{2p_c\over
 m_B+m_{V}}V^{BV}(m_T^2)\right]  ~{\rm for}\ h=\pm1, \cr} \label{eq:X-2.2}
\en
for the decays $\bar B\to TV$, where $\overline X^{(\bar{B}P,T)}$ and $\overline{X}_h^{( \bar BV,T)}$ are expressed in the $B$ rest frame. Note that in the factorization limit, the factorizable amplitude $X^{(\bar BM_1,T)}\equiv \la T |J^{\mu}|0\ra\la
M|J'_{\mu}|\ov B \ra$ vanishes as the tensor meson cannot be produced through the $V-A$ or tensor current. Nevertheless, beyond the factorization approximation, contributions proportional to the decay constant $f_T$ of the tensor meson defined in Eq. (\ref{eq:fT}) can be produced from vertex, penguin and
spectator-scattering corrections.

To evaluate the helicity amplitudes of $B\to TV$, we work in the rest frame of the $B$ meson and assume that the tensor (vector) meson moves along the $-z$ ($z$) axis. The momenta  are thus given by
\be
p^\mu_B=(m_B,0,0,0), \qquad p_T^\mu=(E_T, 0,0,-p_c), \qquad p_V^\mu=(E_V,0,0,p_c).
\en
The polarization tensor $\epsilon^{\mu\nu}_{(\lambda)}$ of the massive tensor meson with helicity $\lambda$ can be constructed in terms of the polarization vectors of a massive vector state
\be
\epsilon^{*\mu}_T(0)=(p_c,0,0, -E_T)/m_T, \qquad \epsilon^{*\mu}_T(\pm)=(0,\mp 1,-i,0)/\sqrt{2}\,.
\en
For the vector meson moving along the $z$ direction, its polarization vectors are
\be
\epsilon^{*\mu}_V(0)=(p_c,0,0,E_V)/m_V, \qquad \epsilon^{*\mu}_V(\pm)=(0,\mp 1,i,0)/\sqrt{2}\,,
\en
where we have followed the Jackson convention, namely, in the
$\overline B$ rest frame, one of the vector or tensor mesons
is moving along the $z$ axis of the coordinate system and the
other along the $-z$ axis, while the $x$ axes of both daughter
particles are parallel \cite{T'Jampens}.
The longitudinal ($h=0$) and transverse ($h=\pm 1$) components of factorization amplitudes $X^{( \bar BT,V)}_h$ then have the expressions
 \be \label{eq:Xh}
 X_0^{(\ov BT,V)} &=& {if_{V}\over 2m_{T}^2}p_c\sqrt{2\over 3}\left[
 (m_B^2-m_{V}^2-m_{T}^2)(m_B+m_{T})A_1^{BT}(m^2_V)-{4m_B^2p_c^2\over
 m_B+m_{T}}A_2^{BT}(m^2_V)\right], \non \\
 X_\pm^{(\ov BT,V)} &=& -if_{V}m_Bm_{V}{p_c\over \sqrt{2}m_T}\left[
 \left(1+{m_{T}\over m_B}\right)A_1^{BT}(m^2_V)\mp{2p_c\over
 m_B+m_{T}}V^{BT}(m^2_V)\right].
 \en
Likewise, the factorizable $B\to TP$ amplitude can be simplified to
\be \label{eq:XTP}
X^{(\bar B T,P)} = -i2\sqrt{2\over 3}\,f_{P}{m_B\over m_T}p_c^2A_0^{ B T}(m_{P}^2).
\en

The flavor operators $a_i^{p,h}$ are basically the Wilson coefficients
in conjunction with short-distance nonfactorizable corrections such
as vertex corrections and hard spectator interactions. In general,
they have the expressions \cite{BBNS,BN}
 \be \label{eq:ai}
  a_i^{p,h}(M_1M_2) &=&
 \left(c_i+{c_{i\pm1}\over N_c}\right)N_i^h(M_2)  + {c_{i\pm1}\over N_c}\,{C_F\alpha_s\over
 4\pi}\Big[V_i^h(M_2)+{4\pi^2\over N_c}H_i^h(M_1M_2)\Big]+P_i^{p,h}(M_2),~~~~
 \en
where $i=1,\cdots,10$,  the upper (lower) signs apply when $i$ is
odd (even), $c_i$ are the Wilson coefficients,
$C_F=(N_c^2-1)/(2N_c)$ with $N_c=3$, $M_2$ is the emitted meson
and $M_1$ shares the same spectator quark with the $B$ meson. The
quantities $V_i^h(M_2)$ account for vertex corrections,
$H_i^h(M_1M_2)$ for hard spectator interactions with a hard gluon
exchange between the emitted meson and the spectator quark of the
$B$ meson and $P_i(M_2)$ for penguin contractions.  The expression
of the quantities $N_i^h(M_2)$, which are relevant to the factorization amplitudes, reads
\be \label{eq:Ni}
 N_i^h(V) &=& \cases{0 & $i=6,8$, \cr
                 1 & else, \cr} \nonumber\\
 N_i^h(T) &=& 0, \qquad N_i(P) = 1.
\en

\vskip 0.2in \noindent {\it \underline{Vertex corrections}} \vskip 0.1in

The vertex corrections are given by
\begin{equation}\label{vertex0}
   V^0_i(M_{2}) = \left\{\,\,
   \begin{array}{ll}
    {\displaystyle C(M_2)\int_0^1\!dx\,\Phi_\parallel^{M_2}(x)\,
     \Big[ 12\ln\frac{m_b}{\mu} - 18 + g(x) \Big]}  & \qquad
     (i=\mbox{1--4},9,10)\,, \\[0.4cm]
   {\displaystyle C(M_2)\int_0^1\!dx\,\Phi_\parallel^{M_2}(x)\,
     \Big[ - 12\ln\frac{m_b}{\mu} + 6 - g(1-x) \Big]}  & \qquad
     (i=5,7)\,, \\[0.4cm]
   {\displaystyle \frac{1}{C(M_2)}\int_0^1\!dx\, \Phi_{m_2}(x)\,\Big[ -6 + h(x) \Big]}
     & \qquad (i=6,8)\,,
   \end{array}\right.
\end{equation}
\begin{equation}\label{vertexpm}
  V^\pm_i(M_2) = \left\{\,\,
   \begin{array}{ll}
    {\displaystyle D(M_2)\int_0^1\!dx\, \Phi^{M_2}_\pm(x) \,
     \Big[ 12\ln\frac{m_b}{\mu} - 18 + g_T(x) \Big]}  & \
     (i=\mbox{1--4},9,10)\,, \\[0.4cm]
   {\displaystyle D(M_2)\int_0^1\!dx\, \Phi^{M_2}_\mp(x) \,
     \Big[ - 12\ln\frac{m_b}{\mu} + 6 - g_T(1-x) \Big]} & \
     (i=5,7)\,, \\[0.4cm]
  0 & \ (i=6,8)\,,
   \end{array}\right.
\end{equation}
with
 \begin{eqnarray}
 g(x)&=& 3\Bigg( \frac{1-2x}{1-x} \ln x -i\pi\Bigg)\nonumber\\
  && + \Big[ 2{\rm Li}_2(x) -\ln^2x +\frac{2\ln x}{1-x} -(3+2i\pi)\ln x -
 (x \leftrightarrow 1-x)\Big]\,, \nonumber\\
 h(x)&=&  2{\rm Li}_2(x) -\ln^2x -(1+2i\pi)\ln x -
 (x \leftrightarrow 1-x)\,, \nonumber\\
 g_T(x)&=& g(x) + \frac{\ln x}{\bar x}\,,
 \end{eqnarray}
and
 \begin{eqnarray}
  C(P)=C(V) = D(V)=1,\quad C(T)=\sqrt{\frac{2}{3}},\quad D(T)=\frac{1}{\sqrt{2}},\quad D(P)=0,
 \end{eqnarray}
where $\bar x=1-x$, $\Phi^M_\parallel$ is a twist-2 light-cone distribution
amplitude of the meson $M$, $\Phi_m$ (for the longitudinal
component), and $\Phi_\pm$ (for transverse components) are twist-3
ones. Specifically, $\Phi_m=\Phi_t,~\Phi_v,~\Phi_p$ for $M=T,V,P$, respectively. The expressions of $C(T)$ and $D(T)$ are obtained by comparing Eqs. (\ref{eq:MT-+simple})-(\ref{eq:VVproL--simple}).

\vskip 0.2in \noindent{\it \underline{Hard spectator terms}} \vskip 0.1in

$H^h_i(M_1 M_2)$ arise from  hard spectator interactions with a hard
gluon exchange between the emitted  meson and the spectator quark of
the $\overline B$ meson. $H^0_i(M_1 M_2)$ have the expressions:
\begin{eqnarray}\label{eq:sepc01}
  H^0_i(M_1 M_2)= \pm {if_B f_{M_1} f_{M_2} \over X^{(\overline{B} M_1,
  M_2)}_0}\,{m_B\over\lambda_B} C(M_2) \int^1_0 d u d v \,
 \Bigg[ C(M_1) \frac{\Phi^{M_1}_\parallel(u) \Phi^{M_2}_\parallel(v)}{\bar u \bar v}
 + r_\chi^{M_1}
  \frac{\Phi_{m_1} (u) \Phi^{M_2}_\parallel(v)}{ C(M_1) \bar u v}\Bigg],
 \hspace{0.1cm}
 \end{eqnarray}
for $i=1-4,9,10$,
\begin{eqnarray}\label{eq:spec02}
  H^0_i(M_1 M_2)= \mp {if_B f_{M_1} f_{M_2} \over X^{(\overline{B} M_1, M_2)}_0}
  \,{m_B\over\lambda_B}C(M_2) \int^1_0 d u d v \,
 \Bigg[ C(M_1) \frac{\Phi^{M_1}_\parallel(u) \Phi^{M_2}_\parallel(v)}{\bar u  v}
 + r_\chi^{M_1}
  \frac{\Phi_{m_1} (u) \Phi^{M_2}_\parallel(v)}{C(M_1)\bar u \bar v}\Bigg],
 \hspace{0.1cm}
 \end{eqnarray}
for $i=5,7$, and $H^0_i(M_1 M_2)=0$ for $i=6,8$, where the upper signs are for $TV$ modes and the lower ones for $TP$ modes. The transverse hard spectator terms $H^\pm_i(M_1
M_2)$ read
 \begin{eqnarray}
 H^-_i(M_1 M_2) &=&  {\sqrt{2} i f_B
 f^{\perp}_{M_1} f_{M_2} m_{M_2} \over m_B X^{({\overline B} {M_1}, M_2)}_-}
 \,{m_B\over\lambda_B}\int^1_0 dudv\,
 {\Phi^{M_{1}}_\perp(u)\Phi_-^{M_2}(v)\over \bar u^2 v}, \label{eq:H1m} \\
 H^+_i(M_1 M_2) &=& -
 \frac{\sqrt{2} i f_B f_{M_1} f_{M_2} m_{M_1} m_{M_2}}{m_B^2
  X^{({\overline B} M_1, M_2)}_+}\,{m_B\over\lambda_B}\int^1_0 dudv\, {(\bar u-v)
  \Phi_+^{M_1}(u)\Phi_+^{M_2}(v)\over \bar u^2\bar v^2}, \label{eq:H1p}
  \en
 for  $i=1-4,9,10$,
and
 \begin{eqnarray}
 H^-_i(M_1 M_2) &=&  - {\sqrt{2} i f_B
 f^{\perp}_{M_1} f_{M_2} m_{M_2} \over m_B X^{({\overline B} {M_1}, M_2)}_-}
 \,{m_B\over\lambda_B}\int^1_0 dudv\,
 {\Phi^{M_{1}}_\perp(u)\Phi_+^{M_2}(v)\over \bar u^2\bar v},\label{eq:H5m}
  \\
 H^+_i(M_1 M_2) &=& -
 \frac{\sqrt{2} i f_B f_{M_1} f_{M_2} m_{M_1} m_{M_2}}{  m_B^2
  X^{({\overline B} M_1, M_2)}_+}\,{m_B\over\lambda_B}\int^1_0 dudv\, {(u-v)
  \Phi_+^{M_1}(u)\Phi_-^{M_2}(v)\over \bar u^2 v^2}, \label{eq:H5p}
  \en
for $i=5,7$, and
 \begin{eqnarray} \label{eq:H6}
 H^-_i(M_1 M_2) &=& - {i f_B
 f_{M_1} f_{M_2} m_{M_2} \over \sqrt{2} m_B X^{({\overline B} {M_1}, M_2)}_-}
 {m_Bm_{M_1}\over m_{M_2}^2}\,
 \,{m_B\over\lambda_B}\int^1_0 dudv\,
 {\Phi^{M_{1}}_+(u)\Phi_\perp^{M_2}(v)\over v\bar u\bar v}, \label{eq:H6m} \\
 H^+_i(M_1 M_2) &=& 0,
 \en
for $i=6,8$. Since we consider only $TP$ and $TV$ modes in the present work, it is obvious that $M_1M_2=TV$ or $VT$ for the transverse components.

\vskip 0.2in \noindent{\it \underline{Penguin terms}} \vskip 0.1in

At order $\alpha_s$, corrections from penguin contractions are
present only for $i=4,6$. For $i=4$ we obtain
\begin{eqnarray}\label{eq:PK}
   P_4^{p,h}(M_2) &=& \frac{C_F\alpha_s}{4\pi N_c}\,\Bigg\{
    c_1  \bigg[ G^h_{M_2}(s_p)+g_{M_2}\bigg] \!
    + c_3 \!\bigg[  G^h_{M_2}(s_s) + G^h_{M_2}(1) +2g^h_{M_2}\bigg] \non \\ && + (c_4+c_6)\!
    \sum_{i=u}^b  \left[G^h_{M_2}(s_i)+g^{\prime h}_{M_2}\right]
%    && + \frac{3}{2}(c_8+c_{10})\!
%    \sum_{i=u}^b  e_i G^h_{M_2}(s_i)
%    + \frac{3}{2}c_9 [ e_{q'} G_{M_2}^h (s_{q'})- \frac{1}{3} %G_{M_2}^h (s_{b})]
     -2 c_{8g}^{\rm eff} G^h_g \Bigg\} \,,
\end{eqnarray}
where $s_i=m_i^2/m_b^2$ and the function $G^h_{M_2}(s)$ is given by
\begin{eqnarray}  \label{eq:GK}
 G^h_{M_2}(s) &=&
 4\int^1_0 du\,\Phi^{M_2,h}(u)\int^1_0 dx\,x\bar x
  \ln[s-\bar u x\bar x-i\epsilon], \nonumber \\
  g^h_{M_2} &=& \left( {4\over 3}\ln \frac{m_b}{\mu}+{2\over 3}\right)\int^1_0\Phi^{M_2,h}(x)dx, \non \\
   g^{\prime h}_{M_2} &=& {4\over 3}\ln \frac{m_b}{\mu}\int^1_0\Phi^{M_2,h}(x)dx,
 \end{eqnarray}
with $\Phi^{M_2,0}=C(M_2)\Phi^{M_2}_\|$, $\Phi^{M_2,\pm}=D(M_2)\Phi^{M_2}_\pm$.
For $i=6$, the result for the penguin contribution is
\begin{eqnarray} \label{eq:P6}
   P_6^{p,h}(M_2)=\frac{C_F\alpha_s}{4\pi N_c}\,\Bigg\{ \!
    c_1 \hat G^h_{M_2}(s_p)
    + c_3\bigg[ \hat G^h_{M_2}(s_s) + \hat G^h_{M_2}(1)
    \bigg]
    + (c_4+c_6)\sum_{i=u}^b  \hat G^h_{M_2}(s_i) \! \Bigg\}.
    \hspace{0.5cm}
\end{eqnarray}
In analogy with (\ref{eq:GK}), the function $\hat G_{M_2}(s)$ is
defined as
 \begin{eqnarray} \label{eq:P6.1}
 \hat G^0_{M_2}(s) &=&  \frac{4}{C(M_2)}\int^1_0 du\, \Phi_{m_2}(u) \int^1_0 dx\,x\bar x
  \ln[s-\bar u x\bar x-i\epsilon], \nonumber \\
 \hat G^\pm_{M_2}(s) &=& 0\,.
 \end{eqnarray}
Therefore, the transverse penguin contractions vanish for $i=6,8$:
$P_{6,8}^{\pm,p}=0$. Note that we have factored out the $r_\chi^{M_2}$ term in
Eq. (\ref{eq:P6}) so that when the vertex correction $V_{6,8}$ is neglected, $a_6^0$ will
contribute to the decay amplitude in the product $r_\chi^{M_2}a_6^0\approx r_\chi^{M_2}P_6^0$.

For $i=8,10$ we find
\begin{equation}
   P_8^{p,h}(M_2) =  \frac{\alpha_{\rm em}}{9\pi N_c}\,(c_1+N_c c_2)\,
   \hat G^h_{M_2}(s_p) \,,
\end{equation}
\begin{equation}\label{PKEW}
   P_{10}^{p,h}(M_2) = \frac{\alpha_{\rm em}}{9\pi N_c} \Bigg\{
   (c_1+N_c c_2) \bigg[ G^h_{M_2}(s_p)+2g_{M_2}\bigg]
   - 3c^{\rm eff}_\gamma  G_g^h \bigg\}.
\end{equation}
For $i=7,9$,
 \be \label{radphoton}
  P_{7,9}^{-,p}(M_2)
 =  -{\alpha_{\rm em}\over 3\pi}C_{7\gamma}^{\rm eff}{m_Bm_b\over m_{M_2}^2}
  +{2\alpha_{\rm em}\over 27\pi}(c_1+N_c c_2)
   \left[\delta_{pc}\ln{m_c^2\over\mu^2}+\delta_{pu}\ln{\nu^2\over \mu^2}+1\right],
 \en
for $M_2=\rho^0, \omega, \phi$, and vanish otherwise. Here the first term is an electromagnetic penguin contribution to the transverse helicity amplitude enhanced by a factor of $m_Bm_b/m_{M_2}^2$, as first pointed out in  \cite{BenekeEWP}. Note that the quark loop contains an ultraviolet divergence for both transverse and longitudinal components
which must be subtracted in accordance with the scheme used to define the Wilson coefficients. The scale and scheme dependence after subtraction is required to cancel the scale and scheme dependence of the electroweak
penguin coefficients. Therefore, the scale $\mu$ in the above equation is the same as the one appearing in the expressions for the penguin corrections, e.g. Eq. (\ref{eq:GK}). On the other hand, the scale $\nu$ is referred  to the scale of the decay constant $f_{M_2} (\nu)$ as the operator $\bar q\gamma^\mu q$ has a non-vanishing anomalous dimension in the presence of electromagnetic interactions \cite{BN}. The $\nu$ dependence of Eq.~(\ref{radphoton}) is compensated by that of $f_{M_2} (\nu)$.

The relevant integrals for the dipole operators $O_{g,\gamma}$ are
  \begin{eqnarray}
&&  G^0_g = C(M_2) \int^1_0 du\,{\Phi^{M_2}_\|(u)\over \bar u}\,, \\
&&  G^\pm_g =D(M_2)\int^1_0 {du\over \bar u}\,\Bigg[  \frac{1}{2}
 \bigg(\bar u \Phi_-^{M_2} (u) - u\Phi_+^{M_2}(u) \bigg) -\bar u \Phi_\pm^{M_2} (u) +
 \frac{1}{2}
 \bigg(\bar u \Phi_-^{M_2}(u) + u\Phi_+^{M_2}(u) \bigg) \Bigg]. \nonumber
 \label{eq:cg}
 \end{eqnarray}
Using Eq.~(\ref{eq:T-ww}), $G_g^\pm$ can be further reduced to
\begin{eqnarray}
G_g^+ &=& D(M_2)\int_0^1 du \Big[ \Phi_-^{M_2}(u) -\Phi_+^{M_2}(u) \Big] =0, \nonumber\\
G_g^- &=& 0.
\end{eqnarray}
Hence,  $G_g^{\pm}$ in Eq.~(\ref{eq:cg}) are actually equal to zero.
It was first pointed out by Kagan \cite{Kagan} that the dipole
operators $Q_{8g}$ and $Q_{7\gamma}$ do not contribute to the
transverse penguin amplitudes at ${\cal O}(\alpha_s)$ due to angular
momentum conservation.

\vskip 0.2in \noindent{\it \underline{Annihilation topologies}} \vskip 0.1in
 The weak annihilation contributions to the decay  $\overline B\to M_{1}M_2$ (with $M _1 M_2 \equiv VT$ or $TV$) can be described in terms of the building blocks $b_i^{p,h}$ and $b_{i,{\rm EW}}^{p,h}$
\begin{eqnarray}\label{eq:h1ksann}
\frac{G_F}{\sqrt2} \sum_{p=u,c} \! \lambda_p\, \!\langle M_{1}M_2
|{\cal T_B}^{p,h} |\overline B^0\rangle &=&
i\frac{G_F}{\sqrt{2}}\sum_{p=u,c} \lambda_p
 f_B f_{M_1} f_{M_{2}}\sum_i (d_ib_i^{p,h}+d'_ib_{i,{\rm EW}}^{p,h}).
\end{eqnarray}
The building blocks have the expressions
 \be \label{eq:bi}
 b_1 &=& {C_F\over N_c^2}c_1A_1^i, \qquad\quad b_3={C_F\over
 N_c^2}\left[c_3A_1^i+c_5(A_3^i+A_3^f)+N_cc_6A_3^f\right], \non \\
 b_2 &=& {C_F\over N_c^2}c_2A_1^i, \qquad\quad b_4={C_F\over
 N_c^2}\left[c_4A_1^i+c_6A_2^f\right], \non \\
 b_{\rm 3,EW} &=& {C_F\over
 N_c^2}\left[c_9A_1^{i}+c_7(A_3^{i}+A_3^{f})+N_cc_8A_3^{i}\right],
 \non \\
 b_{\rm 4,EW} &=& {C_F\over
 N_c^2}\left[c_{10}A_1^{i}+c_8A_2^{i}\right],
 \en
where for simplicity we have omitted the superscripts $p$ and $h$ in above
expressions.  The subscripts 1,2,3 of $A_n^{i,f}$ denote the annihilation
amplitudes induced from $(V-A)(V-A)$, $(V-A)(V+A)$ and $(S-P)(S+P)$ operators,
respectively, and the superscripts $i$ and $f$ refer to gluon emission from the
initial and final-state quarks, respectively. Following \cite{BN} we choose the
convention that
$M_2$ contains an antiquark from the weak vertex with longitudinal fraction
$\bar v$, while $M_1$ contains a quark from the weak vertex with momentum
fraction $u$. The explicit expressions of weak
annihilation amplitudes are:
\begin{eqnarray}
A_1^{i,\,0}(M_{1} M_{2}) &=&
 \sqrt{\frac{2}{3}}\pi\alpha_s \int_0^1\! du \,dv\, \Bigg\{
 \,\Phi_\parallel^{M_1}(u) \Phi_\parallel^{M_2}(v)\,
 \left[\frac{1}{u(1-\bar u v)}+\frac{1}{u\bar v^2}\right]
 \nonumber\\
 && \ \ \ \ \ \  \ \ \  \ \ \ \ \ \ \ \ \
 \mp
 \frac{3}{2}\, r_\chi^{M_1}r_\chi^{M_2} \,\Phi_{m_1}(u)\, \Phi_{m_2}(v)
\frac{2}{u\bar{v}} \Bigg\}\,, \label{eq:A1i0} \\
 A_1^{i,\,-}(M_1 M_2)  &=& -
 \pi\alpha_s {\sqrt{2}m_{M_1}m_{M_2}\over m_B^2}\int_0^1\! du \,dv\,
 \Bigg\{
 \,\Phi^{M_1}_-(u)\,\Phi_-^{M_2} (v)\left[  \frac{\bar u+\bar v}{u^2 \bar{v}^2}
 +{1\over (1-\bar u v)^2}\right]
 \Bigg\}\,,  \\
  A_1^{i,\,+}(M_1 M_2)  &=& -
 \pi\alpha_s {\sqrt{2}m_{M_1}m_{M_2}\over m_B^2}\int_0^1\! du \,dv\,
 \Bigg\{
 \,\Phi^{M_1}_+(u)\,\Phi_+^{M_2} (v)\left[ {2\over u\bar v^3}- \frac{v}{(1-\bar uv)^2}
 -{v\over \bar v^2(1-\bar u v)}\right]
 \Bigg\}\,,\hspace{1cm}\label{eq:A1ip}
 \end{eqnarray}

\begin{eqnarray}
A_2^{i,\,0}(M_{1} M_{2}) &=&
 \sqrt{\frac{2}{3}}\pi\alpha_s \int_0^1\! du \,dv\, \Bigg\{
 \Phi_\parallel^{M_1}(u) \Phi_\parallel^{M_2}(v)\,
 \left[\frac{1}{\bar v(1-\bar u v)}+\frac{1}{u^2\bar v}\right]
 \nonumber\\
 && \ \ \ \ \ \  \ \ \  \ \ \ \ \ \ \ \ \
 \mp
  \frac{3}{2}\, r_\chi^{M_1}r_\chi^{M_2} \,\Phi_{m_1}(u)\, \Phi_{m_2}(v)
\frac{2}{u\bar{v}} \Bigg\}\,, \label{eq:A2i0} \\
 A_2^{i,\,-}(M_1 M_2)  &=& -
 \pi\alpha_s {\sqrt{2}m_{M_1}m_{M_2}\over m_B^2}\int_0^1\! du \,dv\,
 \Bigg\{\Phi^{M_1}_+(u)\,\Phi_+^{M_2} (v)
 \non \\
 && \times \left[  \frac{u+v}{u^2 \bar{v}^2}
 +{1\over (1-\bar u v)^2}\right]
 \Bigg\}\,,  \\
   A_2^{i,\,+}(M_1 M_2)  &=& -
 \pi\alpha_s {\sqrt{2}m_{M_1}m_{M_2}\over m_B^2}\int_0^1\! du \,dv\,
 \Bigg\{ \Phi^{M_1}_-(u)\,\Phi_-^{M_2} (v)
 \non \\
 && \times\left[ {2\over u^3\bar v}
   - \frac{\bar u}{(1-\bar uv)^2}
 -{\bar u\over u^2(1-\bar u v)}\right]
 \Bigg\}\,, \hspace{1cm}\label{eq:A2ip}
 \end{eqnarray}

\begin{eqnarray}
A_3^{i,\,0}(M_1 M_2) &=&
 \pi\alpha_s \int_0^1\! du \,dv\, \Bigg\{ \frac{C(M_2)}{C(M_1)}
 r_\chi^{M_1}\Phi_{m_1}(u) \Phi_\parallel^{M_2}(v)\,
 \frac{2\bar u}{u\bar v(1-\bar u v)}
 \nonumber\\
 && \ \ \ \ \ \  \ \ \  \ \ \ \ \ \ \ \ \
 +
 \frac{C(M_1)}{C(M_2)} r_\chi^{M_2} \,\Phi_\parallel^{M_1}(u)\, \Phi_{m_2}(v)
\frac{2v}{u\bar{v}(1-\bar uv)} \Bigg\}\,, \label{eq:A3i0}
 \end{eqnarray}

\begin{eqnarray}
 A_3^{i,\,-}(M_1 M_2)  &=& -
 \frac{\pi\alpha_s}{\sqrt{2}} \int_0^1\! du \,dv\,
 \Bigg\{   -{m_{M_2}\over m_{M_1}} r_\chi^{M_1}
 \,\Phi^{M_1}_\perp(u)\,\Phi_-^{M_2} (v)\frac{2}{u\bar v(1-\bar uv)}\non \\
 && \ \ \ \ \ \ \ \ \ \ \ \ \ \ \ \ \ \ \ \ \ \
 +  {m_{M_1}\over m_{M_2}} r_\chi^{M_2}
 \,\Phi^{M_1}_+(u)\,\Phi_\perp^{M_2} (v)\frac{2}{u\bar v(1-\bar
 uv)}
 \Bigg\}\,, \hspace{1cm}\label{eq:A3im}
 \end{eqnarray}

\begin{eqnarray}
A_3^{f,\,0}(M_1 M_2) &=&
  \pi\alpha_s \int_0^1\! du \,dv\, \Bigg\{ \frac{C(M_2)}{C(M_1)}
  r_\chi^{M_1} \,\Phi_{m_1}(u) \Phi_\parallel^{M_2}(v)\,
 \frac{2(1+\bar v)}{u\bar{v}^2}
 \nonumber\\
 && \ \ \ \ \ \  \ \ \  \ \ \ \ \ \ \ \ \
 - \frac{C(M_1)}{C(M_2)}
  r_\chi^{M_2} \,\Phi_\parallel^{M_1}(u)\, \Phi_{m_2}(v)
\frac{2(1+u)}{u^2\bar{v}} \Bigg\}\,, \label{eq:A3f0}
 \end{eqnarray}

\begin{eqnarray}
 A_3^{f,\,-}(M_1 M_2)  &=& -
 \frac{\pi\alpha_s}{\sqrt{2}} \int_0^1\! du \,dv\, \Bigg\{
 \frac{m_{M_2}}{m_{M_1}} r_\chi^{M_1} \,\Phi^{M_1}_{\perp}(u)\,\Phi_-^{M_2} (v)
 \frac{2}{u^2 \bar{v}}
 \nonumber\\
 && \ \ \ \ \ \  \ \ \  \ \ \ \ \ \ \ \ \ \ \
  + \frac{m_{M_1}}{m_{M_2}} r_\chi^{M_2}\,\Phi_+^{M_1}(u)\, \Phi^{M_2}_{\perp}(v) \,
 \frac{2}{u\bar v^2 }
 \Bigg\}\,, \hspace{1cm}\label{eq:A3fm}
 \end{eqnarray}
and $A_1^{f,h}=A_2^{f,h}=A_3^{i,+}=A_3^{f,+}=0$. Here in the helicity amplitudes with $h=0$, the upper signs correspond to $(M_1, M_2)=(T,V)$, $(V,T)$, and $(V,P)$ and the lower ones to $(M_1, M_2)=(P,V)$. When  $(M_1, M_2)=(V, P)$, one has to add an overall minus sign to $A_2^{i,0}$. For $(M_1, M_2)=(P, V)$, one has to change the sign of the second term of $A_2^{i,0}$. Note that in this paper, we adopt the notations $A_j^{(i,f), 0} \equiv A_j^{(i,f)}$ for the $TP$ modes.

Since the annihilation contributions $A_{1,2}^{i,\pm}$ are
suppressed by a factor of $m_1m_2/m_B^2$ relative to other terms,
in the numerical analysis we will consider only the annihilation
contributions due to $A_3^{f,0}$, $A_3^{f,-}$, $A_{1,2,3}^{i,0}$ and $A_3^{i,-}$.

Finally, two remarks are in order: (i) Although the parameters $a_i(i\neq 6,8)$
and $a_{6,8}r_\chi$ are formally renormalization scale and $\gamma_5$ scheme
independent, in practice there exists some residual scale dependence in
$a_i(\mu)$ to finite order. To be specific, we shall evaluate the vertex
corrections to the decay amplitude at the scale $\mu=m_b$.  In
contrast, as stressed in \cite{BBNS}, the hard spectator and annihilation
contributions should be evaluated at the hard-collinear scale
$\mu_h=\sqrt{\mu\Lambda_h}$ with $\Lambda_h\approx 500 $ MeV. (ii) Power
corrections in QCDF always involve troublesome endpoint divergences. For
example, the annihilation amplitude has endpoint divergences even at twist-2
level and the hard spectator scattering diagram at twist-3 order is power
suppressed and possesses soft and collinear divergences arising from the soft
spectator quark. Since the treatment of endpoint divergences is model
dependent, subleading power corrections generally can be studied only in a
phenomenological way. We shall follow \cite{BBNS} to model the endpoint
divergence $X\equiv\int^1_0 dx/\bar x$ in the annihilation and hard spectator
scattering diagrams as
 \be \label{eq:XA}
 X_A=\ln\left({m_B\over \Lambda_h}\right)(1+\rho_A e^{i\phi_A}), \qquad
 X_H=\ln\left({m_B\over \Lambda_h}\right)(1+\rho_H e^{i\phi_H}),
 \en
with the unknown real parameters $\rho_{A,H}$ and $\phi_{A,H}$. For simplicity,
we shall assume that $X_A^h$ and $X_H^h$ are helicity independent; that is,
$X_A^-=X_A^+=X_A^0$ and $X_H^-=X_H^+=X_H^0$.

\section{Numerical results}
Let the general amplitude of $\ov B\to TP$ be
\be
\A_{\ov B\to TP}={G_F\over \sqrt{2}}\, \left(a X^{(\ov BT,P)} + \bar{a} \overline{X}^{(\ov BP,T)}\right).
\en
Its decay rate is given by
\be
\Gamma_{\ov B\to TP}={p_c\over 8\pi m_B^2}\left|\A_{\ov B\to TP}\right|^2.
\en
It follows from Eqs. (\ref{eq:X-1.2}) and (\ref{eq:XTP}) that
\be \label{eq:TPrate}
\Gamma_{\ov B\to TP}={G_F^2\over 6\pi}\,{f_P^2 p_c^5\over m_T^2}
\left|a A_0^{BT}(m_P^2)
 +
 \sqrt{\frac{3}{2}}\, \bar{a}\frac{f_T}{f_P}\frac{m_T}{p_c}
 \frac{F_1^{BP}(m_T^2)}{A_0^{BT}(m_P^2)} \right|^2,
\en
where $p_c$ is the center-of-mass momentum of the final-state particle $T$ or $P$. Note that the coefficient $\bar{a}$ vanishes in naive factorization.

The decay amplitude of $\ov B\to TV$ can be decomposed into three components,
one for each helicity of the final state: $\A_0,\A_+,\A_-$. The transverse
amplitudes defined in  the transversity basis are related to the helicity ones
via \be
 \A_{\parallel}=\frac{\A_++\A_-}{\sqrt2}, \qquad
 \A_{\bot}&=&\frac{\A_+-\A_-}{\sqrt2}.
 \label{eq:Atrans}
 \en
The decay rate can be expressed in
terms of these amplitudes as
 \be \label{eq:TVgeneralrate}
 \Gamma_{\ov B\to TV}=\frac{p_c}{8\pi m_B^2}(|\A_0|^2+|\A_+|^2+|\A_-|^2)
       =\frac{p_c}{8\pi m_B^2}(|\A_L|^2+|\A_{\parallel}|^2+|\A_{\bot}|^2).
 \en
Writing the general helicity amplitudes  as
\be
 \A_0(\ov B\to TV)
 &=&
 {G_F\over \sqrt{2}}\,
 \left( b^0 X_0^{(\ov BT,V)} + \bar{b}^0 \overline{X}_0^{(\ov BV,T)} \right), \\
 \A_\pm(\ov B\to TV)
 &=&
 {G_F\over \sqrt{2}}\,
 \left( b^\pm X_\pm^{(\ov BT,V)} + \bar{b}^\pm \overline{X}_\pm^{(\ov BV,T)} \right),
\en
where $X_{0,\pm}^{(\ov BT,V)}$ and $\overline{X}_{0,\pm}^{(\ov BV,T)}$ are given in Eqs. (\ref{eq:Xh}) and (\ref{eq:X-2.2}), respectively, and it is understood that the relevant CKM factors should be put back by the end of calculations,
the decay rate has the following explicit expression
\be \label{eq:TVrate}
\Gamma_{\ov B\to TV} = {G_F^2 \over 48\pi}\,{f_V^2\over m_T^4}(\alpha p_c^7+\beta p_c^5+\gamma p_c^4+\lambda p_c^3),
\en
with
\be
\alpha &=& 8{m_B^2\over (m_B+m_T)^2} |b^0 {\bf A}_2^{BT}|^2, \non \\
\beta &=& 6{m_V^2m_T^2\over (m_B+m_T)^2}
\left( |b^+|^2 ({\bf  V}_+^{BT})^2 +|b^-|^2 ({\bf  V}_-^{BT})^2 \right)
 -4(m_B^2-m_V^2-m_T^2)|b^0|^2 {\bf A}_1^{BT} {\bf A}_2^{BT},
\non \\
\gamma &=& 6{m_V^2m_T^2\over m_B}
 (|b^-|^2 {\bf A}_{1,-}^{BT} {\bf  V}_-^{BT}-|b^+|^2 {\bf A}_{1,+}^{BT} {\bf  V}_+^{BT}) ,  \\
\lambda &=& {(m_B+m_T)^2\over 2m_B^2}
 \Bigg[3(|b^+|^2 ({\bf A}_{1,+}^{BT})^2 +|b^-|^2 ({\bf A}_{1,-}^{BT})^2)m_V^2m_T^2
+|b^0|^2(m_B^2-m_V^2-m_T^2)^2 ({\bf A}_1^{BT})^2\Bigg] , \nonumber
\en
where we have adopted the shorthand notations,
\begin{eqnarray}
{\bf A}_1^{BT} &\equiv& A_1^{BT}(m_V^2) +\frac{\bar b^0}{b^0} \sqrt{\frac{3}{2}} \frac{f_T}{f_V}
\frac{m_T}{m_V} \frac{m_T}{p_c}  \frac{m_B+m_V}{m_B+m_T} A_1^{BV}(m_T^2),
 \\
{\bf A}_2^{BT} &\equiv& A_2^{BT}(m_V^2) +\frac{\bar b^0}{b^0} \sqrt{\frac{3}{2}} \frac{f_T}{f_V}
\frac{m_T}{m_V} \frac{m_T}{p_c}  \frac{m_B+m_T}{m_B+m_V} A_2^{BV}(m_T^2),
 \\
{\bf A}_{1,\pm}^{BT} &\equiv& A_1^{BT}(m_V^2) +\frac{\bar b^\pm}{b^\pm} \sqrt{2} \frac{f_T}{f_V}
\frac{m_T}{m_V} \frac{m_T}{p_c}  \frac{m_B+m_V}{m_B+m_T} A_1^{BV}(m_T^2),
 \\
{\bf V}_\pm^{BT} &\equiv& V^{BT}(m_V^2) +\frac{\bar b^\pm}{b^\pm} \sqrt{2} \frac{f_T}{f_V}
\frac{m_T}{m_V} \frac{m_T}{p_c}  \frac{m_B+m_T}{m_B+m_V} V^{BV}(m_T^2).
\end{eqnarray}
Note that Eqs. (\ref{eq:TPrate}) and (\ref{eq:TVrate}) are in agreement with \cite{Munoz97} for the special case that $a=b^0=b^\pm=1$ and $\bar a=\bar b^0=\bar b^\pm=0$.  As stressed in \cite{Munoz97}, the $p_c^5$ dependence in Eq. (\ref{eq:TPrate}) indicates that only the $L=2$ wave is allowed for the $TP$ system, while in the $TV$ modes the $L=1,2$ and 3 waves are simultaneously allowed, as expected.

%%%%%%%%%%%%%%%%%%%%%%%
\begin{table}[t]
\caption{{\it CP}-averaged branching fractions (in units of $10^{-6}$) and direct \CP asymmetries (\%) for $B\to PT$ decays with $\Delta S=1$. The parameters $\rho_A$ and $\phi_A$ are taken from Eq. (\ref{eq:rhoATP}). The theoretical errors correspond to the uncertainties due to the variation of Gegenbauer moments, decay constants, quark masses, form factors, the $\lambda_B$ parameter for the $B$ meson wave function and the power-correction parameters $\rho_{A,H}$, $\phi_{A,H}$. Then they are added in quadrature. The experimental data are taken from \cite{HFAG} and the model predictions of \cite{Kim:2003} are for $1/N_c^{\rm eff}=0.3$\,.}
\label{tab:PTpenguin}
\begin{ruledtabular}
\begin{tabular}{l c c c c r }
%Mode & QCDF  & KLO \cite{Kim} & MQ \cite{Munoz} & Expt.  & ${\cal %A}_{CP}$ \\ \hline
 &  \multicolumn{4}{c}{$\B$}
 &    \\ \cline{2-5}
\raisebox{2.0ex}[0cm][0cm]{Decay}  & QCDF  & KLO \cite{Kim:2003} & MQ \cite{Munoz} & Expt. & \raisebox{2.0ex}[0cm][0cm]{ $A_{\rm CP}$ }  \\ \hline
 $B^-\to \bar K_2^*(1430)^0\pi^-$   & $3.1^{+8.3}_{-3.1}$ & & & $5.6^{+2.2}_{-1.4}$ & $1.6^{+2.2}_{-1.8}$ \\
 $B^-\to K_2^*(1430)^-\pi^0$ & $2.2^{+4.7}_{-1.9}$ & 0.090 & 0.15 & & $0.2^{+17.8}_{-14.8}$ \\
 $\ov B^0\to K_2^*(1430)^-\pi^+$ & $3.3^{+8.5}_{-3.2}$ & & & $<6.3$ & $1.7^{+4.2}_{-5.2}$   \\
 $\ov B^0\to \bar K_2^*(1430)^0\pi^0$ & $1.2^{+4.3}_{-1.3}$ & 0.084 & 0.13 & $<4.0$  & $7.1^{+23.5}_{-24.1}$ \\
 $B^-\to a_2(1320)^0 K^- $ & $4.9^{+8.4}_{-4.2}$ & 0.311 & 0.39 & $<45$ & $-27.1^{+33.3}_{-41.1}$ \\
 $B^-\to a_2(1320)^- \bar K^0  $ & $8.4^{+16.1}_{-~7.2}$ & 0.011 & 0.015 & & $-0.6^{+0.4}_{-0.8}$  \\
 $\ov B^0\to a_2(1320)^+ K^- $ & $9.7^{+17.2}_{-~8.1}$ & 0.584 & 0.73 & & $-21.5^{+28.9}_{-35.0}$  \\
 $\ov B^0\to a_2(1320)^0 \ov K^0 $ & $4.2^{+8.3}_{-3.5}$ & 0.005 & 0.014 & & $6.7^{+6.5}_{-6.9}$ \\
 $B^-\to f_2(1270) K^- $ & $3.8^{+7.8}_{-3.0}$ & 0.344 & & $1.06^{+0.28}_{-0.29}$  & $-39.5^{+49.4}_{-25.5}$ \\
 $\ov B^0\to f_2(1270) \ov K^0 $ & $3.4^{+8.5}_{-3.1}$ &  0.005 & & $2.7^{+1.3}_{-1.2}$ & $-7.3^{+8.4}_{-7.9}$     \\
 $B^-\to f'_2(1525) K^- $ & $4.0^{+7.4}_{-3.6}$ &  0.004 & & $<7.7$ & $-0.6^{+4.3}_{-6.0}$   \\
 $\ov B^0\to f'_2(1525) \ov K^0 $ & $3.8^{+7.3}_{-3.5}$ & $7\times 10^{-5}$ & & & $0.8^{+1.2}_{-0.7}$    \\
 $B^-\to K_2^*(1430)^-\eta$ & $6.8^{+13.5}_{-~8.7}$ & 0.031 & 1.19  & $9.1\pm3.0$   & $1.5^{+7.4}_{-5.6}$ \\
 $B^-\to K_2^*(1430)^-\eta'$ & $12.1^{+20.7}_{-12.1}$ & 1.405 & 2.70 & $28.0^{+5.3}_{-5.0}$ & $-1.7^{+3.2}_{-3.9}$  \\
 $\ov B^0\to \bar K_2^*(1430)^0\eta$ & $6.6^{+13.5}_{-~8.7}$ & 0.029 & 1.09  & $9.6\pm2.1$ & $3.2^{+16.5}_{-~4.8}$ \\
 $\ov B^0\to \bar K_2^*(1430)^0\eta'$ & $12.4^{+21.3}_{-12.4}$ & 1.304 & 2.46 & $13.7^{+3.2}_{-3.1}$ & $-2.2^{+3.3}_{-4.0}$   \\
\end{tabular}
\end{ruledtabular}
\end{table}
%%%%%%%%%%%%%%%%%%%%%%%%%%%

%%%%%%%%%%%%%%%%%%%%%%%%%%%%%%
\begin{table}[t]
\caption{Same as Table \ref{tab:PTpenguin} except for  $B\to PT$ decays with $\Delta S=0$. }
\label{tab:PTtree}
\begin{ruledtabular}
\begin{tabular}{l c c c c r}
 &  \multicolumn{4}{c}{$\B$}
 &    \\ \cline{2-5}
\raisebox{2.0ex}[0cm][0cm]{Decay}  & QCDF  & KLO \cite{Kim:2003} & MQ \cite{Munoz} & Expt. & \raisebox{2.0ex}[0cm][0cm]{ $A_{\rm CP}$ }  \\ \hline
 $B^-\to  a_2(1320)^0 \pi^-$ & $3.0^{+1.4}_{-1.2}$ & 2.602 &  4.38 & & $9.6^{+47.9}_{-46.6}$ \\
 $B^-\to a_2(1320)^- \pi^0 $ & $0.24^{+0.79}_{-0.31}$ & 0.001 &  0.015 & & $-24.3^{+124.3}_{-~75.7}$  \\
 $\ov B^0\to  a_2(1320)^+ \pi^-$ & $5.2^{+1.8}_{-1.8}$ & 4.882 & 8.19 & & $37.3^{+23.9}_{-40.4}$ \\
 $\ov B^0\to  a_2(1320)^- \pi^+$ & $0.21^{+0.43}_{-0.17}$ & & & & $-26.6^{+111.6}_{-~82.9}$ \\
 $\ov B^0\to  a_2(1320)^0 \pi^0$ & $0.24^{+0.42}_{-0.19}$ & 0.0003 & 0.007 & & $-86.2^{+128.9}_{-~26.4}$ \\
 $B^-\to a_2(1320)^- \eta $ & $0.11^{+0.28}_{-0.11}$ & 0.294 &  45.8 & & $27.6^{+~73.4}_{-127.6}$ \\
 $B^-\to a_2(1320)^- \eta' $ & $0.11^{+0.47}_{-0.12}$ & 1.310 &  71.3 & & $31.3^{+~61.6}_{-131.3}$ \\
 $\ov B^0\to  a_2(1320)^0 \eta$ & $0.06^{+0.16}_{-0.05}$ & 0.138 & 25.2 & & $-76.7^{+\,100}_{-~19.2}$ \\
 $\ov B^0\to  a_2(1320)^0 \eta'$ & $0.05^{+0.22}_{-0.04}$ & 0.615 & 43.3 & & $-66.0^{+154.0}_{-~41.1}$ \\
 $B^-\to f_2(1270) \pi^-$ & $2.7^{+1.4}_{-1.2}$ & 2.874 & & $1.57^{+0.69}_{-0.49}$   & $60.2^{+27.1}_{-72.3}$ \\
 $\ov B^0\to  f_2(1270) \pi^0$ & $0.15^{+0.42}_{-0.14}$ & $0.0003$ & & & $-37.2^{+103.8}_{-~85.5}$ \\
 $\ov B^0\to  f_2(1270) \eta$ & $0.17^{+0.23}_{-0.12}$ & 0.152 & & & $69.7^{+~25.7}_{-102.7}$ \\
 $\ov B^0\to  f_2(1270) \eta'$ & $0.13^{+0.22}_{-0.13}$  & 0.680 & & & $82.3^{+22.9}_{-94.8}$ \\
 $B^-\to f'_2(1525) \pi^- $ & $0.009^{+0.024}_{-0.009}$ & 0.037 & & & 0   \\
 $\ov B^0\to  f'_2(1525) \pi^0$ & $0.005^{+0.012}_{-0.005}$ & $4\times 10^{-6}$ & & & 0  \\
 $\ov B^0\to  f'_2(1525) \eta$ & $0.002^{+0.006}_{-0.003}$ & 0.002 & & & 0 \\
 $\ov B^0\to  f'_2(1525) \eta'$ & $0.008^{+0.008}_{-0.005}$ & 0.009 & & & 0 \\
 $B^-\to K_2^*(1430)^-K^0$ & $0.44^{+0.74}_{-0.41}$ & $4\times 10^{-5}$ & $7.8\times 10^{-4}$ & & $30.3^{+51.2}_{-33.7}$ \\
 $B^-\to K_2^*(1430)^0K^-$ & $0.12^{+0.52}_{-0.12}$ & & & & $-0.26^{+0.23}_{-0.27}$ \\
 $\ov B^0\to K_2^*(1430)^-K^+$ & $0.03^{+0.07}_{-0.02}$ & & & & $-15.0^{+22.7}_{-25.1}$ \\
 $\ov B^0\to K_2^*(1430)^+K^-$ & $0.13^{+0.16}_{-0.10}$ & & & & $18.6^{+27.2}_{-27.4}$ \\
 $\ov B^0\to \bar K_2^*(1430)^0 K^0$ & $0.54^{+0.88}_{-0.49}$ & $3\times 10^{-5}$ & $7.2\times 10^{-4}$ & & $-2.1^{+4.1}_{-2.0}$ \\
 $\ov B^0\to K_2^*(1430)^0 \bar K^0$ & $0.22^{+0.54}_{-0.22}$ & & & & $-14.0^{+13.7}_{-60.1}$  \\
\end{tabular}
\end{ruledtabular}
\end{table}
%%%%%%%%%%%%%%%%%%%%%%

%%%%%%%%%%%%%%%%%%%%%%%
\begin{table}[h]
\caption{{\it CP}-averaged branching fractions (in units of $10^{-6}$), direct \CP asymmetries (\%) and the longitudinal polarization fractions $f_L$ for $B\to VT$ decays with $\Delta S=1$. The parameters $\rho_A$ and $\phi_A$ are taken from Eq. (\ref{eq:rhoVT}).}
\label{tab:VTpenguin}
\begin{ruledtabular}
\begin{tabular}{l c c c c c c r}
%Mode & QCDF  & KLO \cite{Kim} & MQ \cite{Munoz} & Expt  & ${\cal %A}_{CP}$ \\ \hline
 &  \multicolumn{4}{c}{$\B$} & \multicolumn{2}{c}{$f_L$}
     \\ \cline{2-5} \cline{6-7}
\raisebox{2.0ex}[0cm][0cm]{Decay}  & QCDF  & KLO \cite{Kim:2003} & MQ \cite{Munoz} & Expt. &  QCDF & Expt. & \raisebox{2.0ex}[0cm][0cm]{$A_{\rm CP}$}   \\ \hline
 $B^-\to \bar K_2^*(1430)^0\rho^-$ & $18.6^{+50.1}_{-17.2}$ & & & & $0.63^{+0.10}_{-0.09}$ & & $-1.0^{+0.8}_{-1.0}$   \\
 $B^-\to K_2^*(1430)^-\rho^0$ & $10.4^{+18.8}_{-~9.2}$ & 0.253 & 0.74 &  & $0.66^{+0.06}_{-0.07}$ & & $2.1^{+11.1}_{-~9.9}$ \\
 $\ov B^0\to K_2^*(1430)^-\rho^+$ & $19.8^{+52.0}_{-18.2}$ & & & & $0.64^{+0.07}_{-0.03}$ & & $-1.5^{+2.6}_{-2.0}$  \\
 $\ov B^0\to \bar K_2^*(1430)^0\rho^0$ & $9.5^{+33.4}_{-~9.5}$ & 0.235 & 0.68 & & $0.64^{+0.15}_{-0.37}$ & & $-4.0^{+14.1}_{-10.8}$ \\
 $B^-\to K_2^*(1430)^-\omega$ & $7.5^{+19.7}_{-~7.0}$ & 0.112 & 0.06 & $21.5\pm4.3$  & $0.64^{+0.08}_{-0.07}$ & $0.56\pm0.11$ & $2.0^{+12.2}_{-10.5}$ \\
 $\bar B^0\to \bar K_2^*(1430)^0\omega$ & $8.1^{+21.7}_{-~7.6}$ & 0.104 & 0.053 & $10.1\pm2.3$ & $0.66^{+0.11}_{-0.15}$ &  $0.45\pm0.12$ & $4.4^{+10.9}_{-10.0}$  \\
 $B^-\to K_2^*(1430)^-\phi$ & $7.4^{+25.8}_{-~5.2}$ & 2.180 & 9.24   & $8.4\pm2.1$   & $0.85^{+0.16}_{-0.77}$ & $0.80\pm0.10$ & $0.1^{+1.2}_{-0.5}$ \\
 $\bar B^0\to \bar K_2^*(1430)^0\phi$ & $7.7^{+26.9}_{-~5.5}$ & 2.024 & 8.51 & $7.5\pm1.0$ & $0.86^{+0.16}_{-0.77}$ & $0.901^{+0.059}_{-0.069}$ & $0.09^{+0.82}_{-0.21}$\\
 $B^-\to a_2(1320)^0 K^{*-}$ & $2.9^{+11.7}_{-~2.5}$ & 1.852 & 2.80 & & $0.73^{+0.22}_{-0.33}$ & & $-15.0^{+56.0}_{-15.0}$  \\
 $B^-\to a_2(1320)^- \ov K^{*0}$ & $6.1^{+23.8}_{-~5.4}$ & 4.495 & 8.62 & & $0.79^{+0.20}_{-0.64}$ & & $-0.1^{+1.3}_{-0.3}$  \\
 $\ov B^0\to  a_2(1320)^+ K^{*-}$ & $6.1^{+24.3}_{-~5.3}$ & 3.477 & 7.25 & & $0.77^{+0.19}_{-0.46}$ & & $-13.3^{+38.2}_{-~7.0}$  \\
 $\ov B^0\to  a_2(1320)^0 \ov K^{*0}$ & $3.4^{+12.4}_{-~2.8}$ & 2.109 & 4.03 &  & $0.82^{+0.14}_{-0.67}$ & & $1.2^{+~7.0}_{-13.3}$ \\
 $B^-\to  f_2(1270) K^{*-}$ & $8.3^{+17.3}_{-~6.7}$ & 2.032 & & & $0.93^{+0.07}_{-0.63}$ & & $-8.1^{+13.7}_{-~7.1}$  \\
 $\ov B^0\to  f_2(1270) \ov K^{*0}$ & $9.1^{+18.8}_{-~7.3}$ &  2.314 & & & $0.94^{+0.06}_{-0.69}$ & & $-0.08^{+4.3}_{-3.1}$ \\
 $B^-\to  f'_2(1525) K^{*-}$ & $12.6^{+24.0}_{-11.1}$ &  0.025 & & & $0.65^{+0.28}_{-0.38}$ & & $0.6^{+2.5}_{-2.9}$ \\
 $\ov B^0\to  f'_2(1525) \ov K^{*0}$ & $13.5^{+25.4}_{-11.9}$ & $0.029$ & & & $0.66^{+0.27}_{-0.38}$ & & $0.2^{+0.3}_{-0.4}$     \\
\end{tabular}
\end{ruledtabular}
\end{table}
%%%%%%%%%%%%%%%%%%%%%%%%%%%

%%%%%%%%%%%%%%%%%%%%%%%%
\begin{table}[t]
\caption{Predicted branching fractions (in units of $10^{-6}$), direct \CP asymmetries (\%) and the longitudinal polarization fractions $f_L$ for $B\to VT$ decays with $\Delta S=0$. }
\label{tab:VTtree}
\begin{ruledtabular}
\begin{tabular}{l c c c c r}
 &  \multicolumn{3}{c}{$\B$}
 &    \\ \cline{2-4}
\raisebox{2.0ex}[0cm][0cm]{Decay}  & QCDF  & KLO \cite{Kim:2003} & MQ \cite{Munoz} & \raisebox{2.0ex}[0cm][0cm]{$f_L$} & \raisebox{2.0ex}[0cm][0cm]{ $A_{\rm CP}$}   \\ \hline
 $B^-\to  a_2(1320)^0 \rho^-$ & $8.4^{+4.7}_{-2.9}$ & 7.342 &  19.34 & $0.88^{+0.05}_{-0.14}$ & $31.0^{+16.0}_{-45.5}$  \\
 $B^-\to a_2(1320)^- \rho^0 $ & $0.82^{+2.30}_{-0.95}$ & 0.007 &  0.071 & $0.56^{+0.20}_{-0.31}$ & $-13.7^{+74.8}_{-83.2}$  \\
 $\ov B^0\to  a_2(1320)^+ \rho^-$ & $11.3^{+5.3}_{-4.6}$  & 14.686 & 36.18 & $0.91^{+0.03}_{-0.10}$ & $7.6^{+10.2}_{-23.7}$ \\
 $\ov B^0\to  a_2(1320)^- \rho^+$ & $1.2^{+2.6}_{-1.0}$ &  &  & $0.64^{+0.16}_{-0.05}$ & $49.0^{+24.0}_{-66.8}$  \\
 $\ov B^0\to  a_2(1320)^0 \rho^0$ & $0.39^{+1.35}_{-0.20}$ & 0.003 & 0.03 & $0.91^{+0.07}_{-0.60}$ & $55.2^{+~31.9}_{-144.7}$ \\
 $B^-\to a_2(1320)^- \omega $ & $0.38^{+1.84}_{-0.36}$  & 0.010 &  0.14 & $0.73^{+0.09}_{-0.46}$ & $-36.2^{+127.8}_{-~57.2}$  \\
 $B^-\to a_2(1320)^- \phi $ & $0.003^{+0.013}_{-0.001}$ & 0.004 &  0.019 & $0.93^{+0.34}_{-0.00}$ & $0.06^{+0.07}_{-0.07}$  \\
 $\ov B^0\to  a_2(1320)^0 \omega$ & $0.25^{+1.14}_{-0.19}$ & 0.005 & 0.07 & $0.78^{+0.05}_{-0.42}$ &  $60.5^{+~25.7}_{-141.2}$ \\
 $\ov B^0\to  a_2(1320)^0 \phi$ & $0.001^{+0.006}_{-0.001}$ & 0.002 & 0.009 & $0.93^{+0.03}_{-0.56}$ & $0.06^{+0.07}_{-0.07}$ \\
 $B^-\to f_2(1270) \rho^-$ & $7.7^{+4.8}_{-2.9}$ & 8.061 & & $0.90^{+0.04}_{-0.18}$ & $-18.2^{+41.1}_{-20.0}$    \\
 $\ov B^0\to  f_2(1270) \rho^0$ & $0.42^{+1.90}_{-0.44}$ & 0.004 & & $0.82^{+0.11}_{-0.86}$ & $38.1^{+~49.4}_{-113.3}$ \\
 $\ov B^0\to  f_2(1270) \omega$ & $0.69^{+0.97}_{-0.36}$ & 0.005 &  & $0.91^{+0.07}_{-0.40}$ & $-73.3^{+105.1}_{-~11.0}$ \\
 $\ov B^0\to  f_2(1270) \phi$ & $0.001^{+0.007}_{-0.000}$ & 0.002 & & $0.92^{+0.04}_{-0.59}$ & $0.07^{+0.76}_{-0.78}$ \\
 $B^-\to f'_2(1525) \rho^- $ & $0.07^{+0.11}_{-0.04}$ & 0.103 & & $0.96^{+0.03}_{-0.48}$ & $-0.02^{+0.08}_{-0.07}$  \\
 $\ov B^0\to  f'_2(1525) \rho^0$ & $0.03^{+0.06}_{-0.02}$ & $5\times 10^{-5}$ & & $0.96^{+0.03}_{-0.48}$ & $-0.02^{+0.08}_{-0.07}$ \\
 $\ov B^0\to  f'_2(1525) \omega$ & $0.03^{+0.04}_{-0.01}$ & $6\times 10^{-5}$ & & $0.95^{+0.04}_{-0.51}$ & $-0.03^{+0.09}_{-0.08}$ \\
 $\ov B^0\to  f'_2(1525) \phi$ & $0.006^{+0.034}_{-0.005}$ & $2\times 10^{-5}$ &  & 1 & 0 \\
 $B^-\to K_2^*(1430)^-K^{*0}$ & $0.56^{+1.09}_{-0.38}$ & 0.014 & 0.59 & $0.85^{+0.09}_{-0.57}$ & $-14.6^{+14.5}_{-10.7}$ \\
 $B^-\to K_2^*(1430)^0K^{*-}$ & $2.1^{+4.2}_{-1.8}$ & & & $0.54^{+0.06}_{-0.05}$ & $10.1^{+16.0}_{-~8.2}$ \\
 $\ov B^0\to K_2^*(1430)^-K^{*+}$ & $0.06^{+0.09}_{-0.03}$ & & & 1 & $-58.3^{+135.1}_{-~43.9}$ \\
 $\ov B^0\to K_2^*(1430)^+K^{*-}$ & $0.43^{+0.54}_{-0.31}$  & & & 1 & $15.0^{+40.9}_{-39.2}$  \\
 $\ov B^0\to \bar K_2^*(1430)^0 K^{*0}$ & $0.44^{+0.88}_{-0.30}$ & 0.026 & 0.55 & $0.90^{+0.08}_{-0.73}$ & $-2.1^{+~4.0}_{-10.4}$  \\
 $\ov B^0\to K_2^*(1430)^0 \bar K^{*0}$ & $1.1^{+2.9}_{-1.0}$ & & & $0.60^{+0.14}_{-0.23}$ & $-2.3^{+0.1}_{-0.1}$ \\
\end{tabular}
\end{ruledtabular}
\end{table}

\subsection{$\ov B\to PT$ decays}
As noticed in \cite{CCBud},
since the penguin annihilation effects are different for $B\to VP$ and $B\to PV$ decays,
the penguin annihilation parameters $X_A^{VP}$ and $X_A^{PV}$ are not necessarily the same. Indeed, a fit to the $B\to VP,PV$ decays yields $\rho_A^{VP}\approx 1.07$, $\phi_A^{VP}\approx -70^\circ$ and $\rho_A^{PV}\approx 0.87$, $\phi_A^{PV}\approx -30^\circ$ \cite{CCBud}. Likewise, for $B_{u,d}\to TP$ decays we find that
the data of $B_{u,d}\to TP$ can be described by the penguin annihilation parameters
\be \label{eq:rhoATP}
\rho_A^{TP}=0.83, \qquad \phi_A^{TP}=-70^\circ, \qquad
\rho_A^{PT}=0.75, \qquad \phi_A^{PT}=-30^\circ\,.
\en

For $B\to T$ transition form factors, the LEET or pQCD  predictions are favored by the experimental data of $B\to f_2(1270)K$ and $B\to f_2(1270)\pi$, while the CLFQ or ISGW2 model results are preferred by the measurements of $B\to K_2^*(1430)\eta^{(')},K_2^*(1430)\omega,K_2^*(1430)\phi$. For example, the branching fractions (in units of $10^{-6}$; only the central values are quoted here) for the $f_2\pi^-$, $f_2\ov K^0$ and $f_2 K^-$ modes are found to be $8.1$ (2.7), $5.0$ (3.4) and $6.4$ (3.8), respectively, using the CLFQ (LEET) model for $B\to f_2$ transition form factors.  The corresponding experimental  values are $1.57^{+0.69}_{-0.49}$, $2.7^{+1.3}_{-1.2}$, and $1.06^{+0.28}_{-0.29}$. Therefore, it is evident that the data favor LEET over the CLFQ model for decays involving $B\to f_2$ transitions. Likewise, the branching fractions  for the modes $K_2^{*-}\phi$, $K_2^{*-}\eta$ and $\ov K^{*0}_2\eta'$ are calculated  to be 7.4 (4.7), 6.8 (4.7) and 12.4 (8.4), respectively, using the CLFQ (LEET) model for $B\to K_2^*$ transition form factors.  The corresponding experimental  values are $8.4\pm2.1$, $9.1\pm3.0$, and $13.7^{+3.2}_{-3.1}$. It is clear that the CLFQ model works better for decays involving $B\to K_2^*$ transitions.
In this work we shall use the $B\to K_2^*(1430)$ form factors obtained in the CLFQ model and $B\to a_2(1270)$ and $B\to f_2$ ones from LEET (see Table \ref{tab:FF}).

Branching fractions and \CP asymmetries for  $B\to TP$ decays are shown in Tables \ref{tab:PTpenguin} and \ref{tab:PTtree}. The theoretical errors correspond to the uncertainties due to the variation of (i) the Gegenbauer moments,
the  decay constants, (ii) the heavy-to-light form factors and the strange quark mass, (iii) the wave function of the $B$ meson characterized by the
parameter $\lambda_B$, and (iv) the power corrections due to weak annihilation and hard
spectator interactions described by the parameters $\rho_{A,H}$, $\phi_{A,H}$. We allow the variation of  $\rho_{A}$ and $\phi_{A}$ to be $\pm 0.4$ and $\pm 50^\circ$, respectively, and put $\rho_H$ and $\phi_H$ in the respective ranges $0\leq \rho_H\leq 1$ and $0\leq \phi_H\leq 2\pi$.
To obtain the errors shown in these tables, we first scan randomly the points in the
allowed ranges of the above-mentioned parameters and then add errors in quadrature. Power corrections beyond the heavy
quark limit generally give the major theoretical uncertainties.

%====================================================================
\begin{figure}[t]
\begin{center}
\includegraphics[width=0.80\textwidth]{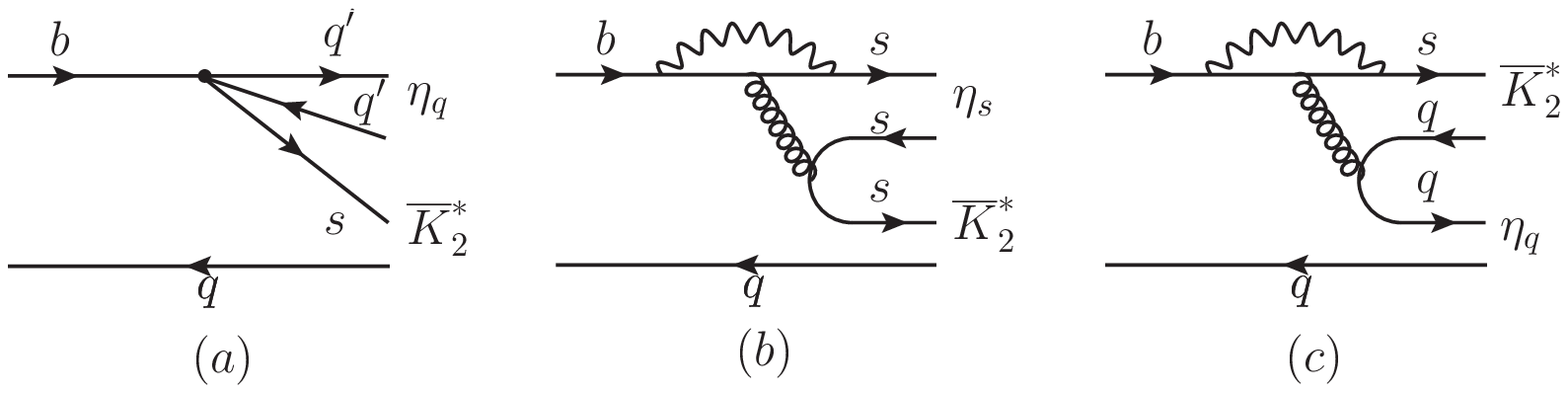}
\vspace{0.0cm}
\caption{Three different penguin contributions to $\ov B\to \ov K_2^*\eta^{(')}$. Fig. 1(a) is induced by the penguin operators $O_{3,5,7,9}$.
} \label{fig:K2eta} \end{center}
\end{figure}
%=====================================================================

For $\ov B\to \ov K_2^*\eta^{(')}$ decays, there exist
three different penguin contributions as depicted in Fig. \ref{fig:K2eta}: (i) $b\to sq\bar q\to s\eta_{q}$, (ii) $b\to ss\bar s\to s\eta_s$, and (iii) $b\to s q\bar q\to q \bar K_2^*$, corresponding to Figs. 1(a), 1(b) and 1(c), respectively. The dominant contributions come from Figs. 1(b) and 1(c). Since the relative sign of the $\eta_s$ state with respect to the $\eta_q$ is negative for the $\eta$ and positive for the $\eta'$ (see Eq. (\ref{eq:qsmixing})), it is evident that the interference between Figs. 1(b) and 1(c) is destructive for $K^*_2\eta$  and constructive for $K_2^*\eta'$. This explains why $K_2^*\eta'$ has a rate larger than $K_2^*\eta$.
It was known that the predicted rates in naive factorization are too small by one order of magnitude, of order $1.0\times 10^{-6}$ for $\B(\ov B^0\to \bar K_2^{*0}\eta)$ and $2.5\times 10^{-6}$ for $\B(\ov B^0\to \bar K_2^{*0}\eta')$ \cite{Munoz,Cheng:2010sn}.
\footnote{The rate of $\ov B^0\to \bar K_2^{*0}\eta$ was predicted to be very suppressed in \cite{Kim:2003} (see Table \ref{tab:PTpenguin}) due to the use of a wrong matrix element for $\la \eta^{(')}|\bar s\gamma_5 s|0\ra$ \cite{Cheng:2010sn}.}
One reason is that the factorizable contribution to Fig. 1(c) vanishes in the naive factorization approach.
The rates of $K_2^*\eta^{(')}$ are greatly enhanced in QCDF owing to the large power corrections from penguin annihilation and the sizable nonfactorizable contributions to Fig. 1(c).

From Tables \ref{tab:PTpenguin} and \ref{tab:PTtree} we see that the predicted branching fractions for penguin-dominated $B\to TP$ decays in QCDF are larger than those of \cite{Kim:2003} and \cite{Munoz} by one to two orders of magnitude through the aforementioned two  mechanisms for enhancement, while the predicted rates in QCDF are consistent with \cite{Kim:2003} for the leading tree-dominated modes such as $a_2^0\pi^-,a_2^+\pi^-,f_2\pi^-$. Note that the branching fractions of $B^-\to \bar K_2^{*0}\pi^-$ and $\ov B^0\to K_2^{*-}\pi^+$ vanish in naive factorization, while experimentally it is $(5.6^{+2.2}_{-1.4})\times 10^{-6}$ for the former. The QCDF calculation indicates that the nonfactorizable contributions arising from vertex, penguin and spectator corrections are sizable to account for the data.

Just as $\ov B\to a_0\pi$ with $a_0=a_0(980)$ or $a_0(1450)$ and $\ov B\to b_1(1235)\pi$ decays, we see from Table \ref{tab:PTtree} that for $\ov B\to a_2(1320)\pi$ decays, the $a_2^-\pi^+$ and $a_2^-\pi^0$ modes are highly suppressed relative to $a_2^+\pi^-$ and $a_2^0\pi^-$, respectively. Since (see Appendix B)
\be
{\cal A}_{\bar B^0\to a_2^- \pi^+}
   &\propto &  \sum_{p=u,c}V_{pb}V^*_{pd}
 \left[ \delta_{pu}\,\alpha_1 + \alpha_4^p
     \right]
      \overline{X}^{(\overline{B} \pi, a_2)},
    \nonumber\\
 {\cal A}_{\bar B^0\to a_2^+ \pi^-}
   &\propto& \sum_{p=u,c}V_{pb}V^*_{pd}
   \left[ \delta_{pu}\,\alpha_1 + \alpha_4^p
     \right]
    X^{(\overline B a_2, \pi)},
\en
it is tempting to argue that $\Gamma(\ov B^0\to a_2^+\pi^-)\gg \Gamma(\ov B^0\to a_2^-\pi^+)$ is a natural consequence of naive factorization as the tensor meson cannot be created from the $V-A$ current. However, the suppression of $a_2^-\pi^+$ relative to $a_2^+\pi^-$ in QCDF stems from a different reasoning. The amplitude $\overline{X}^{(\overline{B} \pi, a_2)}$ does not vanish in QCDF owing to the nonfactorizable corrections. Indeed,
$\overline{X}^{(\overline{B} \pi, a_2)}=0.80$ and $X^{(\overline B a_2, \pi)}=0.69$ are numerically comparable. Therefore, one may wonder  how to see the aforementioned suppression ? The key is the quantity $N_i(M_2)$ appearing in the expression for the effective parameter $a_i$ [see Eq. (\ref{eq:ai})]. This quantity vanishes for the tensor meson [cf Eq. (\ref{eq:Ni})]. As a result, the parameter $a_1(\pi a_2)$ is not of order unity as it receives contributions only from vertex corrections and hard spectator interactions, both suppressed by factors of $\alpha_s/(4\pi)$.  Numerically, we have $a_1(\pi a_2)=-0.035+i0.014$\,. By contrast, $a_1(a_2\pi)$ is of order unity. This explains why $\ov B^0\to a_2^+\pi^-$ has a rate greater than $a_2^-\pi^+$ and why $B^-\to a_2^-\pi^0$ is suppressed relative to $a_2^0\pi^-$.
\footnote{The same argument also explains the suppression of $\ov B^0\to b_1^-\pi^+$ relative to $b_1^+\pi^-$ in QCDF \cite{Cheng:2007mx}.}
The same pattern also occurs in $\ov B\to a_2\rho$ decays, see Table \ref{tab:VTtree}.

The branching fractions of $B\to a_2\eta^{(')}$ of order $10^{-7}$ in QCDF are in sharp contrast to the predictions of
\cite{Munoz}, ranging from $25\times 10^{-6}$ to $70\times 10^{-6}$ (see Table \ref{tab:PTtree}). It seems to us that it is extremely unlikely that the rate of $a_2^-\eta^{(')}$ can be greater than $a_2^-\pi^0$ by four orders of magnitude as claimed in \cite{Munoz}. It appears that the former should be slightly smaller than the latter in rates. This can be tested in the future. It is also interesting to notice that while $\ov B\to \bar K_2^* K$ decays are very suppressed in naive factorization, their branching fractions are a few $\times 10^{-7}$ in QCDF. Finally, it is worth remarking that $\ov B^0\to K_2^{*+}K^-$ and $\ov B^0\to K_2^{*-}K^+$ can only proceed through weak annihilation.

\subsection{$B\to TV$ decays}
Branching fractions, direct \CP asymmetries and the longitudinal polarization fractions for  $B\to TV$ decays are shown in Tables \ref{tab:VTpenguin} and \ref{tab:VTtree}.
Thus far only four of the $B\to TV$ decays have been measured: $B^-\to K_2^{*-}(\phi,\omega)$ and $\ov B^0\to \ov K_2^{*0}(\phi,\omega)$. They can be used to fix the penguin annihilation parameters. From Eqs. (\ref{eq:K2comega}) and (\ref{eq:K2cphi}) we have
\be \label{eq:AmpTV}
  \sqrt{2}{\cal A}_{B^-\to K^{*-}_2\omega}^h
   &\approx&  \sqrt{2}{\cal A}_{\ov B^0\to \ov K^{*0}_2\omega}^h
   \approx \Bigg\{ \Big[
 \alpha_4^{p,h}
 +\beta_3^{p,h}  \Big]   \overline{X}^{(\overline B \omega,\ov K^{*}_2)}_h
   + \Big[2\alpha_3^{p,h}\Big]
   X^{(\overline B\ov K^{*}_2,\omega)}_h\Bigg\},    \non \\
  {\cal A}_{B^-\to K^{*-}_2\phi}^h
   &\approx&  {\cal A}_{\ov B^0\to \ov K^{*0}_2\phi}^h
   \approx
    \Big[  \alpha_3^{p,h}+ \alpha_4^{p,h}
    +\beta_3^{p,h} +\beta^{p,h}_{3,{\rm EW}} \Big]
 X^{(\overline B \ov K^{*}_2,\phi)}_h.
\en
Since  $\overline{X}^{(\overline B \omega,\ov K^{*}_2)}_0/X^{(\overline B \ov K^{*}_2,\phi)}_0=0.56$\,, it is expected that $\B(B^-\to K^{*-}_2\omega)/\B(B^-\to K^{*-}_2\phi)\approx |\overline{X}^{(\overline B \omega,\ov K^{*}_2)}_0/X^{(\overline B \ov K^{*}_2,\phi)}_0|^2/2\approx 0.15$\,, provided that the penguin-annihilation parameters are the same for $K_2^*\omega$ and $K_2^*\phi$, i.e. $\rho_A^{K_2^*\omega}=\rho_A^{\phi K_2^*}$ and likewise for $\phi_A$. However, it is the other way around experimentally: the rate of $K_2^*\omega$ is larger than that of $K_2^*\phi$.
Since in the $B\to VV$ sector, $\B(B^-\to K^{*-}\omega)/\B(B^-\to K^{*-}\phi)\approx 0.3$ \cite{HFAG}, it is thus puzzling as why $K_2^*\omega$ behaves so differently from $K^*\omega$ in terms of branching fractions.
It is clear from Eq. (\ref{eq:AmpTV}) that the $B\to K_2^*\phi$ decay receives penguin annihilation via $\rho_A^{TV}$ and $\phi_A^{TV}$, while $B\to K_2^*\omega$ is governed by $\rho_A^{VT}$ and $\phi_A^{VT}$.
Therefore, we should have $\rho_A^{VT}\gg\rho_A^{TV}$ in order to account for their rates (see Eq. (\ref{eq:rhoVT}) below).

The branching fractions of the tree-dominated modes $a_2\phi,~f_2\phi,~f'_2\phi$ are very small, of order $10^{-9}$ (see Table \ref{tab:VTtree}), as they proceed only through QCD and electroweak penguins.

For charmless $\ov B\to TV$
decays, it is naively expected that the helicity
amplitudes $\A_h$ (helicities $h=0,-,+$ ) for both tree- and penguin-dominated $\ov B \to TV$ decays respect the
hierarchy pattern
\be \label{eq:hierarchy}
\A_0:\A_-:\A_+=1:\left({\Lambda_{\rm QCD}\over m_b}\right):\left({\Lambda_{\rm QCD}\over
m_b}\right)^2.
\en
Hence,  they are dominated by the longitudinal polarization
states and satisfy the scaling law, namely \cite{Kagan},
 \be \label{eq:scaling}
f_T\equiv 1-f_L={\cal O}\left({m^2_{V,T}\over m^2_B}\right), \qquad {f_\bot\over f_\parallel}=1+{\cal
O}\left({m_{V,T}\over m_B}\right),
 \en
with $f_L,f_\bot$, $f_\parallel$ and $f_T$ being the longitudinal, perpendicular, parallel and transverse polarization fractions, respectively, defined as
 \be
 f_\alpha\equiv \frac{\Gamma_\alpha}{\Gamma}
                     =\frac{| \A_\alpha|^2}{|\A_0|^2+|\A_\parallel|^2+|\A_\bot|^2},
 \label{eq:f}
 \en
with $\alpha=L,\parallel,\bot$.

The so-called polarization puzzle in $B\to VV$ decays is the enigma of why the transverse polarization fraction $f_T$ in the penguin-dominated channels such as $B\to \phi K^*,\rho K^*$ is comparable to $f_L$, namely, $f_T/f_L\sim 1$. This poses an interesting challenge for
any theoretical interpretation. For $B\to TV$ decays, the experimental measurement indicates that $f_T/f_L\ll 1$ for $B\to \phi K_2^*(1430)$, whereas $f_T/f_L\sim 1$ for $B\to \omega K_2^*(1430)$, even though both are penguin-dominated.

Consider the ratio of negative- and longitudinal-helicity amplitudes
\be \label{eq:ampratio}
\left.{ \A_-\over \A_0}\right|_{B^-\to
K_2^{*-}\omega} &\approx& \left({\alpha_4^{c,-}+\beta_{3}^-
\over \alpha_4^{c,0}+\beta_{3}^0
}\right)_{\omega K_2^*}\left({ {\ov X}^-_{\omega K_2^*}\over \ov X^0_{\omega K_2^*}}\right), \non \\
\left.{ \A_-\over \A_0}\right|_{B^-\to
K_2^{*-}\phi} &\approx& \left({\alpha_4^{c,-}+\beta_{3}^-
\over \alpha_4^{c,0}+\beta_{3}^0
}\right)_{K_2^*\phi}\left({ X^-_{K_2^*\phi}\over  X^0_{ K_2^*\phi}}\right).
\en
The longitudinal polarization fraction  can be approximated as
\begin{eqnarray}
f_L(\omega K_2^*) &\simeq& 1- \frac{|\alpha_4^{c,-} +\beta_3^-|^2 |\ov X^-_{\omega K_2^*}|^2 }
{\sum_{h=0,-} |\alpha_4^{c,h}+ \beta_3^h |^2 |\ov X^h_{\omega K_2^*}|^2}, \nonumber\\
f_L(K_2^*\phi) &\simeq& 1- \frac{|\alpha_4^{c,-} +\beta_3^-|^2 |X^-_{K_2^*\phi}|^2}
{\sum_{h=0,-} |\alpha_4^{c,h}+ \beta_3^h |^2 |X^h_{K_2^*\phi}|^2}.
\end{eqnarray}
We have
\be
&& |\ov X_{\omega K_2^*}^0|:|\ov X_{\omega K_2^*}^-|:|\ov X_{\omega K_2^*}^+|=1:0.51:0.04\,, \non \\
&& |X_{K_2^*\phi}^0|:|X_{K_2^*\phi}^-|:|X_{K_2^*\phi}^+|=1:0.38:0.06\,.
\en
In the absence of penguin annihilation, we find $f_L(\omega K_2^*)\simeq f_L(K_2^*\phi)\approx 0.72$. As we have stressed in \cite{Cheng:2008gxa}, in the presence of NLO nonfactorizable corrections e.g. vertex,
penguin and hard spectator scattering contributions, the parameters $a_i^h$ are helicity dependent.
Although the factorizable helicity amplitudes $X^0$, $X^-$ and $X^+$ or $\ov X^0$, $\ov X^-$, $\ov X^+$ respect the scaling law (\ref{eq:hierarchy}) with $\Lambda_{\rm QCD}/m_b$ replaced by $2m_{V,T}/m_B$ for the tensor and vector meson productions, one needs to consider the effects of helicity-dependent Wilson coefficients: $\A_-/\A_0= f(a_i^-)X^-/[f(a_i^0)X^0]$.
The constructive (destructive) interference in the negative-helicity (longitudinal-helicity) amplitude of the penguin-dominated $\ov B\to TV$ decay will render $f(a_i^-)\gg f(a_i^0)$ so that $\A_-$ is comparable to $\A_0$ and the transverse polarization is enhanced. Indeed, we find $f_T(\omega K_2^*)\simeq f_T(K_2^*\phi)\approx 0.28$\,. Therefore, when NLO effects are turned on,  their corrections on $a_i^-$ will render the negative helicity amplitude $\A_-(\ov B \to\bar K_2^{*}\phi)$ comparable to the longitudinal one $\A_0(\ov B \to\bar K_2^{*}\phi)$ so that even at the short-distance level, $f_L$ for $\ov B^0\to \bar K_2^{*}\phi$ is reduced to the level of 70\% and likewise for $\ov B^0\to \bar K_2^{*}\omega$.

As noticed in passing, penguin annihilation is needed in order to account for the observed rates. This is because, in the absence of power corrections,  QCDF  predicts too small rates for penguin-dominated $\ov B\to TV$ and $VV,VA$ decays. For example,  the calculated $\ov B\to \bar K_2^*\phi$ rate is too small by a factor of 2.5 and $\ov B\to \bar K_2^*\omega$ by two orders of magnitude. We shall rely on power corrections from penguin annihilation to enhance the rates.
A nice feature of the $(S-P)(S+P)$ penguin annihilation is that it contributes to $\A_0$ and $\A_-$ with the same order of magnitude \cite{Kagan:2004ia}
\be \label{eq:hierarchy-2}
\A_0^{\rm PA}: \A_-^{\rm PA}:\A_+^{\rm PA}=\left({\Lambda_{\rm QCD}\over m_b}{\ln{m_b\over \Lambda_h}}\right)^2:\left({\Lambda_{\rm QCD}\over m_b}{\ln{m_b\over \Lambda_h}}\right)^2:\left({\Lambda_{\rm QCD}\over
m_b}\right)^4.
\en
The logarithmic divergences are associated with the limit in which both the $s$ and $\bar s$ quarks originating from the gluon are soft \cite{Kagan:2004ia}. As for the power counting, the annihilation topology is of order $1/m_b$ and each remaining factor of $1/m_b$ is associated with a quark helicity flip. The fact that $\A_-^{\rm PA}$ and $\A_0^{\rm PA}$ have the same power counting explains why penguin annihilation is helpful to resolve the polarization puzzle.
The relative size of $\A_-^{\rm PA}$ and $\A_0^{\rm PA}$ depends mainly on the phase $\phi_A$. It turns out that the longitudinal polarization fraction for $B\to K_2^*\phi$ is quite sensitive to the phase $\phi_A^{TV}$, while $f_L(\omega K_2^*)$ is not so sensitive to $\phi_A^{VT}$. For example, $f_L(K_2^*\phi)=0.88,~0.72,~ 0.48$, respectively, for $\phi_A^{TV}=-30^\circ,-45^\circ,-60^\circ$ and $f_L(\omega K_2^*)=0.68,~0.66,~ 0.64$, respectively, for $\phi_A^{VT}=-30^\circ,-45^\circ,-60^\circ$. Hence, we can use the experimental measurements of $f_L$ to fix the phases $\phi_A^{VT}$ and $\phi_A^{TV}$ and branching fractions to pin down the parameters $\rho_A(VT)$ and $\rho_A(TV)$:
\be \label{eq:rhoVT}
\rho_A^{TV}=0.65, \qquad \phi_A^{TV}=-33^\circ, \qquad
\rho_A^{VT}=1.20, \qquad \phi_A^{VT}=-60^\circ\,.
\en
It should be stressed that although the experimental observation
of the longitudinal polarization in $B\to K_2^*\phi$ and $B\to K_2^*\omega$ decays
can be accommodated in the QCDF approach, no dynamical explanation is offered for the smallness of $f_T(K_2^*\phi)$ and the sizable $f_T(\omega K_2^*)$.

For penguin-dominated $B\to TV$ decays, we find $f_L(K_2^*\rho)\sim f_L(K_2^*\omega)\sim 0.65$\,, whereas $f_L(f_2K^*)\sim 0.93$ (cf. Table \ref{tab:VTpenguin}). It will be very interesting to measure $f_L$ for these modes to test the approach of QCDF. Theoretically,
transverse polarization is expected to be small in  tree-dominated $\ov B\to TV$ decays except for the $a_2^-\rho^0,~a_2^-\rho^+,~K_2^{*0}K^{*-}$ and $K_2^{*0}\bar
K^{*0}$ modes.

\section{Conclusions}

We have studied in this work the charmless hadronic $B$ decays with a tensor meson in the final state within the framework of QCD factorization. Due to the $G$-parity of the tensor meson, both the chiral-even and chiral-odd two-parton LCDAs of the tensor meson are antisymmetric under the interchange of momentum fractions of the {\it quark} and {\it anti-quark} in the SU(3) limit. The main results of this work are as follows:

\begin{itemize}

\item  We have worked out the longitudinal and transverse helicity projection operators for the tensor meson. They are very similar to the projectors for the vector meson. Consequently, the nonfactorizable contributions such as vertex, penguin and hard spectator corrections to $B\to T(P,V)$ decays can be directly obtained from $B\to VP,VV$ ones by making some suitable replacement.

\item The factorizable amplitude with a tensor meson emitted vanishes under the factorization hypothesis owing to the fact that a tensor meson cannot be created from the local $V-A$ and tensor currents. As a result, $B^-\to \bar K_2^{*0}\pi^-$ and $\ov B^0\to K_2^{*-}\pi^+$ vanish in naive factorization. The experimental observation of the former implies the importance of nonfactorizable effects.

\item Five different models for $B\to T$ transition form factors were considered. While the predictions of $B\to f_2(1270)$ form factors based on large energy effective theory or pQCD are favored by experiment, the covariant light-front quark model or the ISGW2 model for $B\to K_2^*(1430)$ ones is preferred by the data.

\item For penguin-dominated $B\to TP$ and $TV$ decays, the predicted rates in naive factorization are normally too small by one to two orders of magnitude. In QCDF, they are enhanced by the power corrections from penguin annihilation and nonfactorizable contributions.

\item There exist three distinct types of penguin contributions to $B\to K_2^*\eta^{(')}$: (i) $b\to sq\bar q\to s\eta_{q}$, (ii) $b\to ss\bar s\to s\eta_s$, and (iii) $b\to s q\bar q\to q \bar K_2^*$ with $\eta_q=(u\bar u+d\bar d)/\sqrt{2}$ and $\eta_s=s\bar s$. The dominant effects arise from the last two penguin contributions. The interference, constructive for $K_2^*\eta'$ and destructive for $K_2^*\eta$ between type-(ii) and type-(iii) diagrams, explains why $\Gamma(B\to K_2^*\eta')\gg\Gamma(B\to K_2^*\eta)$.

\item We use the measured rates of $K_2^*\omega$ and $K_2^*\phi$ modes to extract the penguin annihilation parameters $\rho_A^{TV}$ and $\rho_A^{VT}$ and the observed longitudinal polarization fractions $f_L(K_2^*\omega)$ and $f_L(K_2^*\phi)$ to fix the phases $\phi_A^{VT}$ and $\phi_A^{TV}$. The unexpectedly large rate of $B\to K_2^*\omega$ relative to $B\to K_2^*\phi$ implies that $\rho_A^{VT}\gg\rho_A^{VT}$. However, it may be hard to offer more intuitive understanding for the large disparity between $\rho_A^{TV}$ and $\rho_A^{VT}$ in magnitude.

\item The experimental observation that $f_T/f_L\ll 1$ for $B\to \phi K_2^*(1430)$, whereas $f_T/f_L\sim 1$ for $B\to \omega K_2^*(1430)$, can be
{\it accommodated}  in QCDF, but cannot be {\it dynamically explained} at first place. For penguin-dominated $B\to TV$ decays, we find $f_L(K_2^*\rho)\sim f_L(K_2^*\omega)\sim 0.65$ and $f_L(K^*f_2)\sim 0.93$. It will be of great interest to measure $f_L$ for these modes to test QCDF. Theoretically,
transverse polarization is expected to be small in  tree-dominated $\ov B\to TV$ decays except for the $a_2^-\rho^0,~a_2^-\rho^+,~K_2^{*0}K^{*-}$ and $K_2^{*0}\bar
K^{*0}$ modes.

\item For tree-dominated decays, their rates are usually very small except for the $a_2^0(\pi^-,\rho^-),~a_2^+(\pi^-,\rho^-)$ and $f_2(\pi^-,\rho^-)$ modes with branching fractions of order $10^{-6}$ or even bigger.

\end{itemize}

\vskip 1.71cm {\bf Acknowledgments}

One of us (H.Y.C.) wishes to thank C.N. Yang
Institute for Theoretical Physics at SUNY Stony Brook for its
hospitality. This work was supported in part by the National Center for Theoretical Sciences and the National Science Council of R.O.C. under Grant Nos.
NSC97-2112-M-008-002-MY3 and NSC99-2112-M-003-005-MY3.

\appendix
\section{Form factors in the ISGW2 quark model}
Consider the transition $B\to T$ in the ISGW2 quark model \cite{ISGW2}, where the tensor meson $T$
has the quark content $q_1\bar q_2$ with $\bar q_2$ being the
spectator quark. We begin with the definition \cite{ISGW2}
 \be \label{eq:Fn}
 F_n=\left({\tilde m_T\over \tilde
 m_B}\right)^{1/2}\left({\beta_B\beta_T\over
 \beta_{BT}^2}\right)^{n/2}\left[1+{1\over
 18}r^2(t_m-t)\right]^{-3},
 \en
where
 \be
 r^2={3\over 4m_bm_1}+{3m_2^2\over 2\bar m_B\bar
 m_T\beta_{BT}^2}+{1\over \bar m_B\bar m_T}\left({16\over
 33-2n_f}\right)\ln\left[{\alpha_s(\mu_{\rm QM})\over
 \alpha_s(m_1)}\right],
 \en
$\tilde m$ is the sum of the meson's constituent quarks' masses,
$\bar m$ is the hyperfine-averaged mass (for example, $\bar m_B={3\over 4}m_{B^*}+{1\over 4}m_B$), $t_m=(m_B-m_T)^2$ is the
maximum momentum transfer, and
 \be
 \mu_\pm=\left({1\over m_1}\pm {1\over m_b}\right)^{-1},
 \en
with $m_1$ and $m_2$ being the masses of the quarks $q_1$ and
$\bar q_2$, respectively. In Eq. (\ref{eq:Fn}), the values of the
parameters $\beta_B$ and $\beta_T$ are available in \cite{ISGW2}
and $\beta_{BT}^2={1\over 2}(\beta_B^2+\beta_T^2)$.

The form factors defined by Eq. (\ref{eq:BTff}) have the following
expressions in the ISGW2 model:
 \be \label{formISGW2}
 h &=& {m_2\over 2\sqrt{2}\tilde m_B\beta_B}\left[ {1\over m_1}-{m_2\beta^2_B\over 2\mu_-\tilde m_T\beta^2_{BT}}\right]F_5^{(h)},  \non \\
 k &=& {m_2\over \sqrt{2}\beta_B}(1+\tilde \omega)F_5^{(k)},
 \non \\
  b_+ +b_- &=& {m_2^2\over 4\sqrt{2}m_qm_b\tilde m_B\beta_B}\,{\beta^2_T\over \beta^2_{BT}}\left(1-{m_2\over 2\tilde m_B}{\beta^2_T\over \beta^2_{BT}}\right)F_5^{(b_++b_-)},
 \\
 b_+ -b_- &=& -{m_2\over \sqrt{2}m_b\tilde m_T\beta_B}\left[1-{m_2m_b\over 2\mu_+\tilde m_B}{\beta^2_T\over \beta^2_{BT}}+{m_2\over 4m_q}{\beta^2_T\over \beta^2_{BT}}\left(1-{m_2\over 2\tilde m_B}{\beta^2_T\over \beta^2_{BT}}\right]\right)F_5^{(b_+-b_-)},
 \non
 \en
where
 \be
 && F_5^{(h)}=F_5\left({\bar m_B\over\tilde
 m_B}\right)^{-3/2}\left({\bar m_T\over \tilde m_T}\right)^{-1/2},
 \non \\
 && F_5^{(k)}=F_5\left({\bar m_B\over\tilde
 m_B}\right)^{-1/2}\left({\bar m_T\over \tilde m_T}\right)^{1/2},
 \non \\
 && F_5^{(b_++b_-)}=F_5\left({\bar m_B\over\tilde
 m_B}\right)^{-5/2}\left({\bar m_T\over \tilde m_T}\right)^{1/2},
 \non \\
 && F_5^{(b_+-b_-)}=F_5\left({\bar m_B\over\tilde
 m_B}\right)^{-3/2}\left({\bar m_T\over \tilde m_T}\right)^{-1/2},
 \en
and
 \be
 \tilde \omega-1=\,{t_m-t\over 2\bar m_B\bar m_T}.
 \en

In the original version of the ISGW model \cite{ISGW}, the
function $F_n$ has a different expression in its $(t_m-t)$
dependence:
  \be \label{oldFn}
 F_n=\left({\tilde m_T\over \tilde
 m_B}\right)^{1/2}\left({\beta_B\beta_T\over
 \beta_{BT}^2}\right)^{n/2}{\rm exp}\left\{-{m_2\over 4 \tilde m_B\tilde
 m_T}\,{t_m-t\over \kappa^2\beta_{BT}^2}\right\},
 \en
where $\kappa=0.7$ is the relativistic correction factor. The form
factors are then given by
 \be
 h &=& {m_2\over 2\sqrt{2}\tilde m_B\beta_B}\left[ {1\over m_1}-{m_2\over 2\tilde m_T\mu_-}{\beta^2_B\over \beta^2_{BT}}\right]F_5,  \non \\
 k &=& {m_2\over \sqrt{2}\beta_B}F_5,
 \non \\
  b_+ &=& -{m_2\over 2\sqrt{2}m_b\tilde m_T\beta_B}\left[1-{m_2m_b\over 2\mu_+\tilde m_B}{\beta^2_T\over \beta^2_{BT}}+{m_2m_b\over 4\tilde m_B\mu_-}{\beta^2_T\over \beta^2_{BT}}\left(1-{m_2\over 2\tilde m_B}{\beta^2_T\over \beta^2_{BT}}\right)\right]F_5.
 \en
Note that the expressions in Eq.
(\ref{formISGW2}) in the ISGW2 model allow one to determine the
form factor $b_-$, which vanishes in the ISGW model.

\section{Decay amplitudes}

The coefficients of the flavor operators $\alpha_i^{p,(h)}$ can be expressed in terms of $a_i^{p,(h)}$ in the
following:
\begin{eqnarray}\label{eq:alphai--1}
   \alpha_1(M_1 M_2) &=& a_1(M_1 M_2) \,, \nonumber\\
   \alpha_2(M_1 M_2) &=& a_2(M_1 M_2) \,, \nonumber\\
   \alpha_3^p(M_1 M_2) &=& \left\{
    \begin{array}{cl}
     a_3^p(M_1 M_2) - a_5^p(M_1 M_2)
      & \quad \mbox{for~} M_1 M_2=TP \,, \\
     a_3^p(M_1 M_2) + a_5^p(M_1 M_2)
      & \quad \mbox{for~} M_1 M_2=PT  \,,
    \end{array}\right. \nonumber\\
   \alpha_4^p(M_1 M_2) &=& \left\{
    \begin{array}{cl}
     a_4^p(M_1 M_2) - r_{\chi}^{M_2}\,a_6^p(M_1 M_2)
      & \quad \mbox{for~} M_1 M_2=TP  \\
     a_4^p(M_1 M_2) + r_{\chi}^{M_2}\,a_6^p(M_1 M_2)
      & \quad \mbox{for~} M_1 M_2=PT \,,
    \end{array}\right.\\
   \alpha_{3,\rm EW}^p(M_1 M_2) &=& \left\{
    \begin{array}{cl}
     a_9^p(M_1 M_2) - a_7^p(M_1 M_2)
      &  \quad \mbox{for~} M_1 M_2=TP\,, \\
     a_9^p(M_1 M_2) + a_7^p(M_1 M_2)
      &  \quad \mbox{for~} M_1 M_2=PT  \,,
    \end{array}\right. \nonumber\\
   \alpha_{4,\rm EW}^p(M_1 M_2) &=& \left\{
    \begin{array}{cl}
     a_{10}^p(M_1 M_2) - r_{\chi}^{M_2}\,a_8^p(M_1 M_2)
      & \quad \mbox{for~} M_1 M_2=TP\,,  \\
     a_{10}^p(M_1 M_2) + r_{\chi}^{M_2}\,a_8^p(M_1 M_2)
      &  \quad \mbox{for~} M_1 M_2=PT \,,
     \end{array}\right. \nonumber
\end{eqnarray}
for $\overline B\to TP$ decays, and
\begin{eqnarray}\label{eq:alphai--2}
   \alpha_1^{h}(M_1 M_2) &=& a_1^{h}(M_1 M_2) \,, \nonumber\\
   \alpha_2^{h}(M_1 M_2) &=& a_2^{h}(M_1 M_2) \,, \nonumber\\
   \alpha_3^{p,h}(M_1 M_2) &=&
     a_3^{p,h}(M_1 M_2) + a_5^{p,h}(M_1 M_2),
     \\
   \alpha_4^{p,h}(M_1 M_2) &=&
     a_4^{p,h}(M_1 M_2) - r_{\chi}^{M_2}\,a_6^{p,h}(M_1 M_2)\,, \nonumber\\
   \alpha_{3,\rm EW}^{p,h}(M_1 M_2) &=&
     a_9^{p,h}(M_1 M_2) + a_7^{p,h}(M_1 M_2)\,, \nonumber\\
   \alpha_{4,\rm EW}^{p,h}(M_1 M_2) &=&
     a_{10}^{p,h}(M_1 M_2) - r_{\chi}^{M_2}\,a_8^{p,h}(M_1 M_2), \nonumber
\end{eqnarray}
for $\overline B\to TV$ decays with $(M_1 M_2) \equiv (TV)$ or $(VT)$. It should be noted that the order of the arguments of $\alpha_i^p(M_1 M_2)$ and
$a_i^p(M_1 M_2)$ is relevant. The chiral factor $r_\chi^M$ is given by
\begin{equation}\label{rchP}
   r_\chi^{\pi,K}(\mu) = \frac{2m_{\pi, K}^2}{m_b(\mu)\,(m_{q_1}+m_{q_2})(\mu)} \,, \quad
   r_\chi^{\eta_{q,s}^{(\prime)}}(\mu) = \frac{h_{q,s}}{f_{q,s}^{(\prime)}\, m_b(\mu)\,m_{q,s}(\mu)}
\end{equation}
for the pseudoscalar mesons,
\begin{equation}
\label{eq:rchiV}
   r_\chi^V(\mu) = \frac{2m_V}{m_b(\mu)}\,\frac{f_V^\perp(\mu)}{f_V} \,
\end{equation}
for the vector meson, and
\begin{equation}
\label{eq:rchiA}
   r_\chi^T(\mu) = \frac{2m_T}{m_b(\mu)}\,\frac{f_T^\perp(\mu)}{f_T} \,
\end{equation}
for the tensor meson.  See Appendix \ref{app:eta-etap} for further discussions on the parameters $h_{q,s}$ and $f_{q,s}^{(\prime)}$.

In the following decay amplitudes, the order
of the arguments of $\alpha_i^{p(,h)} (M_1 M_2)$ and $\beta_i^{p(,h)}(M_1 M_2)$ is consistent with the order of the arguments of $X_{(h)}^{(\overline B M_1, M_2)}$ or $\overline{X}_{(h)}^{(\overline B M_1, M_2)}$, where
\be \label{eq:beta}
 \beta_i^{p} (T P) =\frac{-i f_B f_{T} f_{P}}{X^{(\overline B T,P)}}b_i^{p} , \quad
 \beta_i^{p} (P T) =\frac{-i f_B f_{T} f_{P}}{\overline{X}^{(\overline B P,T)}}b_i^{p} ,
 ~~ {\rm for}~TP~{\rm modes}, \\
 \beta_i^{p,h} (T V) =\frac{i f_B f_{T}
f_{V}}{X_h^{(\overline B T, V)}}b_i^{p,h} , \quad
 \beta_i^{p,h} (V T) =\frac{i f_B f_{T}
f_{V}}{ \overline{X}_h^{(\overline B V,T)}}b_i^{p,h} ,~~ {\rm for}~TV~{\rm modes}.
\en
The decay amplitudes for $\overline B \to TP, TV$ are summarized as follows:

\subsection{$\overline{B} \to TP$ decays}
\subsubsection{Decay amplitudes with $\Delta S=0$:}
\begin{eqnarray}
\sqrt2\,{\cal A}_{B^-\to f_2 \pi^-}
  &=& \frac{G_F}{\sqrt{2}}\sum_{p=u,c}\lambda_p^{(d)} \nonumber\\
 && \times \Bigg\{ \Big[
    \delta_{pu}\,(\alpha_2 + \beta_2)
    + 2\alpha_3^p + \alpha_4^p + \frac{1}{2}\alpha_{3,{\rm EW}}^p
    - \frac{1}{2}\alpha_{4,{\rm EW}}^p
    + \beta_3^p + \beta_{3,{\rm EW}}^p
    \Big] \overline{X}^{(\overline{B} \pi, f_2^q)}
  \nonumber\\
  & & + \sqrt{2} \left[  \alpha_3^p - \frac{1}{2}\alpha_{3,{\rm EW}}^p
    \right] \overline{X}^{(\overline{B}\pi, f_2^s)}
  \nonumber\\
  & & +\left[ \delta_{pu}\,(\alpha_1 + \beta_2)
    + \alpha_4^p + \alpha_{4,{\rm EW}}^p + \beta_3^p
    + \beta_{3,{\rm EW}}^p \right]  X^{(\overline{B} f_2^q,\pi)} \Bigg\},
\\
 -2\,{\cal A}_{\overline{B}^0\to f_2 \pi^0}
   &=& \frac{G_F}{\sqrt{2}}\sum_{p=u,c}\lambda_p^{(d)}  \Bigg\{
 \Big[
    \delta_{pu}\,(\alpha_2 - \beta_1)
    + 2\alpha_3^p + \alpha_4^p + \frac{1}{2}\alpha_{3,{\rm EW}}^p
    - \frac{1}{2}\alpha_{4,{\rm EW}}^p
    \nonumber\\
    &&\hspace*{1cm}+\, \beta_3^p -\frac{1}{2}\beta_{3,{\rm EW}}^p
    -\frac{3}{2}\beta_{4,{\rm EW}}^p  \Big]  \overline{X}^{(\overline{B} \pi, f_2^q)}
  + \sqrt{2}  \left[
    \alpha_3^p - \frac{1}{2}\alpha_{3,{\rm EW}}^p
 \right]  \overline{X}^{(\overline{B}\pi, f_2^s)}
   \nonumber\\
  & & + \left[ \delta_{pu}\,(-\alpha_2 - \beta_1)
    + \alpha_4^p -\frac{3}{2}\alpha_{3,{\rm EW}}^p -\frac{1}{2}\alpha_{4,{\rm EW}}^p \right.\nonumber\\
    & & \hspace*{1cm} \, \left. + \beta_3^p -\frac{1}{2}\beta_{3,{\rm EW}}^p -\frac{3}{2}\beta_{4,{\rm EW}}^p
    \right]X^{(\overline{B} f_2^q,\pi)} \Bigg\}  \hspace*{-0.025cm},
\end{eqnarray}

\begin{eqnarray}
2\,{\cal A}_{\bar B^0\to f_2 \eta}
  &=& \frac{G_F}{\sqrt{2}}\sum_{p=u,c}\lambda_p^{(d)}  \Bigg\{
  \Big[
    \delta_{pu}\,(\alpha_2 + \beta_1 )
    + 2\alpha_3^p + \alpha_4^p + \frac{1}{2}\alpha_{3,{\rm EW}}^p
    - \frac{1}{2}\alpha_{4,{\rm EW}}^p
    \nonumber\\[-0.1cm]
    &&\hspace*{1cm}+\, \beta_3^p + 2\beta_4^p - \frac{1}{2}\beta_{3,{\rm EW}}^p
    + \frac{1}{2}\beta_{4,{\rm EW}}^p
    \Big] X^{(\overline B f_{2}^q,\eta_q)}
  \nonumber\\
  &+& \sqrt2 \left[
    \alpha_3^p - \frac{1}{2}\alpha_{3,{\rm EW}}^p
   \right] X^{(\overline B f_{2}^q,\eta_s)}
 + \sqrt{2} \left[
    \delta_{pc}\,\alpha_2 + \alpha_3^p \right]  X^{(\overline B f_{2}^q,\eta_c)}
  - 2 i f_B f_{2}^s f_\eta^s \left[
    b_4^p -\frac{1}{2} b_{4,{\rm EW}}^p  \right]_{f_2^s \eta_s}
  \nonumber \\
  &+& \Big[
    \delta_{pu}\,(\alpha_2 + \beta_1)
    + 2\alpha_3^p + \alpha_4^p + \frac{1}{2}\alpha_{3,{\rm EW}}^p
    - \frac{1}{2}\alpha_{4,{\rm EW}}^p
    \nonumber\\[-0.1cm]
    &&\hspace*{1cm}+\, \beta_3^p + 2\beta_4^p - \frac{1}{2}\beta_{3,{\rm EW}}^p
    + \frac{1}{2}\beta_{4,{\rm EW}}^p \Big]  \overline{X}^{(\overline B\eta_q, f_{2}^q)}
  \nonumber\\
  &+& \sqrt2  \left[
    \alpha_3^p - \frac{1}{2}\alpha_{3,{\rm EW}}^p  \right] \overline{X}^{(\overline B\eta_q, f_{2}^s)}
  - 2 i  f_B f_{2}^s f_\eta^s  \left[
    b_4^p -\frac{1}{2} b_{4,{\rm EW}}^p  \right]_{\eta_s f_2^s} \Bigg\},
\end{eqnarray}
the amplitudes for $\overline{B}\to f_2 \eta^\prime$ can be obtained from $\overline{B}\to f_2 \eta$ with the replacement $(f_2,\eta) \to (f_2,\eta^\prime)$,
\begin{eqnarray}
   \sqrt2\,{\cal A}_{B^-\to a_2^0 \pi^- }
   &=& \frac{G_F}{\sqrt{2}}\sum_{p=u,c}\lambda_p^{(d)}  \Bigg\{
 \left[ \delta_{pu}\,(\alpha_2 - \beta_2)
    - \alpha_4^p + \frac{3}{2}\alpha_{3,{\rm EW}}^p
    + \frac{1}{2}\alpha_{4,{\rm EW}}^p  - \beta_3^p - \beta_{3,{\rm EW}}^p
    \right] \overline{X}^{(\overline{B} \pi, a_2)} \nonumber\\
   & & + \left[ \delta_{pu}\,(\alpha_1 + \beta_2)
    + \alpha_4^p + \alpha_{4,{\rm EW}}^p + \beta_3^p
    + \beta_{3,{\rm EW}}^p \right]  X^{(\overline{B} a_2,\pi)} \Bigg\},
  \\
\sqrt2\,{\cal A}_{B^-\to a_2^- \pi^0}
   &=&  \frac{G_F}{\sqrt{2}}\sum_{p=u,c}\lambda_p^{(d)}
   \Bigg\{ \left[ \delta_{pu}\,(\alpha_2 - \beta_2)
    - \alpha_4^p + \frac{3}{2}\alpha_{3,{\rm EW}}^p
    + \frac{1}{2}\alpha_{4,{\rm EW}}^p  - \beta_3^p - \beta_{3,{\rm EW}}^p
    \right] X^{(\overline B a_2, \pi)} \nonumber\\
   &+&  \left[ \delta_{pu}\,(\alpha_1 + \beta_2)
    + \alpha_4^p + \alpha_{4,{\rm EW}}^p + \beta_3^p
    + \beta_{3,{\rm EW}}^p \right] \overline{X}^{(\overline B \pi, a_2)}\Bigg\}  , \\
{\cal A}_{\bar B^0\to a_2^- \pi^+}
   &=& \frac{G_F}{\sqrt{2}}\sum_{p=u,c}\lambda_p^{(d)} \Bigg\{
 \left[ \delta_{pu}\,\alpha_1 + \alpha_4^p
    + \alpha_{4,{\rm EW}}^p + \beta_3^p + \beta_4^p
    - \frac{1}{2}\beta_{3,{\rm EW}}^p - \frac{1}{2}\beta_{4,{\rm EW}}^p \right]
      \overline{X}^{(\overline{B} \pi, a_2)}
    \nonumber\\
   & & + \left[ \delta_{pu}\,\beta_1 + \beta_4^p
    + \beta_{4,{\rm EW}}^p \right]  X^{(\overline{B} a_2,\pi)} \Bigg\},
  \\
 {\cal A}_{\bar B^0\to a_2^+ \pi^-}
   &=& \frac{G_F}{\sqrt{2}}\sum_{p=u,c}\lambda_p^{(d)}
   \Bigg\{ \left[ \delta_{pu}\,\alpha_1 + \alpha_4^p
    + \alpha_{4,{\rm EW}}^p + \beta_3^p + \beta_4^p
    - \frac{1}{2}\beta_{3,{\rm EW}}^p - \frac{1}{2}\beta_{4,{\rm EW}}^p \right]
    X^{(\overline B a_2, \pi)}
    \nonumber\\
   &+& \left[ \delta_{pu}\,\beta_1 + \beta_4^p
    + \beta_{4,{\rm EW}}^p \right]  \overline{X}^{(\overline B \pi, a_2)}\Bigg\}  , \\
-2\,{\cal A}_{\bar B^0\to a_2^0 \pi^0}
   &=&  \frac{G_F}{\sqrt{2}}\sum_{p=u,c}\lambda_p^{(d)}  \Bigg\{
 \Big[ \delta_{pu}\,(\alpha_2 - \beta_1)
    - \alpha_4^p + \frac{3}{2}\alpha_{3,{\rm EW}}^p
    + \frac{1}{2}\alpha_{4,{\rm EW}}^p - \beta_3^p - 2\beta_4^p
    \nonumber\\[-0.1cm]
    &&\hspace*{1cm}
    +\, \frac{1}{2}\beta_{3,{\rm EW}}^p - \frac{1}{2}\beta_{4,{\rm EW}}^p \Big]
     \overline{X}^{(\overline{B} \pi, a_2)}
    \nonumber\\
   & & +  \Big[ \delta_{pu}\,(\alpha_2 - \beta_1)
    - \alpha_4^p + \frac{3}{2}\alpha_{3,{\rm EW}}^p
    + \frac{1}{2}\alpha_{4,{\rm EW}}^p - \beta_3^p - 2\beta_4^p
    \nonumber\\[-0.1cm]
    &&\hspace*{1cm}
    + \,\frac{1}{2}\beta_{3,{\rm EW}}^p - \frac{1}{2}\beta_{4,{\rm EW}}^p \Big]
     X^{(\overline{B} a_2,\pi)} \Bigg\},
\end{eqnarray}
\begin{eqnarray}
 \sqrt2\,{\cal A}_{B^-\to a_2^- \eta^{(\prime)}}
   &=&  \frac{G_F}{\sqrt{2}}\sum_{p=u,c}\lambda_p^{(d)}
   \Bigg\{ \bigg[ \delta_{pu}\,(\alpha_2 + \beta_2)
     + 2\alpha_3^{p}+ \alpha_4^{p} + \frac{1}{2}\alpha_{3,{\rm EW}}^{p} \non \\
   && -
   \frac{1}{2}\alpha_{4,{\rm EW}}^{p}  + \beta_3^{p} + \beta_{3,{\rm EW}}^{p} \bigg]
   X^{(\overline B a_2, \eta_q^{(\prime)})} \nonumber\\
   & & +  \left[ \delta_{pu}\,(\alpha_1 + \beta_2)
    + \alpha_4^{p} + \alpha_{4,{\rm EW}}^{p} + \beta_3^{p}
    + \beta_{3,{\rm EW}}^{p} \right] \overline{X}^{(\overline B \eta_q^{(\prime)}, a_2)} \nonumber\\
    & & + \sqrt{2}
    \left[ \alpha_3^{p} - \frac{1}{2}\alpha_{3,{\rm EW}}^{p} \right]
    X^{(\overline B a_2, \eta_s^{(\prime)})}
    + \sqrt{2}
    \left[ \delta_{pc}\, \alpha_2 +\alpha_3^{p} \right]
    X^{(\overline B a_2, \eta_c^{(\prime)})}  \Bigg\}  , \\
   -2\,{\cal A}_{\bar B^0\to a_2^0\eta^{(\prime)}}
   &=& \frac{G_F}{\sqrt{2}}\sum_{p=u,c}\lambda_p^{(d)}
   \Bigg\{ \Big[ \delta_{pu}\,(\alpha_2 - \beta_1)
    + 2\alpha_3^{p}+ \alpha_4^{p} + \frac{1}{2}\alpha_{3,{\rm EW}}^{p}
    - \frac{1}{2}\alpha_{4,{\rm EW}}^{p}
    \nonumber\\[-0.1cm]
    &&
 + \beta_3^{p}-\, \frac{1}{2}\beta_{3,{\rm EW}}^{p} -
\frac{3}{2}\beta_{4,{\rm EW}}^{p} \Big]
    X^{(\overline Ba_2, \eta_q^{(\prime)})}
    \nonumber\\
   &&  +\Big[ \delta_{pu}\,(-\alpha_2 - \beta_1)
    + \alpha_4^{p}- \frac{3}{2}\alpha_{3,{\rm EW}}^{p}
    - \frac{1}{2}\alpha_{4,{\rm EW}}^{p}
    \nonumber\\
    &&
 + \beta_3^{p}- \,\frac{1}{2}\beta_{3,{\rm EW}}^{p} -
\frac{3}{2}\beta_{4,{\rm EW}}^{p} \Big]
     \overline{X}^{(\overline B\eta_q^{(\prime)}, a_2)} \nonumber\\
 &&  + \sqrt{2}
    \left[ \alpha_3^{p} - \frac{1}{2}\alpha_{3,{\rm EW}}^{p} \right]
    X^{(\overline B a_2, \eta_s^{(\prime)})}
    + \sqrt{2}
    \left[ \delta_{pc}\, \alpha_2 +\alpha_3^{p} \right]
    X^{(\overline B a_2, \eta_c^{(\prime)})}
  \Bigg\}\,,
\end{eqnarray}
\be
  {\cal A}_{B^-\to K^{*-}_2 K^0}
   &=&  \frac{G_F}{\sqrt{2}}\sum_{p=u,c}\lambda_p^{(d)}
    \Big[ \delta_{pu}\beta_2+ \alpha_4^p - \frac{1}{2}\alpha_{4,{\rm
    EW}}^p +\beta_3^p +\beta^p_{3,{\rm EW}} \Big]   X^{(\overline B \overline K_2^*,K)},   \\
   {\cal A}_{B^-\to K^{*0}_2 K^-}
   &=&  \frac{G_F}{\sqrt{2}}\sum_{p=u,c}\lambda_p^{(d)}
    \Big[ \delta_{pu}\beta_2+ \alpha_4^p - \frac{1}{2}\alpha_{4,{\rm
    EW}}^p +\beta_3^p +\beta^p_{3,{\rm EW}} \Big]   \overline{X}^{(\overline B \ov K,K_2^*)}, \\
    {\cal A}_{\ov B^0\to K^{*-}_2 K^+}
   &=&  \frac{G_F}{\sqrt{2}}\sum_{p=u,c}\lambda_p^{(d)}\Bigg\{
    \Big[ \delta_{pu}\beta_1+\beta_4^p +\beta^p_{4,{\rm EW}} \Big]   X^{(\overline B \ov K_2^*,K)} \nonumber\\
   & & -icf_Bf_{K_2^*}f_K\left[b_4^p-{1\over 2}b^p_{4,{\rm EW}}\right]_{K K_2^*}\Bigg\}, \\
     {\cal A}_{\ov B^0\to K^{*+}_2 K^-}
   &=&  \frac{G_F}{\sqrt{2}}\sum_{p=u,c}\lambda_p^{(d)}\Bigg\{
    \Big[ \delta_{pu}\beta_1+\beta_4^p +\beta^p_{4,{\rm EW}} \Big]   \overline{X}^{(\overline B \ov K, K_2^*)}
    \nonumber\\
   & &-icf_Bf_{K_2^*}f_K\left[b_4^p-{1\over 2}b^p_{4,{\rm EW}}\right]_{K_2^* K}\Bigg\},  \\
     {\cal A}_{\ov B^0\to \ov K^{*0}_2 K^0}
   &=&  \frac{G_F}{\sqrt{2}}\sum_{p=u,c}\lambda_p^{(d)}\Bigg\{
    \Big[ \alpha_4^p-{1\over 2}\alpha^p_{4,{\rm EW}}+\beta^p_3
    +\beta_4^p -{1\over 2}\beta^p_{3,{\rm EW}} -{1\over 2}\beta^p_{4,{\rm EW}}
    \Big]   X^{(\overline B \ov K_2^*,K)},  \non \\
    & & - ic f_Bf_{K_2^*}f_K\left[b_4^p-{1\over 2}b^p_{4,{\rm EW}}\right]_{K\ov K_2^*}\Bigg\},  \\
     {\cal A}_{\ov B^0\to K^{*0}_2 \ov K^0}
   &=&  \frac{G_F}{\sqrt{2}}\sum_{p=u,c}\lambda_p^{(d)}\Bigg\{
    \Big[ \alpha_4^p-{1\over 2}\alpha^p_{4,{\rm EW}}+\beta^p_3
    +\beta_4^p -{1\over 2}\beta^p_{3,{\rm EW}} -{1\over 2}\beta^p_{4,{\rm EW}}
    \Big]   \overline{X}^{(\overline B \bar K,K_2^*)}, \non \\
    & & - ic f_Bf_{K_2^*}f_K\left[b_4^p-{1\over 2}b^p_{4,{\rm EW}}\right]_{K^*_2\ov K}\Bigg\},
 \en
with $c=1$.

\subsubsection{Decay amplitudes with $\Delta S=1$}
\begin{eqnarray}
   \sqrt2\,{\cal A}_{B^-\to f_2 K^-} &=& \frac{G_F}{\sqrt{2}}\sum_{p=u,c}\lambda_p^{(s)}
 \Bigg\{
   \left[
    \delta_{pu}\, \alpha_2
    + 2\alpha_3^p + \frac{1}{2}\alpha_{3,{\rm EW}}^p \right] \overline{X}^{(\overline{B}\bar K, f_2^q)}
  \nonumber\\
  & & + \sqrt{2}\Big[\delta_{pu}\, \beta_2 +
    \alpha_3^p + \alpha_4^p - \frac{1}{2}\alpha_{3,{\rm EW}}^p
    -\frac{1}{2}\alpha_{4,{\rm EW}}^p +\beta_3^p +\beta_{3,{\rm EW}}^p \Big]  \overline{X}^{(\overline{B}\bar K, f_2^s)}
  \nonumber\\
  &+&  \left[ \delta_{pu}\,(\alpha_1+\beta_2)+
    \alpha_4^p+\alpha_{4,{\rm EW}}^p
    + \beta_3^p +\beta_{3,{\rm EW}}^p\right] X^{(\overline{B} f_2^q,\bar K)} ,
 \\
   \sqrt2\,{\cal A}_{\bar B^0\to f_2 K^0 }
 &=& \frac{G_F}{\sqrt{2}}\sum_{p=u,c}\lambda_p^{(s)}
 \Bigg\{   \left[
    \delta_{pu}\,\alpha_2
    + 2\alpha_3^p + \frac{1}{2}\alpha_{3,{\rm EW}}^p \right]
    \overline{X}^{(\overline{B}\bar K, f_2^q)}
  \nonumber\\
  &+& \sqrt{2} \Big[
    \alpha_3^p + \alpha_4^p- \frac{1}{2}\alpha_{3,{\rm EW}}^p
    -\frac{1}{2}\alpha_{4,{\rm EW}}^p + \beta_3^p - \frac{1}{2}\beta_{3,{\rm EW}}^p \Big]
    \overline{X}^{(\overline{B}\bar K, f_2^s)}
  \nonumber\\
  &+&  \left[\alpha_4^p -\frac{1}{2}\alpha_{4,{\rm EW}}^p
    + \beta_3^p -\frac{1}{2}\beta_{3,{\rm EW}}^p\right] X^{(\overline{B} f_2^q,\bar K)},
\end{eqnarray}
\be
  \sqrt{2}{\cal A}_{B^-\to a_2^0 K^- }
   &=&  \frac{G_F}{\sqrt{2}}\sum_{p=u,c}\lambda_p^{(s)} \Bigg\{
    \Big[ \delta_{pu}(\alpha_1+\beta_2)+ \alpha_4^{p}
    +\alpha_{4,{\rm EW}}^{p} +\beta_3^{p} +\beta^{p}_{3,{\rm EW}} \Big]   X^{(\overline B a_2,\ov K)}  \non \\
   && +\Big[\delta_{pu}\alpha_2 +{3\over 2}\alpha^{p}_{3,{\rm EW}}\Big]
   \overline{X}^{(\overline B\ov K,a_2)}\Bigg\},
       \\
 {\cal A}_{B^-\to  a_2^- K^0}
   &=&  \frac{G_F}{\sqrt{2}}\sum_{p=u,c}\lambda_p^{(s)}
    \Big[ \delta_{pu}\beta_2+ \alpha_4^{p}  - \frac{1}{2}\alpha_{4,{\rm
    EW}}^{p}  +\beta_3^{p} +\beta^{p}_{3,{\rm EW}} \Big]   X^{(\overline Ba_2, \ov K)},
       \\
 {\cal A}_{\ov B^0\to a_2^+ K^-}
   &=&  \frac{G_F}{\sqrt{2}}\sum_{p=u,c}\lambda_p^{(s)}
    \Big[
 \delta_{pu}\alpha_1^h+ \alpha_4^{p} + \alpha_{4,{\rm EW}}^{p}
 +\beta_3^{p} -{1\over 2}\beta^{p}_{3,{\rm EW}} \Big]
       X^{(\overline Ba_2, \ov K)},
       \\
  \sqrt{2}{\cal A}_{\ov B^0\to  a_2^0 K^0}
   &=&  \frac{G_F}{\sqrt{2}}\sum_{p=u,c}\lambda_p^{(s)} \Bigg\{
    \Big[ -\alpha_4^{p} + {1\over 2}\alpha_{4,{\rm EW}}^{p} -\beta_3^{p}
    + {1\over 2}\beta^{p}_{3,{\rm EW}} \Big]
       X^{(\overline B a_2,\ov K)}  \non \\
   &+& \Big[\delta_{pu}\alpha_2+ {3\over 2}\alpha^{p}_{3,{\rm EW}}\Big]
     \overline{X}^{(\overline B \ov K,a_2)}\Bigg\},
 \en
\begin{eqnarray}
   \sqrt2\,{\cal A}_{B^-\to K_2^{*-} \eta^{(\prime)}}
   &=&   \frac{G_F}{\sqrt{2}}\sum_{p=u,c}\lambda_p^{(s)} \Bigg\{
   \left[
    \delta_{pu}\, \alpha_2
    + 2\alpha_3^p + \frac{1}{2}\alpha_{3,{\rm EW}}^p \right]  X^{(\overline B\ov K_2^*, \eta^{(\prime)}_q)}
  \nonumber\\
  &+& \sqrt{2} \Big[\delta_{pu}\, \beta_2 +
    \alpha_3^p + \alpha_4^p - \frac{1}{2}\alpha_{3,{\rm EW}}^p
    -\frac{1}{2}\alpha_{4,{\rm EW}}^p +\beta_3^p +\beta_{3,{\rm EW}}^p \Big]
  X^{(\overline B\ov K_2^*, \eta^{(\prime)}_s)}
  \nonumber\\
  &+& \sqrt{2}  \left[
    \delta_{pc}\,\alpha_2 + \alpha_3^p \right]  X^{(\overline B\ov K_2^*, \eta^{(\prime)}_c)}
  \nonumber\\
  &+&  \left[ \delta_{pu}\,(\alpha_1+\beta_2)+
    \alpha_4^p+\alpha_{4,{\rm EW}}^p
    + \beta_3^p +\beta_{3,{\rm EW}}^p\right] X^{(\overline B  \eta^{(\prime)}_q, \ov K_2^*)}
    \Bigg\},
\\
   \sqrt2\,{\cal A}_{\bar B^0\to\bar K_2^{*0}\eta^{(\prime)}}
   &=& \frac{G_F}{\sqrt{2}}\sum_{p=u,c}\lambda_p^{(s)} \Bigg\{
 \left[
    \delta_{pu}\,\alpha_2
    + 2\alpha_3^p + \frac{1}{2}\alpha_{3,{\rm EW}}^p  \right] X^{(\overline B\ov K_2^*, \eta^{(\prime)}_q)}
  \nonumber\\
  &+& \sqrt{2}  \Big[
    \alpha_3^p + \alpha_4^p- \frac{1}{2}\alpha_{3,{\rm EW}}^p
    -\frac{1}{2}\alpha_{4,{\rm EW}}^p + \beta_3^p - \frac{1}{2}\beta_{3,{\rm EW}}^p
   \Big]  X^{(\overline B\ov K_2^*, \eta^{(\prime)}_s)}
  \nonumber\\
  &+& \sqrt{2} \left[
    \delta_{pc}\,\alpha_2 + \alpha_3^p \right] X^{(\overline B\ov K_2^*, \eta^{(\prime)}_c)}
  \nonumber\\
  &+&   \left[\alpha_4^p -\frac{1}{2}\alpha_{4,{\rm EW}}^p
    + \beta_3^p -\frac{1}{2}\beta_{3,{\rm EW}}^p\right] \overline{X}^{(\overline B  \eta^{(\prime)}_q, \ov K_2^*)}
  \Bigg\},
\end{eqnarray}
and the amplitudes for $\overline{B}\to \overline{K}_2^* \pi$ can be obtained from $\overline{B}\to \overline{K} a_2$ with the replacement $(\overline{K}, a_2) \to (\overline{K}_2^*, \pi)$.

\subsection{$\overline{B} \to TV$ decays}
\subsubsection{Decay amplitudes with $\Delta S=0$:}

The amplitudes for $\overline{B}\to f_2 \rho$ can be obtained from $\overline{B}\to f_2 \pi$ with the replacement $(f_2,\pi) \to (f_2,\rho)$,
\begin{eqnarray}
2\,{\cal A}_{\bar B^0\to f_2 \omega}
  &=& \frac{G_F}{\sqrt{2}}\sum_{p=u,c}\lambda_p^{(d)}  \Bigg\{
  \Big[
    \delta_{pu}\,(\alpha_2^h + \beta_1^h )
    + 2\alpha_3^{p,h} + \alpha_4^{p,h} + \frac{1}{2}\alpha_{3,{\rm EW}}^{p,h}
    - \frac{1}{2}\alpha_{4,{\rm EW}}^{p,h}
    \nonumber\\[-0.1cm]
    &&\hspace*{0.3cm}+\, \beta_3^{p,h} + 2\beta_4^{p,h} - \frac{1}{2}\beta_{3,{\rm EW}}^{p,h}
    + \frac{1}{2}\beta_{4,{\rm EW}}^{p,h}
    \Big] X^{(\overline B f_{2}^q,\omega)}
  \nonumber\\
  &+& \Big[
    \delta_{pu}\,(\alpha_2^h + \beta_1^h)
    + 2\alpha_3^{p,h} + \alpha_4^{p,h} + \frac{1}{2}\alpha_{3,{\rm EW}}^{p,h}
    - \frac{1}{2}\alpha_{4,{\rm EW}}^{p,h} +\, \beta_3^{p,h} + 2\beta_4^{p,h}
    \nonumber\\[-0.1cm]
    &&\hspace*{0.3cm} - \frac{1}{2}\beta_{3,{\rm EW}}^{p,h}
    + \frac{1}{2}\beta_{4,{\rm EW}}^{p,h} \Big]  \overline{X}^{(\overline B\omega, f_{2}^q)}
  + \sqrt2  \left[
    \alpha_3^{p,h} - \frac{1}{2}\alpha_{3,{\rm EW}}^{p,h}  \right] \overline{X}^{(\overline B\omega, f_{2}^s)}
  \Bigg\},
\end{eqnarray}
\begin{eqnarray}
2\,{\cal A}_{\bar B^0\to f_2 \phi}
  &=& \frac{G_F}{\sqrt{2}}\sum_{p=u,c}\lambda_p^{(d)}  \Bigg\{
   \sqrt2 \left[
    \alpha_3^{p,h} - \frac{1}{2}\alpha_{3,{\rm EW}}^{p,h}
   \right] X^{(\overline B f_{2}^q, \phi)}
  + 2 i f_B f_{2}^s f_\phi \left[
    b_4^{p,h} -\frac{1}{2} b_{4,{\rm EW}}^{p,h}  \right]_{f_2^s \phi}
  \nonumber \\
  &+& 2 i  f_B f_{2}^s f_\phi  \left[
    b_4^{p,h} -\frac{1}{2} b_{4,{\rm EW}}^{p,h}  \right]_{\phi f_2^s} \Bigg\},
\end{eqnarray}
the amplitudes for $\overline{B}\to a_2 \rho$ can be obtained from $\overline{B}\to a_2 \pi$ with the replacement $(a_2,\pi) \to (a_2,\rho)$,
\begin{eqnarray}
 \sqrt2\,{\cal A}_{B^-\to a_2^- \omega}^h
   &=&  \frac{G_F}{\sqrt{2}}\sum_{p=u,c}\lambda_p^{(d)}
   \Bigg\{ \bigg[ \delta_{pu}\,(\alpha_2^h + \beta_2^h)
     + 2\alpha_3^{p,h}+ \alpha_4^{p,h} + \frac{1}{2}\alpha_{3,{\rm EW}}^{p,h} \non \\
   && - \frac{1}{2}\alpha_{4,{\rm EW}}^{p,h}  + \beta_3^{p,h} + \beta_{3,{\rm EW}}^{p,h}
    \bigg] X^{(\overline B a_2, \omega)}_h \nonumber\\
   &+&  \left[ \delta_{pu}\,(\alpha_1^h + \beta_2^h)
    + \alpha_4^{p,h} + \alpha_{4,{\rm EW}}^{p,h} + \beta_3^{p,h}
    + \beta_{3,{\rm EW}}^{p,h} \right] \overline{X}^{(\overline B \omega, a_2)}_h\Bigg\}  , \\
   -2\,{\cal A}_{\bar B^0\to a_2^0\omega}^h
   &=& \frac{G_F}{\sqrt{2}}\sum_{p=u,c}\lambda_p^{(d)}
   \Bigg\{ \Big[ \delta_{pu}\,(\alpha_2^h - \beta_1^h)
    + 2\alpha_3^{p,h}+ \alpha_4^{p,h} + \frac{1}{2}\alpha_{3,{\rm EW}}^{p,h}
    - \frac{1}{2}\alpha_{4,{\rm EW}}^{p,h}
    \nonumber\\[-0.1cm]
    &&\hspace*{1cm}
 + \beta_3^{p,h}-\, \frac{1}{2}\beta_{3,{\rm EW}}^{p,h} -
\frac{3}{2}\beta_{4,{\rm EW}}^{p,h} \Big]
    X^{(\overline Ba_2, \omega)}_h
    \nonumber\\
   && \hspace*{1cm} +\Big[ \delta_{pu}\,(-\alpha_2^h - \beta_1^h)
    + \alpha_4^{p,h}- \frac{3}{2}\alpha_{3,{\rm EW}}^{p,h}
    - \frac{1}{2}\alpha_{4,{\rm EW}}^{p,h}
    \nonumber\\[-0.1cm]
    &&\hspace*{1cm}
 + \beta_3^{p,h}- \,\frac{1}{2}\beta_{3,{\rm EW}}^{p,h} -
\frac{3}{2}\beta_{4,{\rm EW}}^{p,h} \Big]
     \overline{X}^{(\overline B\omega, a_2)}_h \Bigg\}\,,   \\
 {\cal A}_{B^-\to a_2^- \phi}^h
   &=&  \frac{G_F}{\sqrt{2}}\sum_{p=u,c}\lambda_p^{(d)}
   \Bigg\{ \left[
    \alpha_3^{p,h} - \frac{1}{2}\alpha_{3,{\rm EW}}^{p,h}
    \right] X^{(\overline B a_2, \phi)}_h \Bigg\}\,,  \\
   -\sqrt{2}\,{\cal A}_{\bar B^0\to a_2^0\phi}^h
   &=& \frac{G_F}{\sqrt{2}}\sum_{p=u,c}\lambda_p^{(d)}
   \Bigg\{ \left[
    \alpha_3^{p,h} - \frac{1}{2}\alpha_{3,{\rm EW}}^{p,h}
    \right] X^{(\overline B a_2, \phi)}_h \Bigg\}\,,
\end{eqnarray}
and the amplitudes for $\overline{B}\to \overline{K}_2^* K^*$ can be obtained from
$\overline{B}\to \overline{K}_2^* K$ with the replacement $(\overline{K}^*_2, K) \to (\overline{K}_2^*, K^*)$ and $c=-1$.

\subsubsection{Decay amplitudes with $\Delta S=1$}

The amplitudes for $\overline{B}\to f_2 \overline K^*$ can be obtained from $\overline{B}\to f_2 \overline K$ with the replacement $(f_2, \overline{K}) \to (f_2, \overline{K}^*)$, and the ones for $\overline{B}\to \overline{K}_2^* \rho$ and $\overline{B}\to \overline{K}^* a_2$ can be obtained from $\overline{B}\to \overline{K} a_2$ with the replacement  $(\overline{K}, a_2) \to (\overline{K}_2^*, \rho)$ and $(\overline{K}, a_2) \to (\overline{K}^*, a_2)$, respectively,
\be
  \sqrt{2}{\cal A}_{B^-\to K^{*-}_2\omega}^h
   &=&  \frac{G_F}{\sqrt{2}}\sum_{p=u,c}\lambda_p^{(s)} \Bigg\{
 \Big[ \delta_{pu}(\alpha_1^h+\beta_2^h)+ \alpha_4^{p,h} +\alpha_{4,{\rm EW}}^{p,h}
 +\beta_3^{p,h} +\beta^{p,h}_{3,{\rm EW}} \Big]   \overline{X}^{(\overline B \omega,\ov K^{*}_2)}_h  \non \\
   &+& \Big[\delta_{pu}\alpha_2^h+2\alpha_3^{p,h}+{1\over 2}\alpha^{p,h}_{3,{\rm EW}}\Big]
   X^{(\overline B\ov K^{*}_2,\omega)}_h\Bigg\}, \label{eq:K2comega}   \\
  \sqrt{2}{\cal A}_{\ov B^0\to \ov K^{*0}_2\omega}^h
   &=&  \frac{G_F}{\sqrt{2}}\sum_{p=u,c}\lambda_p^{(s)} \Bigg\{
    \Big[ \alpha_4^{p,h} - {1\over 2}\alpha_{4,{\rm
    EW}}^{p,h} +\beta_3^{p,h} - {1\over 2}\beta^{p,h}_{3,{\rm EW}} \Big]
   \overline{X}^{(\overline B \omega,\ov K^{*}_2)}_h  \non \\
   &+& \Big[\delta_{pu}\alpha_2^h+ 2\alpha_3^{p,h}+ {1\over 2}\alpha^{p,h}_{3,{\rm EW}}\Big]
   X^{(\overline B \ov K^{*}_2,\omega)}_h \Bigg\},  \label{eq:K2nomega}  \\
  {\cal A}_{B^-\to K^{*-}_2\phi}^h
   &=&  \frac{G_F}{\sqrt{2}}\sum_{p=u,c}\lambda_p^{(s)}
    \Big[ \delta_{pu}\beta_2^h+ \alpha_3^{p,h}+ \alpha_4^{p,h} -{1\over 2}\alpha_{3,{\rm EW}}^{p,h} \non \\
&& -{1\over 2}\alpha_{4,{\rm
    EW}}^{p,h}
    +\beta_3^{p,h} +\beta^{p,h}_{3,{\rm EW}} \Big]
 X^{(\overline B \ov K^{*}_2,\phi)}_h, \label{eq:K2cphi} \\
    {\cal A}_{\ov B^0\to \ov K^{*0}_2\phi}^h
   &=&  \frac{G_F}{\sqrt{2}}\sum_{p=u,c}\lambda_p^{(s)}
    \Big[\alpha_3^{p,h}+ \alpha_4^{p,h} -{1\over 2}\alpha_{3,{\rm EW}}^{p,h} -{1\over 2}\alpha_{4,{\rm
    EW}}^{p,h} \non \\
    && +\beta_3^{p,h} -{1\over 2}\beta^{p,h}_{3,{\rm EW}} \Big]
 X^{(\overline B \ov K^{*}_2,\phi)}_h. \label{eq:K2nphi}
 \en

\section{Explicit expressions of annihilation amplitudes}
The general expressions of the helicity-dependent annihilation amplitudes are
given in Eqs. (\ref{eq:A1i0})-(\ref{eq:A3fm}). They can be further
simplified by considering the asymptotic distribution amplitudes
for $\Phi_V,\Phi_v,\Phi_T$ and $\Phi_t$:
 \be
&& \Phi^V_{\parallel,\perp}(u)=6u\bar u, \qquad
 \Phi_v(u)=3(2u-1), \non \\ && \Phi^T_{\parallel,\bot}(u)=30u\bar u(2u-1), \qquad \Phi_t(u)=5(1-6u+6u^2), \non \\
&& \Phi^M_+(u)=\int^1_u dv{\Phi^M_\parallel(v)\over v}, \qquad
\Phi^M_-(u)=\int^u_0 dv{\Phi^M_\parallel(v)\over \bar v}.
 \en
We find
\begin{eqnarray}
  A_3^{f,\,0} (V\,T) &\approx&  -30\sqrt{2\over 3} \pi\alpha_s
 \left[
 \left(6{X_A^0}^2-23X_A^0+22\right) r_\chi^{V} + {3\over 2}(2X_A^0-1)(X_A^0-3)r_\chi^T
 \right]
 ,\label{eq:vt-A3f0}
 \\
  A_3^{f,\,-} (V\, T) &\approx& -{10\over\sqrt{2}} \pi\alpha_s
 \left[
  \left(6{X_A^-}^2-23X_A^-+17\right)\frac{m_T}{m_{V}}r_\chi^V +
  9(2X_A^--3)(X_A^--2)\frac{m_V}{m_T}r_\chi^T
 \right], \label{eq:vt-A3fm}
  \\
 A_3^{f,\,0}( T\, V) & =& -A_3^{f,\,0}(V\,T), \qquad
  A_3^{f,\,-}( T\, V) = -A_3^{f,\,-}(V\, T), \label{eq:tv-A3f}
 \\
   A_3^{i,\,0} (V\, T) &\approx&  30\sqrt{2\over 3} \pi\alpha_s
 \left[ - 3\left({X_A^0}^2-4X_A^0 -4 +\pi^2 \right)r_\chi^{V} +
 \frac{3}{2}\left({X_A^0}^2-2X_A^0-6 + \frac{\pi^2}{3} \right) r_\chi^T \right]
   ,\label{eq:vt-A3i0} \\
  A_3^{i,\,-} (V\, T) &\approx& - {30\over\sqrt{2}}\pi\alpha_s
 \Bigg[
  - \left({X_A^-}^2-2X_A^--2\right)\frac{m_T}{m_V}r_\chi^V +
  \left(3{X_A^-}^2-12X_A^-+2\pi^2\right)\frac{m_V}{m_T}r_\chi^T
 \Bigg] , \label{eq:vt-A3im}
 \\
  A_3^{i,\,0}( T\, V) &=& A_3^{i,\,0}(V\, T),
\qquad
  A_3^{i,\,-}( T\, V) = A_3^{i,\,-}(V\, T),\label{eq:vt-A3i}
 \\
  A_{1}^{i,\,0} (V\,T) &\approx&  30\sqrt{2\over 3} \pi\alpha_s
 \left[ 3X_A^0+4-\pi^2+{3\over 2}(X_A^0-3)(X_A^0-2)r_\chi^V r_\chi^T \right],\label{eq:tv-A3i0}
 \\
   A_{1}^{i,\,0} (T\,V) &\approx&  -30\sqrt{2\over 3} \pi\alpha_s
 \left[X_A^0+29-3\pi^2+ {3\over 2}(X_A^0-3)(X_A^0-2)r_\chi^V r_\chi^T \right],
  \\
 A_2^{i,\,0}( V\, T) &=& -A_1^{i,\,0}(T\,V), \qquad
  A_2^{i,\,0}( T\, V) = -A_1^{i,\,0}(V\, T), \label{eq:tv-Ai2}
\end{eqnarray}
for $TV$ modes, and
\begin{eqnarray}
   A_3^{f} (P\,T) &\approx& \sqrt{2\over 3} \pi\alpha_s
 \left[  10 X_A (6X_A-11) r_\chi^{P} + 45 (2X_A-1)(X_A-3)r_\chi^T \right],\label{eq:pt-A3f}
 \\
  A_3^{i} (P\, T) &\approx&  30\sqrt{2\over 3} \pi\alpha_s
 \left[ \left({X_A}^2-4X_A  +4 + \frac{\pi^2}{3}\right)r_\chi^{P}
 + \frac{3}{2}\left({X_A}^2-2X_A-6 + \frac{\pi^2}{3}\right) r_\chi^T \right],\label{eq:pt-A3i}
 \\
  A_{1}^{i} (P\,T) &\approx&  10\sqrt{2\over 3} \pi\alpha_s
 \left[ 3(3X_A +4-\pi^2) + {3\over 2} X_A (X_A-3) r_\chi^P r_\chi^T \right], \label{eq:pt-a1i}
 \\
  A_{1}^{i} (T\,P) &\approx&  -10\sqrt{2\over 3} \pi\alpha_s
 \left[ 3(X_A + 29 - 3\pi^2) + {3\over 2} X_A (X_A-3) r_\chi^P r_\chi^T \right], \label{eq:tp-a1i}
 \\
 A_3^{f}( P\, T) & =& A_3^{f}(T\,P), \qquad
 A_3^{i}( T\, P) = A_3^{i}(P\, T),
 \\
 A_2^{i}( T\, P) &=& A_1^{i}(P\, T),  \qquad
 A_2^{i}( P\, T) = A_1^{i}(T\, P),
 \end{eqnarray}
for $TP$ modes.
As pointed out in \cite{Cheng:2008gxa}, since the annihilation contributions $A_{1,2}^{i,\pm}$ are
suppressed by a factor of $m_1m_2/m_B^2$ relative to other terms,
in numerical analysis we will consider only the annihilation
contributions due to $A_3^{f,0}$, $A_3^{f,-}$, $A_{1,2,3}^{i,0}$ and $A_3^{i,-}$.

The logarithmic divergences that occurred in weak annihilation in above equations
are described by the variable $X_A$
 \be
 \int_0^1 {du\over u}\to X_A, \qquad \int_0^1 du {\ln u\over u}\to -{1\over 2} (X_A)^2.
 \en
Following \cite{BBNS}, these variables are parameterized in Eq. (\ref{eq:XA}) in terms of
the unknown real parameters $\rho_A$ and $\phi_A$. For simplicity,  we shall assume in practical calculations that $X_A^h$ are helicity independent, $X_A^-=X_A^+=X_A^0$.

\section{The $\eta-\eta'$ system}\label{app:eta-etap}

For the $\eta$ and $\eta'$ particles,
it is more convenient to work with the flavor states   $q\bar q\equiv (u\bar u+d\bar
d)/\sqrt{2}$, $s\bar s$ and $c\bar c$ labeled by the $\eta_q$, $\eta_s$ and $\eta_{c}^0$, respectively. Neglecting the small mixing with $\eta_c^0$, we write
\begin{equation}\label{eq:qsmixing}
   \left( \begin{array}{c}
    |\eta\rangle \\ |\eta'\rangle
   \end{array} \right)
   = \left( \begin{array}{ccc}
    \cos\phi & -\sin\phi \\
    \sin\phi & \cos\phi
   \end{array} \right)
   \left( \begin{array}{c}
    |\eta_q\rangle \\ |\eta_s\rangle
   \end{array} \right) \;,
\end{equation}
where $\phi=(39.3\pm1.0)^\circ$ \cite{FKS} is the $\eta-\eta'$ mixing angle in the $\eta_q$ and $\eta_s$ flavor basis. \footnote{A different mixing angle $\phi=(35.9\pm3.4)^\circ$ was obtained recently in the analysis of \cite{Pham} based on vector meson radiative decays.}
Decay constants $f^{q}_{\eta^{(')}}$, $f_{\eta^{(')}}^{s}$ and $f_{\eta^{(')}}^c$ are defined by
\be
\la 0|\bar q\gamma_\mu\gamma_5q|\eta^{(')}\ra=i{1\over\sqrt{2}}f_{\eta^{(')}}^q  p_\mu, \quad \la 0|\bar s\gamma_\mu\gamma_5s|\eta^{(')}\ra=if_{\eta^{(')}}^s p_\mu, \quad \la 0|\bar c\gamma_\mu\gamma_5c|\eta^{(')}\ra=if_{\eta^{(')}}^c p_\mu,
\en
while the widely studied decay constants
$f_q$ and $f_s$ are defined as \cite{FKS}
\begin{eqnarray}
   \langle 0|\bar q\gamma^\mu\gamma_5 q|\eta_q\rangle
   &=& \frac{i}{\sqrt2}\,f_q\,p^\mu \;, \qquad
   \langle 0|\bar s\gamma^\mu\gamma_5 s|\eta_s\rangle
   = i f_s\,p^\mu \;.\label{deffq}
\end{eqnarray}
The ansatz made by Feldmann, Kroll and Stech (FKS) \cite{FKS} is that the decay constants in the quark flavor basis follow the same pattern of $\eta-\eta'$ mixing given in Eq. (\ref{eq:qsmixing})
\begin{eqnarray} \label{eq:ansatz}
\left(
\begin{array}{cc}
f_\eta^q & f_\eta^s \\
f_{\eta'}^q & f_{\eta'}^s
\end{array} \right)=
\left( \begin{array}{ccc}
    \cos\phi & -\sin\phi \\
    \sin\phi & \cos\phi
   \end{array} \right) \left(
\begin{array}{cc}
f_q & 0 \\
0 & f_s
\end{array} \right).
\end{eqnarray}
Empirically, this ansatz works very well \cite{FKS}. Theoretically, it has been shown recently that this assumption can be justified in the large-$N_c$ approach  \cite{Mathieu:2010ss}.

It is useful to consider the matrix elements of pseudoscalar densities \cite{BenekeETA}
\be \label{eq:hetaqs}
2m_q\la 0 |\bar q\gamma_5q|\eta^{(')}\ra={i\over\sqrt{2}}h^q_{\eta^{(')}}, \qquad
2m_s\la 0 |\bar s\gamma_5s|\eta^{(')}\ra={i}h^s_{\eta^{(')}},
\en
and define the parameters $h_q$ and $h_s$ in analogue to $f_q$ and $f_s$
\begin{eqnarray}
   2m_q\langle 0|\bar q\gamma_5 q|\eta_q\rangle
   = \frac{i}{\sqrt2}\,h_q \;, \qquad
   2m_s \langle 0|\bar s\gamma_5 s|\eta_s\rangle
   &=& i h_s\;,\label{deffq-2}
\end{eqnarray}
and relate them to $h_{\eta,\eta'}^{q,s}$ by the similar FKS ansatz as in Eq. (\ref{eq:ansatz}).

In this work, we shall follow \cite{BN} to use
\be   \label{eq:fetaqs}
&& h_{\eta}^q=0.0013\,{\rm GeV}^3, \quad h_{\eta}^s=-0.0555\,{\rm GeV}^3, \quad h_{\eta'}^q=0.0011\,{\rm GeV}^3, \quad h_{\eta'}^s=0.068\,{\rm GeV}^3, \non \\
&& f_{\eta}^q=109\,{\rm
MeV}, \qquad~~ f_{\eta}^s=-111\,{\rm MeV}, \qquad~ f_{\eta'}^q= 89\,{\rm
MeV}, \qquad\quad~ f_{\eta'}^s=136\,{\rm MeV}, \non \\
&& f_{\eta}^c=-2.3\,{\rm MeV}, \qquad f_{\eta'}^c=-5.8\,{\rm MeV}, \qquad~~m_{\eta_q}=741\,{\rm MeV}, \qquad m_{\eta_s}=802\,{\rm MeV},
\en
where we have used the perturbative result \cite{fetac}
\be \label{eq:fetac}
f_{\eta^{(')}}^c=-{m_{\eta^{(')}}^2\over 12 m_c^2}\,{f_{\eta^{(')}}^q\over\sqrt{2}}.
\en

The form factors for $B\to \eta^{(')}$ transitions obtained in QCD sum rules are \cite{Ball:Beta}
\be
F^{B\eta}_{0,1}(0) &=& 0.229\pm0.024\pm0.011, \non \\
F^{B\eta'}_{0,1}(0) &=& 0.188\pm0.002B_2^g\pm0.019\pm0.009,
\en
where the flavor-singlet contribution to the $B\to\eta^{(')}$ form factors is characterized by the parameter $B_2^g$, a gluonic Gegenbauer moment. It appears that the singlet contribution to the form factor is small unless $B_2^g$ assumes extreme values $\sim 40$ \cite{Ball:Beta}. Using the relation
\be \label{eq:FFeta}
F^{B\eta}_{0,1}=F^{B\eta_q}_{0,1}\cos\phi, \quad
F^{B\eta'}_{0,1}=F^{B\eta_q}_{0,1}\sin\phi,
\en
we obtain $F^{B\eta_q}_{0,1}(0)=0.296\pm 0.028$ as shown in Table \ref{tab:input}. The momentum dependence of the form factor can be found in \cite{Ball:Beta}.

%%%%%%%%%%%%%%%%%%%%%%%%%%%%%%%%%%%%%%%%%%%%%%%%%%%%%%%%

\end{document}